\documentclass{lmcs}

\usepackage[all]{xy}
\usepackage{adjustbox}
\usepackage{ebproof}
\usepackage{enumitem}
\usepackage{float}
\usepackage{hyperref, colortbl}
\usepackage{amssymb}
\usepackage{mathpartir}
\usepackage{mathtools}
\usepackage{multicol}
\usepackage{tikz}
\usepackage{multirow}
\usepackage{stmaryrd}
\usepackage{textcomp}
\usepackage{xcolor}
\usepackage{xspace}
\usepackage[nameinlink]{cleveref}
\usetikzlibrary{shapes.geometric}
\usetikzlibrary{positioning}
\usetikzlibrary{calc}
\usetikzlibrary{arrows.meta}
\usepackage{hyperref}

\usepackage{iwilare}

\createrule{J}{$\J$}
\createrule{refl}{\textsf{refl}}

\createrule{axiom}{\textsf{axiom}}
\createrule{axiomminus}{$\textsf{axiom}^-$}
\createrule{axiomplus}{$\textsf{axiom}^+$}
\createrule{reindexwop}{$\textsf{reindex}'$}

\createrule{cut}{\textsf{cut}}
\createrule{hyp}{\textsf{hyp}}
\createrule{id}{\textsf{id}}
\createrule{op}{\textsf{op}}
\createrule{refl}{\textsf{refl}}
\createrule{reindex}{\textsf{reindex}}
\createrule{struct}{\textsf{struct}}
\createrule{subst}{\textsf{subst}}

\usepackage{tikz}
\usepackage{everypage}

\createrule{le}{$\le$}
\createrule{leinv}{$\le^{-1}$}
\createrule{leminus}{$\le^-$}
\createrule{leplus}{$\le^+$}
\createrule{lerefl}{${\le}{\text -}\refl$}
\createrule{lefull}{$\le^\Phi$}
\createrule{lereflterm}{${\le}{\text -}\refl_t$}
\createrule{leterm}{$\le^\Phi_t$}

\createrule{implvarchange}{$\Rightarrow^+_-$}

\createrule{existstermminus}{$\exists^-_t$}
\createrule{existstermdelta}{$\exists^\Delta_t$}
\createrule{existstermplus}{$\exists^+_t$}

\createrule{foralltermminus}{$\forall^-_t$}
\createrule{foralltermdelta}{$\forall^\Delta_t$}
\createrule{foralltermplus}{$\forall^+_t$}

\createrule{foralldelta}{$\forall^\Delta$}

\createrule{iff}{$\Leftrightarrow$}

\createrule{exists}{$\exists$}
\createrule{forall}{$\forall$}
\createrule{and}{$\land$}
\createrule{or}{$\lor$}
\createrule{top}{$\top$}
\createrule{bot}{$\bot$}

\createrule{dinatcollapse}{$\contract{A}$}
\createrule{dinatlift}{$\lift{A,B}$}
\createrule{upgrade}{\textsf{upgrade-to-nat}}

\createrule{impl}{$\Rightarrow$}
\createrule{implL}{$\Rightarrow_L$}
\createrule{implR}{$\Rightarrow_R$}
\createrule{implplusl}{$\Rightarrow^+_L$}
\createrule{implplusr}{$\Rightarrow^+_R$}
\createrule{implminusl}{$\Rightarrow^-_L$}
\createrule{implminusr}{$\Rightarrow^-_R$}
\createrule{implinvert}{$\Rightarrow^{\pm}$}

\Crefname{defi}{Definition}{Definition}
\Crefname{prop}{Proposition}{Proposition}
\Crefname{thm}{Theorem}{Theorem}
\Crefname{cor}{Corollary}{Corollary}
\Crefname{lem}{Lemma}{Lemma}
\Crefname{exa}{Example}{Example}
\Crefname{rem}{Remark}{Remark}

\def\RuledinatliftSPECIALA{\hyperref[rule:dinatlift]{{\textnormal{($\lift{A}$)}}}}

\usepackage{eso-pic}
\usepackage{xcolor}
\begin{document}

\title[Doctrinal Semantics of Directed First-Order Logic]{Doctrinal Semantics of Directed First-Order Logic}
\thanks{Loregian was supported by the Estonian Research
Council grant PRG1210. Veltri was supported by the Estonian Research Council grant PSG749.}

\author[A.~Laretto]{Andrea Laretto\lmcsorcid{0000-0002-6413-5794}}[a]
\author[F.~Loregian]{Fosco Loregian\lmcsorcid{0000-0003-3052-465X}}[a]
\author[N.~Veltri]{Niccolò Veltri\lmcsorcid{0000-0002-7230-3436}}[a]

\address{Tallinn University of Technology, Tallinn, Estonia}
\email{andrea.laretto@taltech.ee, fosco.loregian@taltech.ee, niccolo.veltri@taltech.ee}
\begin{abstract}
We present a first-order logic equipped with an ``asymmetric'' directed notion of equality, which can be thought of as rewrites between terms, allowing for types to be interpreted as preorders. The logic is equipped with a precise syntactic system of polarities, inspired by dinaturality, that keeps track of the occurrence of variables (positive/negative/both). We use this to give a characterization of directed equality as a relative left adjoint, generalizing the idea by Lawvere of equality as left adjoint; intuitively, the relativeness is used to capture the syntactic restriction that avoids symmetry of equality. The semantics of this logic and its system of variances is captured categorically using the notion of directed doctrine, which we prove sound and complete with respect to the syntax. Moreover, we prove that the classical fragment of our directed logic is complete with respect to a standard semantics in preorders.
\end{abstract}

\maketitle

\section{Introduction}

In a series of seminal papers, Lawvere~\cite{Lawvere1969adjointness,Lawvere1969diagonal,Lawvere1970equality} unveiled the deep connection between logic and the algebraic perspective of category theory~\cite{Awodey1997logic,Leinster2014basic}. One of the key insights of this connection was to view the \emph{logical connectives} of first-order logic, such as truth $\top$, conjunction $\land$, implication $\Rightarrow$, all as instances of the single general notion of \emph{adjunction}~\cite{MacLane1998categories}, an elementary category-theoretical definition. These adjunctions appear in the notion of cartesian closed category~\cite{Lawvere1969adjointness}, which acts as the common framework in which propositions, types, proofs, and programs can be unified via the Curry-Howard-Lambek correspondence~\cite{Lambek1986introduction,Crole1993deriving,Jacobs1999categorical}.

Remarkably, this rephrasing of logical ideas as adjoints extends beyond the propositional connectives: \emph{quantifiers} $\forall$ and $\exists$ can be characterized as right and left adjoints to \emph{weakening}~\cite[4.1.8]{Jacobs1999categorical}; the same idea also applies to \emph{equality predicates} $x = y$, which can be captured as the left adjoint to \emph{contraction}~\cite[4.1.7]{Jacobs1999categorical}, i.e., the act of identifying two variables $x,y$ with a single fresh one $z$ and sending a formula $\varphi(x,y)$ to $\varphi(z,z)$.

In standard presentations of logic, equality is often taken to be just another relation symbol~\cite[5.2.1]{Mimram2020program}, \cite[\S 48]{Church1956introduction}. Under the modern ``post-Lawvere'' perspective of logic~\cite[4.1.5]{Shulman2016categorical}, equality plays a special role similar to that of a \emph{logical connective}, since its behaviour is described using exactly the same kind of rules (via adjoints) that other more fundamenta connectives are equipped with. Concretely, the fact that equality has a categorical universal property means for example that it is characterized up-to-isomorphism~\cite{Leinster2014basic}, and therefore that ``having equality'' is a \emph{property} of a logic, not arbitrary structure~\cite{Jacobs1999categorical}.

\paragraph{(The presheaf hyperdoctrine.)}

The core notion behind the categorical perspective on logic is that of \emph{doctrine}~\cite{Lawvere1969adjointness,Maietti2013quotient}, a concise category-theoretical definition which unifies both syntax and semantics of (first-order) logics. A doctrine is a category $\C$ with finite products, called the ``base'' category, equipped with a functor $P : \C^\op \to \Pos$ to the category $\Pos$ of posets and monotone functions. The intuition is that $\C$ acts as a category of \emph{types and terms}, and $P$ sends the type $A$ to the poset $P(A)$ of formulas in context $A$ with logical entailment as relation. A survey of doctrines and their extensions can be found in~\cite{Maietti2013quotient,Jacobs1999categorical}.

In the very same paper introducing the idea of equality-as-adjoint~\cite{Lawvere1970equality}, Lawvere describes a particularly important example of a doctrine that is \emph{not} equipped with such a notion of equality, namely the \emph{presheaf doctrine} $\Psh : \Cat^\op \> \CAT$, which sends a category $\C \in \Cat$ (the {category of small categories}) to the large category of presheaves $\Psh(\C) := [\Cop,\Set]$.

This doctrine plays a central role in category theory because it should intuitively capture the ``\emph{the logic of categories}'': categories and functors are viewed as \emph{types and terms} of such logic, and, as well-known among category theorists, presheaves can be thought of as \emph{generalized predicates}, in the same way that profunctors are generalized relations~\cite{Bainbridge1976feedback,Lawvere1970equality}.

We report here a particularly suggestive quote by Lawvere~\cite[p. 11]{Lawvere1970equality}, where he comments on the failure of the notion of equality as left adjoint in the setting of presheaves, and what should be done instead:

\begin{quote}
  [...]
This should not be taken as indicative of a lack of vitality of [$\Psh$] as a hyperdoctrine, or even of a lack of a satisfactory theory of equality for it. Rather, it indicates that we have probably been too naive in defining equality in a manner too closely suggested by the classical conception. [...] But present categorical conceptions indicate that [...] the graph of a functor $f : \B\to\C$ should be [...] a binary attribute of mixed variance in $\Psh(\B^\op\x\C)$. Thus in particular ``equality'' should be the functor $\hom_\B$ [...]. The term which would take the place of $\delta$ in such a more enlightened theory of equality would then be the forgetful functor $\Tw(\B) \> \Bop \x \B$ from the ``twisted morphism category'' [...]. Of course to abstract from this example would require at least the addition of a functor $T \xto{\op} T$ to the structure of a [doctrine].\\\hspace*{\fill}\cite[p. 11]{Lawvere1970equality}
\end{quote}

\noindent The key remark is that, if types are now categories, the more natural notion of equality is given by $\hom$-functors $\C^\op\times \C \to \Set$, which play the role of a \emph{directed, asymmetric equality}.

\paragraph{(Directed type theory.)} This perspective of types-as-categories first envisioned by Lawvere has recently gained new popularity under the topic of \emph{directed type theory}~\cite{Nuyts2023higher,Gratzer2024directed,Gratzer2025yoneda,Altenkirch2024synthetic,Chu2024directed,Chu2025dependent,Laretto2026di}.

The motivation behind directed type theory starts from a seminal paper of Hofmann and Streicher~\cite{Hofmann1998groupoid}, in which the types of \MLTT are interpreted as \emph{groupoids} (i.e., categories where every morphism is invertible), laying the conceptual foundation for Homotopy Type Theory \cite{Awodey2009homotopy,Kapulkin2012simplicial,UnivalentFoundationsProgram2013homotopy}: the key idea is that inhabitants of types are given by objects of groupoids, and proofs of equalities are given by the morphisms between such objects, of which there can be more than a unique one.

The need for morphisms to be invertible comes from the inherently \emph{symmetric} nature of equality: for every proof of equality $x = y$ there has to be a proof of the equality $y = x$. The typical syntactic presentation of equality in type theory~\cite{Hofmann1997syntax}, which we recall in \Cref{appendix:symmetric_equality}, consists of an introduction rule $\refl$ for reflexivity of equality $x = x$, and an elimination rule, often called $\J$-rule~\cite{Hofmann1997syntax}, that allows one to contract equalities: this last principle allows for symmetry to be \emph{derived}, hence justifying the need for groupoids.

A natural question arises: why not \emph{categories}, in\-ste\-ad of group\-oids? Such a setting should take the name of \emph{directed} type theory~\cite{North2019towards,Altenkirch2024synthetic,Licata20112}, where types are interpreted as categories, and morphisms capture an \emph{asymmetric} notion of equality (i.e. symmetry is not a theorem, since not all arrows are invertible). Such equalities can then be used to represent intrinsically directed phenomena, such as processes, transitions, or rewrites~\cite{North2019towards}.

\paragraph{(Logic and proof-irrelevance.)} In both traditional and categorical accounts of logic~\cite{Dalen2013logic}, however, equality is typically \emph{proof-irrelevant}, i.e., one does not distinguish between proofs of equality; this is the main distinction between logic and \emph{type theory} in the sense of Jacobs~\cite[8.4.11]{Jacobs1999categorical}. In the directed case, one should talk thus about \emph{directed logic}, i.e., a generalization of first-order logic with a proof-irrelevant \emph{asymmetric} notion of equality.

In the same way that groupoids generalize sets by making equality {proof relevant} (since there might be many morphisms/proofs of equalities), categories generalize \emph{preorders}: in a preorder there is no information on \emph{in what way} two objects $a,b$ are (directionally) connected by a morphism $f\!:\!a\!\>\!b$, but only whether the inequality $a \le b$ holds; this interest in proof-irrelevance is similarly shared in applications of rewriting in computer science, e.g., in rewriting logic~\cite{Meseguer2012twenty}, where either rewrites happen or not. We summarize the relationship between these elementary models in \Cref{table:models}.

\begin{figure}
  \centering
\begin{tabular}{c|c|c}
\multirow{2}{*}{\textbf{Models of logic/TT}} & \multicolumn{2}{c}{\textbf{Proofs of equality}} \\

                                       & \textbf{Irrelevant} & \textbf{Relevant} \\
\cline{2-3}
\hline
\textbf{Symmetric Equality} & \text{Sets} ($\Set$) & \text{Groupoids} ($\Gpd$) \\
\textbf{Directed Equality}  & \text{Preorders} ($\Preord$) & \text{Categories} ($\Cat$) \\
\end{tabular}
\caption{Elementary models of symmetric and directed logic.}
\label{table:models}
\end{figure}

\paragraph{(Polarity/symmetry problems.)}
\label{sec:intro:variance_and_dinaturality}
Directed type theory has proven to be not so straightforward~\cite{North2019towards,Ahrens2023bicategorical,Nuyts2023higher}; we illustrate why the n\"aive approach to the directed setting fails even in the simple case of preorders, which will be the focus of this paper.

For any type $A$ (i.e., a preorder in the semantics) there is a naturally associated type $A^\op$, called the \emph{opposite type of $A$}, where the inhabitants of the type are the same but all directed equalities are reversed, as in the notion of opposite preorder. The \emph{type of directed equalities}, often renamed $\hom$-types \cite{North2019towards}, should then be interpreted via the monotone function ${\le_A} : A^\op \x A \> \I$, which receives a ``contravariant'' argument $a : A^\op$ and a ``covariant'' one $b : A$ and returns ${\le}(a,b) := (1 \text{ if } a \le b, 0 \text{ otherwise})$, where $\I = \set{0 \to 1}$ is the preorder with two elements $0,1 \in \I$ such that $0 \le 1$. A directed logic/type theory should therefore have some notion of ``\emph{polarity}'', which allows variables to be distinguished and appear only in the appropriate position, as treated in \cite{Licata20112,New2023formal,North2019towards,Nuyts2015towards,Laretto2026di}.

However, in the statement for transitivity of directed equality $
x \le y \land y \le z \vdash x \le z$, the variable $y$ appears both on the right side of $x \le y$, with type $A$, and on the right side of $y \le z$, with seemingly different type $A^\op$! A similar problem arises for reflexivity $x \le x$.

Even if we allow variables to appear with both polarities at the same time, the problem then becomes: \emph{how do we make sure that symmetry is not derivable in the syntax}? That is, how should the rules $\refl$ and $\J$ change in such a way that symmetry cannot be derived, while still being able to recover the expected theorems of directed type theory?

\subsection{Contribution}\label{sec:contribution}
In this paper, we advance the ideas of Lawvere of equality-as-adjoint to present syntax and semantics for a first-order proof-irrelevant (i.e. logical) version of directed type theory, which we call \emph{directed first-order logic}. This logic advances the well-known canonical setting of first-order logic by equipping it with a propositional notion of directed equality, which we study here the universal property of.

We describe the key features of our logic and its semantics:

\begin{itemize}[leftmargin=1em]
\item The logic is equipped with a precise system of polarities, inspired by dinaturality~\cite{Dubuc1970dinatural}. Variables are allowed to appear anywhere in formulas, regardless of their polarity: however, the rule to eliminate directed equalities (reminiscent of the $\J$ rule) is equipped with a certain syntactic restriction on the appearance of variables, thus forbidding symmetry.

\item We introduce a generalization of the notion of doctrine, called \emph{directed doctrines}, for which the syntax is sound and complete, with the preorder model $\Preord$ as our main example: types are interpreted by preorders,\,with {$\le$} representing the existence of a directed equality. Moreover, we show by a standard Henkin technique~\cite{Dalen2013logic} that the classical fragment of our directed logic is complete with respect to this specific preorder model.

\item Directed doctrines allow us to give a characterization of directed equality, not as a left adjoint to contraction as in Lawvere's approach, but as a left \emph{relative} adjoint~\cite{Altenkirch2010monads,Ulmer1968properties} to a certain contraction-like operation that collapses two variables of \emph{different} polarity $A$, $A^\op$ into a single one of type $A$, hence generalizing and directifying the notion of symmetric equality as left adjoint to contraction first introduced by Lawvere. Intuitively, relativeness is used to capture the syntactic restriction on the appearance of variables.

\item We devise a way to capture the asymmetry of directed equality using the doctrinal approach~\cite{Maietti2013quotient}: the idea is to separate contexts between positive, negative, and dinatural variables, and then require that doctrines have specific reindexings that capture such polarity. There are other possible approaches to capture variance; the one explored here allows for the above ``contraction-like'' operation to be expressed explicitly in the doctrinal semantics.
\item The usual logical connectives are generalized to a corresponding ``directified'' version: \emph{polarized quantifiers} keep track of the variance of the variable being quantified over, and implication is generalized to \emph{polarized implication}, which reverses the polarity of all positions in the curried formula and is characterized by a \emph{relative coadjunction}~\cite{Arkor2024formal}.
\end{itemize}
\noindent Our precise treatment of polarity and directed equality via doctrines allows us to capture a logical flavour to directedness, thus progressing on the line of work first posed by Lawvere \cite{Lawvere1970equality} on the precise role of variance for the presheaf doctrine. Our work treats the proof-irrelevant case and, instead of considering the \emph{logic of categories}, we give a characterization for the \emph{logic of preorders} by studing the universal property of directed equality. The focus on the proof-irrelevant \emph{logical} case is justified by the use of \emph{posetal} fibrations in categorical logic~\cite{Maietti2015unifying,Maietti2023characterization,Bonchi2021doctrines}, and by the fact that rewriting (e.g. in the logical setting \cite{Meseguer2012twenty}) is also typically proof-irrelevant, since either a rewrite happens or it does not.

\subsection{Related work}\label{sec:related_work}
The failure of equality in the presheaf (hyper)doctrine was recently revived in a paper by Melliès and Zeilberger \cite{Mellies2016bifibrational} under the perspective of linear logic and type refinement systems.

North \cite{North2019towards,Chu2024directed} and Altenkirch and Neumann~\cite{Altenkirch2024synthetic} describe a dependent directed type theory with semantics in $\Cat$, using groupoidal structure to deal with the polarity problems in the rules for $\hom$. Here we instead focus on rewriting via preorders (as in recent work by Neumann~\cite{Neumann2025judgmental}) and the first-order case using doctrines.

Other approaches to directedness based on synthetic intervals and geometric spaces are given in \cite{Riehl2017type,Gratzer2024directed,Weaver2020constructive}; our paper focuses on syntactic aspects of a logic, generalizing the abstract doctrinal approach and focusing on the elementary model of preorders instead of using geometric spaces.

Another approach to directed equality is the judgemental one \cite{Licata20112,Ahrens2023bicategorical}, which however does not allow for contraction rules to be described and a universal property to be given. New and Licata \cite{New2023formal} give a sound and complete presentation for certain double categorical models of which categories (and therefore preorders) are an instance, but at the cost of heavily restricting the syntax (i.e., a symmetricity statement cannot be formulated).

{
Laretto et al.~\cite{Laretto2026di} describe a first-order proof-relevant type theory with semantics in $\Cat$, interpreting entailments as dinatural transformations which do not always compose in the semantics. Because of this non-compositionality, their type theory cannot be given a genuine semantics in a well-established model using, e.g., fibrations or categories with families~\cite{Jacobs1999categorical,Castellan2020categories}. In this work, we take the key idea of using dinaturality for directed type theory and apply it to generalize Lawvere's idea of equality as left adjoint in doctrinal semantics, focusing on the simpler preorder case. To this end we use a proof-irrelevant version of dinatural transformation and preorders, which always compose (hence hexagons always commute trivially), allowing us to capture our syntax using doctrinal semantics. In this framework we can state precisely how the rules of directed equality $\le_A$ correspond to a left-relative adjunction, hence providing a precise tool for the community interested in doctrines and logic (typically with posetal fibers, as in this paper) to study directedness.
Because of our focus on standard models, we reformulate a different approach for the treatment of polarity: instead of classifying variables using predicates as in \emph{op. cit.}, here we use separate contexts for positive/dinatural/negative variables, thus allowing us to extend already existing models to the directed case in a natural way.}

\subsection{Structure of the paper}

We present the syntax of directed first-order logic in \Cref{sec:syntax_of_directed_logic}, showing examples of derivations and theories in \Cref{sec:examples}. The categorical semantics is given in \Cref{sec:doctrinal_semantics}, establishing the notion of directed doctrine to capture polarity and directed equality. Syntax and semantics are finally connected in \Cref{sec:interpretation} and we prove completeness for preorders in \Cref{poset_completeness}, concluding with future work in \Cref{sec:conclusions}.

\section{Syntax of Directed Logic}\label{sec:syntax_of_directed_logic}
We introduce the syntax of directed first-order logic (dFOL) with a natural deduction-style proof system. The main syntactic judgements for types, terms, formulas and entailments are presented in \Cref{fig:syntax:types_terms,fig:syntax:formulas,fig:syntax:entailments}. As a guiding intuition, the reader can refer to \Cref{fig:semantics_overview} to see how the syntax of directed first-order logic is semantically interpreted in the preorder model.

\begin{figure*}[h]
\begin{adju}[1.0]
\begin{tabular}{l@{\ \ }ll}
\textbf{Concept} & \textbf{Preorder model} & \textbf{Judgement} \\\hline
Type $A,B,P,N,...$ & Preorder $\sem{A}$ & $A \type$ \\
Context $\Theta,\Delta,\Gamma$ & Product of preorders & $\Gamma \ctx$ \\
Term $s,t,\eta,\delta,\rho$ & Monotone function $\sem{\Gamma} \to \sem{A}$ & $\Gamma \vdash t : A$ \\
Equality of terms & Equivalence of monotone functions & $\Gamma \vdash t' = t : A$ \\
Polarized context $[\Theta \mid \Delta \mid \Gamma]$ & $\sem{\Theta \mid \Delta \mid \Gamma} := \sem{\Theta}^\op\times(\sem{\Delta}^\op\times\sem{\Delta})\times\sem{\Gamma}$ & \\
Formulas $\varphi,\psi$ & Monotone function $\sem{\varphi} := \sem{\Theta \mid \Delta \mid \Gamma} \> \I$ & $[\Theta \mid \Delta \mid \Gamma]\ \varphi \propx$ \\
Directed equality $\le_A$ & Monotone function $\le_A : \sem{A}^\op \times \sem{A} \> \I$ & \\
Implication formula $\Rightarrow$ & Monotone function $\Implication : \I^\op \times \I \> \I$ & \\
Conjunction formula $\land$ & Monotone function $\Conjunction : \I \times \I \> \I$ &  \\
Propositional context $\Phi$ & Pointwise product $\sem{\Phi} := \sem{\varphi_1} \land \sem{\varphi_2} \land ...$ & $[\Theta \mid \Delta \mid \Gamma]\ \Phi \propctx$ \\
Entailment & $\forall n,d,p.\ \sem{\Phi}(n,d,d,p) \le \sem{\varphi}(n,d,d,p)$ & $[\Theta \mid \Delta \mid \Gamma]\ \Phi \vdash \varphi$ \\
\end{tabular}
\end{adju}
\caption{Intuition for syntax and preorder semantics of directed first-order logic.}
\label{fig:semantics_overview}
\end{figure*}

The types and terms of directed first-order logic are given in \Cref{fig:syntax:types_terms}, and they are a straightforward axiomatization of simply typed $\lambda$-calculus (e.g., \cite{Pitts1995categorical}) with unit, product, and function types.

As customary in logic, we omit weakenings that place formulas in the correct context, which we later make explicit in \Cref{sec:doctrinal_semantics} for the doctrinal semantics. Judgements are mutually defined with the notion of \emph{signature/theory} \cite{Dalen2013logic} to add ``generating'' symbols for, e.g., base types, terms, propositions, and axioms.
\begin{defi}[Judgements and signatures]\label{def:sign_judg}
  We list the main judgements of our logic which are sets inductively defined by the rules in \Cref{fig:syntax:types_terms,fig:syntax:formulas,fig:syntax:entailments}.
  \begin{itemize}[leftmargin=1em]
    \item The set of \emph{type derivations} is $\set{A \type}$ is inductively defined by the judgement ``$A \type$'' given in \Cref{fig:syntax:types_terms}, along with the judgements for \emph{contexts} $\set{\Gamma \ctx}$, i.e. finite lists of types.
    \item The set of \emph{term derivations} $\set{\Gamma \vdash t : A}$, assuming $A \type$, $\Gamma \ctx$.
    \item The set of \emph{term equality judgements} $\set{\Gamma \vdash t = t' : A}$, assuming $\Gamma \vdash t, t' : A$, $A \type$, $\Gamma \ctx$.
    \item The set of \emph{formula derivations} $\set{[\Theta \mid \Delta \mid \Gamma]\propx}$ is given in \Cref{fig:syntax:formulas} and assumes $\Theta,\Delta,\Gamma\ctx$.
    \item The set of \emph{propositional contexts} $\set{[\Theta \mid \Delta \mid \Gamma]\ \Phi \propctx}$ captures finite lists of formulas in the same context, assuming $\Theta, \Delta, \Gamma \ctx$.
    \item The set of \emph{entailments} $\set{[\Theta \mid \Delta \mid \Gamma]\ \Phi \vdash \varphi}$ is defined inductively in \Cref{fig:syntax:entailments} and assumes $[\Theta \mid \Delta \mid \Gamma]\ \Phi \propctx,\ [\Theta \mid \Delta \mid \Gamma]\ \varphi \propx$. We use double lines to indicate bi-directional inference rules.
  \end{itemize}
  For each of the above judgements there is a corresponding notion of signature which adds generating symbols to the logic; we do not explicitly indicate the dependence between judgements and the signatures that come before them.
   \begin{itemize}[leftmargin=1em]
    \item A \emph{type signature} is defined as a set $\Sigma_B$ of symbols for base types.
    \item A \emph{term signature} $\Sigma_F$ is defined as a set of function symbols $\Sigma_F$, along with two functions $\dom,\cod : \Sigma_F \> \set{A \type}$ that return their (co)domain type.
    \item A \emph{term equality signature} $\Sigma_E$ is a set $\Sigma_E$ with (dependent) functions $\mathsf{eqC} : \Sigma_E\!\>\!\set{\Gamma \ctx}$, $\mathsf{eqT} : \Sigma_E\!\>\!\set{A \type}$, and $\mathsf{eqL}, \mathsf{eqR}\!:\!(e\!:\!\Sigma_E)\!\>\!\set{\mathsf{eqC}(e) \vdash t\!:\!\mathsf{eqT}(e)}$.
    \item A \emph{formula signature} $\Sigma_P$ is a set of symbols $\Sigma_P$ with functions $\mathsf{neg},\mathsf{pos} : \Sigma_P \> \set{A \type}$.
    \item \emph{Axioms} are given by a set $\Sigma_A$ of symbols and (dependent) functions ${\mathsf{actx} : \Sigma_A\!\>\!\set{\Gamma \ctx}^3}$, ${\mathsf{hyp}\!:\!(a\!:\!\Sigma_A)} {\>\!\set{[\mathsf{actx}(a)]\ \Phi \propctx}}$, and ${\mathsf{conc} : (a\!:\!\Sigma_A)\!\>\!\set{[\mathsf{actx}(a)]\propx}}$, with $\textsf{actx}$ returning triples of contexts.
   \end{itemize}
\end{defi}

\begin{figure*}

\begin{adju}[1.0]$\begin{array}{c}
  \fbox{$A \type$}
  \quad \inference{C \in \Sigma_B}{C \type}
  \quad \inferenceTwo{A \type}{B \type}{A \times B \type}
  \quad \inferenceTwo{A \type}{B \type}{A \Rightarrow B \type}
  \quad \inference{\phantom{A\!\!\type}}{\top \type}
  \\[1em]
  \fbox{$\Gamma \ctx$}
  \quad \inference{\phantom{A\!\!\type}}{[\,] \ctx}
  \quad \inferenceTwo{\Gamma \ctx}{A \type}{\Gamma,A \ctx}
  \\[1em]
  \fbox{$\Gamma \vdash t : A$}
  \quad \inference{}{\Gamma, x:A, \Gamma' \vdash x:A}
  \quad \inferenceTwo{f \in \Sigma_F}{\Gamma \vdash t : \dom(f)}{\Gamma \vdash f(t) : \cod(f)}
  \quad \inference{\phantom{\Gamma\!\!\!\vdash s : A}}{\Gamma \vdash\ ! : \top}
  \quad \inferenceTwo{\Gamma \vdash s : A}{\Gamma \vdash t : B}{\Gamma \vdash \ang{s,t} : A \times B}
  \\[1em]
  \inference{\Gamma \vdash p : A \times B}{\Gamma \vdash \pi_1(p) : A}
  \quad \inference{\Gamma \vdash p : A \times B}{\Gamma \vdash \pi_2(p) : B}
  \quad \inferenceTwo{\Gamma \vdash s : A \Rightarrow B}{\Gamma \vdash t : A}{\Gamma \vdash s \cdot t : B}
  \quad \inference{\Gamma, x:A \vdash t(x) : B}{\Gamma \vdash \lambda x.t(x) : A \Rightarrow B}
  \\[1em]
  \fbox{$\Gamma \vdash t = t' : A$}
  \quad \inference{\Gamma \vdash t : \top}{\Gamma \vdash t =\ ! : \top}
  \quad \inferenceTwo{\Gamma \vdash s : A}{\Gamma \vdash t : B}{\Gamma \vdash \pi_1(\ang{s,t}) = s : A}
  \quad \inferenceTwo{\Gamma \vdash s : A}{\Gamma \vdash t : B}{\Gamma \vdash \pi_2(\ang{s,t}) = t : B}
  \\[1em]
  \quad \inference{\Gamma \vdash p : A \times B}{\Gamma \vdash \ang{\pi_1(p),\pi_2(p)} = p : A \times B}
  \quad \inferenceTwo{\Gamma, x:A \vdash f(x) : B}{\Gamma \vdash t : A}{\Gamma \vdash (\lambda x.f(x)) \cdot t = f[x \mapsto t] : B}
  \quad \inference{\Gamma \vdash f : A \Rightarrow B}{\Gamma \vdash (\lambda x.f \cdot x) = f : A \Rightarrow B}
  \\[1em]
  \quad \inference{E \in \Sigma_E}{\mathsf{eqC}(E) \vdash \mathsf{eqL}(E) = \mathsf{eqR}(E) : \mathsf{eqT}(E)}
\end{array}$\end{adju}
\caption{Syntax of directed first-order logic -- types and terms.}
\label{fig:syntax:types_terms}
\begin{adju}[1.0]$\begin{array}{c}
\fbox{$[\Theta \mid \Delta \mid \Gamma]\ \varphi \propx$}
\\[1.5em]
\inferenceTwo{[\Theta \mid \Delta \mid \Gamma]\ \psi \propx}{[\Theta \mid \Delta \mid \Gamma]\ \varphi\propx}{[\Theta \mid \Delta \mid \Gamma]\ \psi \land \varphi \propx}
\quad \inferenceTwo{[\Theta \mid \Delta \mid \Gamma]\ \psi \propx}{[\Theta \mid \Delta \mid \Gamma]\ \varphi\propx}{[\Theta \mid \Delta \mid \Gamma]\ \psi \lor \varphi \propx}
\\[1.5em]
\quad \inference{}{[\Theta \mid \Delta \mid \Gamma]\ \bot \propx}
\quad \inference{}{[\Theta \mid \Delta \mid \Gamma]\ \top \propx}
\quad \inferenceTwo{[\Gamma \mid \Delta \mid \Theta]\ \psi \propx}{[\Theta \mid \Delta \mid \Gamma]\ \varphi\propx}{[\Theta \mid \Delta \mid \Gamma]\ \psi \Rightarrow \varphi \propx}
\\[1.5em]
\quad \inferenceTwo{\Theta, \Delta \vdash s : A}{\Gamma, \Delta \vdash t : A}{[\Theta \mid \Delta \mid \Gamma]\ s \le_A t \propx}
\quad
\begin{prooftree}
\hypo{P \in \Sigma_P}
\hypo{\Theta, \Delta \vdash s : \mathsf{neg}(P)}
\hypo{\Gamma, \Delta \vdash t : \mathsf{pos}(P)}
\infer3{[\Theta \mid \Delta \mid \Gamma]\ P(s \mid t) \propx}
\end{prooftree}
\\[1.5em]
\quad \inferenceTwo{p \in \set{-,\Delta,+}}{[\Theta \mid \Delta \mid \Gamma],[x:^p A]\ \varphi(x) \propx}{[\Theta \mid \Delta \mid \Gamma]\ \exists^p x.\varphi(x) \propx}
\quad \inferenceTwo{p \in \set{-,\Delta,+}}{[\Theta \mid \Delta \mid \Gamma],[x:^p A]\ \varphi(x) \propx}{[\Theta \mid \Delta \mid \Gamma]\ \forall^p x.\varphi(x) \propx}
  \\[1.5em]
  \fbox{$[\Theta \mid \Delta \mid \Gamma]\ \Phi \propctx$}
  \quad \inference{\phantom{A\!\!\type}}{[\,] \propctx}
  \quad \inferenceTwo{[\Theta \mid \Delta \mid \Gamma]\ \varphi \propx}{[\Theta \mid \Delta \mid \Gamma]\ \Phi \propctx}{[\Theta \mid \Delta \mid \Gamma]\ \varphi,\Phi \propctx}
\end{array}$\end{adju}
\caption{Syntax of directed first-order logic -- formulas and propositional contexts.}
\label{fig:syntax:formulas}
\end{figure*}

\begin{figure*}
\begin{adju}[1.0]$\begin{array}{c}
\fbox{$[\Theta \mid \Delta \mid \Gamma]\ \Phi \vdash \varphi$}
\\[1.5em]
\begin{prooftree}
\hypo{[\Theta \mid \Delta \mid \Gamma]\ \Psi \vdash \psi}
\hypo{[\Theta \mid \Delta \mid \Gamma]\ \Phi,\psi,\Phi' \vdash \varphi}
\infer2[\Rulecut]{[\Theta \mid \Delta \mid \Gamma]\ \Phi,\Psi,\Phi' \vdash \varphi}
\end{prooftree}
\quad
\begin{prooftree}
\hypo{A \in \Sigma_A}
\infer1[\Ruleaxiom]{[\mathsf{actx}(A)]\ \mathsf{hyp}(A) \vdash \mathsf{conc}(A)}
\end{prooftree}
\\[2.1em]
\begin{prooftree}
\infer0[\Rulehyp]{[\Theta \mid \Delta \mid \Gamma]\ \Phi,\varphi,\Phi' \vdash \varphi}
\end{prooftree}
\quad
\begin{prooftree}
\hypo{\Theta,\Delta & \vdash \eta : N}
\hypo{\Delta & \vdash \delta : D}
\hypo{\Gamma,\Delta & \vdash \rho : P}
\infer[no rule]3{[\Theta,n\!:\!N \mid \Delta,d\!:\!D \mid \Gamma,p\!:\!P]\ \Phi(n,\n d,d,p) \vdash \varphi(n,\n d,d,p)}
\infer1[\Rulereindex]{[\Theta \mid \Delta \mid \Gamma]\ \Phi(\eta,\delta,\delta,\rho) \vdash \varphi(\eta,\delta,\delta,\rho)}
\end{prooftree}
\\[2.1em]
\begin{prooftree}
\hypo{\sigma : \set{1,...,m} \> \set{1,...,n}}
\infer[no rule]1{[\Theta \mid \Delta \mid \Gamma]\ \varphi_{\sigma(1)},...,\varphi_{\sigma(m)} \vdash \psi}
\infer1[\Rulestruct]{[\Theta \mid \Delta \mid \Gamma]\ \varphi_1, ..., \varphi_n \vdash \psi}
\end{prooftree}
\quad
\begin{prooftree}
\infer[no rule]0{[\Theta, x\!:\!N \mid \Delta \mid \Gamma, y\!:\!P]\ {\Phi \propctx,\,\varphi \propx}}
\infer[no rule]1{[\Theta \mid \Delta,x\!:\!N,y\!:\!P \mid \Gamma]\ \Phi(x,y) \vdash \varphi(x,y)}
\infer1[\Ruleupgrade]{[\Theta,x\!:\!N \mid \Delta \mid \Gamma,y\!:\!P]\ \Phi(x,y) \vdash \varphi(x,y)}
\end{prooftree}
\\[2.6em]
\begin{prooftree}
\infer0[\Ruletop]{[\Theta \mid \Delta \mid \Gamma]\ \Phi \vdash \top}
\end{prooftree}
\quad
\begin{prooftree}
\infer0[\Rulebot]{[\Theta \mid \Delta \mid \Gamma]\ \Phi,\bot \vdash \varphi}
\end{prooftree}
\\[1.8em]
\begin{prooftree}
\hypo{[\Theta \mid \Delta \mid \Gamma]\ \Phi \vdash \psi}
\hypo{[\Theta \mid \Delta \mid \Gamma]\ \Phi \vdash \varphi}
\infer[double]2[\Ruleand]{[\Theta \mid \Delta \mid \Gamma]\ \Phi \vdash \psi \land \varphi}
\end{prooftree}
\quad
\begin{prooftree}
\hypo{[\Theta \mid \Delta \mid \Gamma]\ \psi, \Phi \vdash \varphi}
\hypo{[\Theta \mid \Delta \mid \Gamma]\ \phi, \Phi \vdash \varphi}
\infer[double]2[\Ruleor]{[\Theta \mid \Delta \mid \Gamma]\ \psi \lor \phi, \Phi \vdash \varphi}
\end{prooftree}
\\[1.5em]
\begin{prooftree}
\hypo{[\Theta \mid \Delta, z:A \mid \Gamma]\ \takespace{a \le b,\Phi}{\Phi} & \vdash \varphi(\nzz)}
\infer[double]1[\Rulele]{[\Theta,a:A \mid \Delta \mid \Gamma,b:A]\ {a \le b,\Phi} & \vdash \varphi(a,b)}
\end{prooftree}
\quad
\begin{prooftree}
\infer[no rule]0{\hspace{-4em}[\Gamma \mid \Delta \mid \Theta]\ \takespace[l]{\Phi \propctx,\,\varphi \propx}{\psi \propx}}
\infer[no rule]1{\hspace{-4em}[\Theta \mid \Delta \mid \Gamma]\ \Phi \propctx,\,\varphi \propx}
\infer[no rule]1{[\,\emptyctx \mid \Delta,\Theta,\Gamma \mid \emptyctx\,]\ \psi, \Phi & \vdash \varphi}
\infer[double]1[\Ruleimpl]{[\Theta \mid \Delta \mid \Gamma]\ \takespace{\psi, \Phi}{\Phi} & \vdash \psi \Rightarrow \varphi}
\end{prooftree}
\\[3.4em]
\begin{prooftree}
\hypo{p \in \set{-,\Delta,+}}
\hypo{[\Theta \mid \Delta \mid \Gamma]\ \exists^p x.\psi(x),\Phi \vdash \varphi}
\infer[double]2[\Ruleexists]{[\Theta \mid \Delta \mid \Gamma],[x:^p\!A]\ \psi(x),\Phi \vdash \varphi}
\end{prooftree}
\quad \begin{prooftree}
\hypo{p \in \set{-,\Delta,+}}
\hypo{[\Theta \mid \Delta \mid \Gamma],[x:^p\!A]\ \Phi \vdash \varphi(x)}
\infer[double]2[\Ruleforall]{[\Theta \mid \Delta \mid \Gamma]\ \Phi \vdash \forall^p x.\varphi(x)}
\end{prooftree}
\end{array}$\end{adju}
\caption{Syntax of directed first-order logic -- entailments.}
\LinkRulecut
\LinkRuleaxiom
\LinkRulehyp
\LinkRuletop
\LinkRuleand
\LinkRuleor
\LinkRuleop
\LinkRulebot
\LinkRulele
\LinkRuleimpl
\LinkRuleexists
\LinkRuleforall
\LinkRuleupgrade
\LinkRulestruct
\LinkRulereindex
\label{fig:syntax:entailments}
\end{figure*}

\begin{defi}[Language and theory]\label{def:theory}
We define a \emph{language} $\Sigma_L := (\Sigma_B,\Sigma_F,\Sigma_E,\Sigma_P)$ to be a tuple which collects data from the previous signatures. A \emph{theory} $\Sigma = (\Sigma_L,\Sigma_A)$ consists of a language $\Sigma_L$ with a set of axiom symbols $\Sigma_A$ defined as in \Cref{def:sign_judg}; moreover, we ask that $\Sigma_A$ is closed under syntactic entailment \cite[3.2.5]{Jacobs1999categorical}.
\end{defi}

We now formally introduce the concepts of \emph{position, variance, and polarity} with their notation.

\begin{defi}[Positions in a formula]
We use the name \emph{position} to indicate any point in which a (term) variable can appear in a formula, e.g., there are four positions $x,y,z,w$ for variables to appear in the formula ``$x \le y \land P(z,f(w))$''.
\end{defi}
\begin{defi}[Polarity of a position]
Positions have a \emph{polarity}, which can either be \emph{positive} or \emph{negative}: intuitively, a position starts out as positive, and flips between being positive and negative precisely in the following cases:
\begin{enumerate}[leftmargin=1.5em]
\item when it occurs on the left of the formula $x \le y$: e.g., the variable $x$ indicates a negative position in $x \le c$ and $f(x) \le y$;
\item when it occurs on the left of an implication formula $\psi \Rightarrow \varphi$, e.g., the position $x$ is negative in $P(x) \Rightarrow \varphi(y)$ and $Q(f(x)) \Rightarrow \top$;
\item when it occurs on the negative side $n$ of a predicate symbol $P(n \mid p)$, e.g., $x$ is negative in $P(x \mid q)$ and $P(f(g(x),y) \mid f(z))$.
\end{enumerate}
Polarity can be inverted twice, e.g., $x$ occurs \emph{positively} in $x\le y \Rightarrow\!\varphi$ and  $(y \le x \Rightarrow\!\varphi)\!\Rightarrow\!\varphi$. This idea is made precise in the judgement rules for formulas in \Cref{fig:syntax:formulas}.
\end{defi}
As we will see in \Cref{posetal_semantics}, in the preorder semantics this flipping of polarity corresponds with the presence of the opposite preorder $P^\op$ on the left side of functors ${\le_P} : P^\op \x P\>\I$, and $\Implication := {\le_\I} : \I^\op \times \I \> \I$ in particular. The fact that $(P^\op)^\op \equiv P$ justifies how inverting a negative variable makes it positive again.
\begin{defi}[Variance of a variable]
Variables can occur in multiple positions at the same time. The \emph{variance} of a variable $x$ is defined to be either \emph{positive}, \emph{negative}, or \emph{dinatural}:
\begin{itemize}[leftmargin=1em]
  \item \emph{positive} iff every position where $x$ occurs is positive;
  \item \emph{negative} iff every position where $x$ occurs is negative;
  \item \emph{dinatural} {iff\hspace{-0.3pt} every\hspace{-0.3pt} position\hspace{-0.3pt} where\hspace{-0.3pt} $x$\hspace{-0.3pt} occurs\hspace{-0.3pt} is\hspace{-0.3pt} positive\hspace{-0.3pt} or\hspace{-0.3pt} negative (i.e., always).}
\end{itemize}
We will see in \Cref{thm:reindex_lift} how indeed \emph{any} variable can be lifted to be dinatural. In practice, we use the more intuitive idea that a variable $x$ is called ``(strictly) dinatural'' iff it is neither positive nor negative, i.e., it occurs at least once with both polarities. Variables that are positive \emph{or} negative are called \emph{natural}. We denote the set of variances as $\set{-,\Delta,+}$. More in general, given a variance $p \in \set{-,+,\Delta}$, we define $p^\op$ to be the opposite variance in the intuitive way: $\Delta^\op := \Delta$, and $(+)^\op$ := $-$, $(-)^\op$ := $+$.
\end{defi}

We will indicate with $\overline a:A$ in formulas $\varphi(\overline a,a,\overline b,b,...)$ to highlight whenever a dinatural variable is used with a different polarity. These are just for convenience and do not exist syntactically. Positive and negative variables will never have an overline, since they always unambiguously appear with the same variance.

The main technical tool that captures this idea of variance is the notion of \emph{polarized context}, which simply consists of physically dividing the variables in the context based on their variance. This technique is particularly well suited to \emph{doctrinal} semantics, which places specific emphasis on the role of contexts and free variables for formulas \cite{Jacobs1999categorical}.
\begin{defi}[Polarized context]\label{def:polarized_context}
  A \emph{polarized context} is a triple of contexts $[\Theta \mid \Delta \mid \Gamma]$ for which $\Theta,\Delta,\Gamma\ctx$, where, intuitively, $\Theta$ is a list of variables that can be used only negatively, variables in $\Delta$ are dinatural (i.e., can be used either positively or negatively in any position), and variables in $\Gamma$ only positively. The polarized context \makebox{$[\Theta \mid \Delta \mid \Gamma],[x :^p A]$} indicates the context extension with a variable $x\!:\!A$, which is added to the corresponding context with variance $p \in \set{-,\Delta,+}$.
\end{defi}

We denote context concatenation as $\Gamma,\Delta$ for $\Gamma \ctx$, $\Delta \ctx$ and use $\emptyctx$ for the empty context.
We will use the notation $[\Theta,n:N \mid \Delta,d:D \mid \Gamma,p:P]\ \varphi(n,\n d,d,p)$ to indicate (some of) the free variables of the formula $\varphi$, omitting other variables in $\Theta,\Delta,\Gamma$ for brevity.

Polarized contexts are used in the formation rule of base formulas in \Cref{fig:syntax:formulas} for directed equality formulas $s \le_A t$ and base predicates $P(s \mid t)$, for terms $s,t$. The idea is that $s$ is typed in context $\Theta,\Delta$, and $t$ in context $\Gamma,\Delta$: intuitively, positive positions in formulas can be filled \emph{either} by a \emph{positive} variable, or by a \emph{dinatural} one, i.e., in the term ${\Theta,\Delta \vdash x : A}$, either ${(x:A) \in \Theta}$ or ${(x:A) \in \Delta}$. However, a dinatural variable can only be replaced by another dinatural variable, since it must appear twice. This is captured by the following definition for substitution of terms inside formulas, later used in \Cref{sec:doctrinal_semantics}.

\begin{defi}[Substitution of terms in formulas]\label{def:substitution_in_formulas} A \emph{term substitution} $\Gamma \vdash \sigma : \Gamma'$ is defined as a list of terms $\Gamma \vdash t : A_i$ for $A_i \in \Gamma' = [A_1,...,A_n]$. Given three substitutions $(\Theta',\Delta'\vdash \eta : \Theta)$, $(\Delta'\vdash \delta : \Delta)$ and $(\Gamma',\Delta'\vdash \rho : \Gamma)$ we define a multi-variable reindexing function $\textsf{subst}_{\eta,\delta,\rho}$ by induction on formulas, as follows:

\[
\begin{array}{l}
\mathsf{subst}_{\eta,\delta,\rho} : \set{[\Theta \mid \Delta \mid \Gamma]\propx} \to \set{[\Theta' \mid \Delta' \mid \Gamma']\propx} \\
\mathsf{subst}_{\eta,\delta,\rho}(\top) := \top, \mathsf{subst}(\bot) := \bot \\
\mathsf{subst}_{\eta,\delta,\rho}(\psi \land \varphi) := \subst_{\eta,\delta,\rho}(\psi) \land \subst_{\eta,\delta,\rho}(\varphi) \\
\mathsf{subst}_{\eta,\delta,\rho}(\psi \Rightarrow \varphi) := \subst_{\rho,\delta,\eta} (\psi) \Rightarrow \subst_{\eta,\delta,\rho} (\varphi) \\
\mathsf{subst}_{\eta,\delta,\rho}(\exists^p x. \varphi(x,n,d,p)) := \exists^p x. \subst_{\eta,\delta,\rho}(\varphi(x,n,d,p)) \\
\mathsf{subst}_{\eta,\delta,\rho}(\forall^p x. \varphi(x,n,d,p)) := \forall^p x. \subst_{\eta,\delta,\rho}(\varphi(x,n,d,p)) \\
\mathsf{subst}_{\eta,\delta,\rho}(s(n,d) \le t(d,p)) := s(\eta,\delta) \le t(\delta,\rho) \\
\mathsf{subst}_{\eta,\delta,\rho}(P(s(n,d) \mid t(d,p))) := P(s(\eta,\delta) \mid t(\delta,\rho)) \\
\end{array}
\]
In the case of terms we indicate with $s(\eta,\delta)$ and $t(\delta,\rho)$ the usual substitutions of terms in the intuitive way, e.g., if $\Theta',\Delta' \vdash \eta : \Theta$ and $\Delta' \vdash \delta : \Delta$, and a term $n:\Theta,d:\Delta \vdash s(n,d) : N$ then $\Theta',\Delta' \vdash s(\eta,\delta) : N$. Note the inversion of the terms in the case of implication. In the case of polarized quantifiers we substitute under binders in the usual capture-avoidant way.

For example, given a formula \[[\Theta,n\!:\!N\!\mid\!\Delta,d\!:\!D\!\mid\!\Gamma,p\!:\!P]\ \varphi(n,\n d,d,p) \propx,\] we can substitute any given three terms $(\Theta,\Delta \vdash \eta\!:\!N)$, $(\Delta \vdash \delta\!:\!D)$, $(\Gamma,\Delta \vdash \rho\!:\!P)$ for the variables $n,d,p$ in $\varphi$, denoting the resulting formula as $\varphi(\eta,\delta,\delta,\rho)$.

\end{defi}

\begin{exa}[Lifting natural to dinatural]\label{thm:reindex_lift}
Any natural variable can be lifted to dinatural using the substitution function of \Cref{def:substitution_in_formulas} as follows:
\begin{adju}[1.0]
$\hspace{-6.5em}\begin{array}{r@{}l}
\subst_{{\eta,\delta,\rho}} : & \set{[\Theta,n : N \mid \Delta \mid \Gamma,p : P]\propx} \\ \to & \set{[\Theta \mid \Delta,n : N,p : P \mid \Gamma]\propx},
\end{array}
\begin{minipage}{0.22\textwidth}
\text{\textit{where}}
\[\begin{array}{r@{\ }r@{\ }l}
  \eta := & \Theta,\Delta,n\!:\!N,p\!:\!P & \vdash \langle \pi_1,n \rangle : \Theta,N\\
  \delta := & \Delta,n\!:\!N,p\!:\!P & \vdash \pi_1 : \Delta \\
  \rho := & \Gamma,\Delta,n\!:\!N,p\!:\!P & \vdash \langle \pi_1,p \rangle : \Gamma,P
\end{array}\]
\end{minipage}$
\end{adju}
\end{exa}
\begin{exa}[Collapsing two naturals in one dinatural]
\label{thm:reindex_collapse}
Given a formula $\varphi(a,b)$ with two variables $a,b$ of different variance, we can collapse them to a formula $\varphi(\n x,x)$ with a single dinatural variable $x$ by substituting as follows:
\begin{adju}[1.0]
$\hspace{-6.5em}\begin{array}{r@{}l}
\subst_{{\eta,\delta,\rho}} : & \set{[\Theta,a : A \mid \Delta \mid \Gamma,b : A]\propx} \\ \to & \set{[\Theta \mid \Delta,x : A \mid \takespace[l]{\Gamma,b : A}{\Gamma}]\propx},
\end{array}
\begin{minipage}{0.22\textwidth}
\text{\textit{where}}
\[\begin{array}{r@{\ }r@{\ }l}
\eta := &  \Theta,\Delta,a\!:\!A & \vdash \langle \pi_1,a \rangle : \Theta,A \\
\delta := & \Delta,a\!:\!A & \vdash \pi_1 : \Delta \\
\rho := &  \Gamma,\Delta,a\!:\!A & \vdash \langle \pi_1,a \rangle : \Gamma,A \\
\end{array}\]
\end{minipage}$
\end{adju}
This contraction operation is used to characterize $\le_A$ as (relative) adjoint in \Cref{def:sem:directed_doctrine}.
\end{exa}

\subsection{Rules}

\noindent We now describe the main rules of the logic, both in formula construction and entailments.
\paragraph{(Structural rules.)}
Variables can be substituted via $\Rulereindex$, showing just the one-variable case as in \Cref{def:substitution_in_formulas}.
The structural rule $\Ruleupgrade$ allows for dinatural variables that are not \emph{used} with both variances (i.e., are \emph{weakened} as dinatural) to be upgraded as natural; note that the entailment in the hypothesis implicitly uses the operation given in \Cref{thm:reindex_lift} on both sides. Similar rules which instead act on a single variable or entire contexts can be derived from the above in the intuitive way.

\paragraph{(Polarized implications.)}\label{def:polarized_implications}
In directed first-order logic, implication is given a special treatment due to the contravariance of $\Rightarrow$ in its first argument, which we call \emph{polarized implication}. In the preorder semantics, this is motivated by the presence of $\I^\op$ in $\Implication : \I^\op \x \I \> \I$. In the syntax, this is reflected in implication formulas $[\Theta\mid\Delta\mid\Gamma]\ \psi \Rightarrow \varphi \propx$, where we swap the negative and positive context in the formula $[\Gamma\mid\Delta\mid\Theta]\ \psi \propx$ on the left side.

The intuition behind the rule for entailments $\Ruleimpl$ is that one can curry a formula $\psi$ between the left and right side of entailments, at the cost of \emph{inverting} the polarity of all positions in $\psi$: for example, consider a formula $\psi(x,y,z)$, with $x,y$ negative positions and $z$ positive, and take the entailment $\psi(x,\n y,z) \land \Phi(y) \vdash \varphi(z)$ where $x$ has negative variance, $y$ dinatural (negative in $\psi$ and positive in $\Phi$), $z$ positive; by currying $\psi$, we obtain the entailment $\Phi(y) \vdash \psi(x,y,\n z) \Rightarrow \varphi(z)$, where now $x$ is positive, $z$ becomes dinatural (due to the positive appearance in $\varphi$ and the \emph{negative} appearance in $\psi$), and $y$ is now positive.

This idea is concretely implemented by making \emph{all} variables in $\Phi,\psi,\varphi$ \emph{dinatural} in the top side of $\Ruleimpl$: other derived rules for polarity inversion ($\Ruleimplinvert$, $\Ruleimplplusr$, $\RuleimplR$) follow by using the structural rule $\Ruleupgrade$ to make variables natural again.

\paragraph{(Directed equality.)}\label{def:directed_equality}
The formula $a \le_A b$ (where $a,b$ are positions with negative and positive polarity, respectively) is the main construct of dFOL, and intuitively states that $a$ ``rewrites'' to $b$: in the preorder model, this is the fact that the relation $a \le b$ holds. The introduction rule is given by $\Rulelerefl$. We illustrate the intuition behind the rule for directed equality contraction $\Rulele$: the $\Downarrow$ direction states that a directed equality $a \le b$ in context can be contracted \emph{only if} $a,b$ only appear \emph{naturally} in the conclusion; using $\Ruleimpl$, $a$ and $b$ can appear in the hypothesis context only if they appear \emph{negatively}, shown in $\Rulelefull$.

\paragraph{(Polarized quantifiers.)}\label{def:polarized_quantifiers}
Since polarized contexts keep track of the variance of variables, quantifiers also need to keep track of this information: we define polarized quantifiers $\forall^+.\phi(x)$, $\forall^-x.\phi(x)$, $\forall^\Delta x.\phi(\nxx)$, similarly for exists $\exists^p.\phi(x)$ for $p \in \set{-,\Delta,+}$, with rules $\Ruleforall$ and $\Ruleexists$.

Note that $\forall^+,\forall^-$ and $\exists^+,\exists^-$ can be derived in terms of their dinatural counterparts $\forall^\Delta,\exists^\Delta$ using $\Ruleupgrade$; we show this in \Cref{rem:forall_plus_minus}. We keep positive quantifiers as primitive since they immediately place the variable in the correct context and are more convenient to use. Moreover, they are more natural to work with in the semantics, since they are the adjoints to weakening later described in \Cref{def:sem:directed_doctrine}.

\section{Examples}\label{sec:examples}

In this section we give examples of derived rules and use them to showcase the main properties of directed equality and polarized implications/quantifiers.
We leave other examples and complete derivations for all statements shown here in \Cref{appendix:derived_rules}.

\begin{exa}[Derived rules]\label{ex:derived_rules}
The rules in \Cref{fig:derivable_rules} are derivable:
\begin{itemize}[leftmargin=1em]
\item Rule $\Rulelereflterm$ states reflexivity with a general term $t$. Note that $t$ cannot depend on any natural variable because the variables of $t$ have to appear on both sides of $\le$.
\item Rules $\Ruleleminus,\Ruleleplus$ are versions of $\Rulele$ which take into account the polarity of the variable being contracted, since $\Rulele$ always makes them dinatural.
\item Rules $\, \Ruleimplinvert$ intuitively show how polarity inversion follows from $\Ruleimpl,\Ruleupgrade$.
\item Using polarized implications, one can derive a general rule $\Rulelefull$ that allows for variables $a,b$ to appear in $\Phi$. However, both $a,b$ must be used only \emph{with the opposite polarity} because of the application of $\Ruleimpl$. This is reminiscent of a (polarized) Frobenius formulation for equality \cite[3.2.4]{Jacobs1999categorical}. A version with terms $\Ruleleterm$ follows using $\Rulereindex$; the syntactic restriction is embodied by the fact that $a,b\!:\!A$ are still constrained to be \emph{natural} in $\Phi,\varphi$. Both terms $\eta,\rho$ can only be given in a \emph{dinatural} context, since they appear dinaturally in the bottom sequent of $\Ruleleterm$.
\item The inverse direction of $\Rulele$ is logically equivalent to $\Rulelerefl$, for which we report the derivation in \Cref{appendix:rule_j_equivalent}.
\item The rules $\Ruledinatlift,\Ruledinatcollapse$ follow from $\Rulereindex$ and witness the functoriality of \Cref{thm:reindex_lift,thm:reindex_collapse}.
\item Typical rules for quantifiers, e.g., $\Ruleexiststermminus,\Ruleforalltermdelta$, follow from $\Ruleexists,\Ruleforall$ and $\Rulecut,\Rulereindex$.
\end{itemize}
\end{exa}

\begin{figure*}
\begin{center}
\begin{prooftree}
\hypo{\Theta,\Delta \vdash \eta : N}
\hypo{\hspace{-6.1em}[\Theta,n:N \mid \Delta \mid \Gamma]\ \varphi(n) \propx}
\infer[no rule]1{[\Theta \mid \Delta \mid \Gamma]\ \Phi \vdash \varphi(\eta)}
\infer2[\Ruleexiststermminus]{[\Theta \mid \Delta \mid \Gamma]\ \Phi \vdash \exists^- x.\varphi(x)}
\end{prooftree}
\quad
\begin{prooftree}
\hypo{\Delta \vdash \delta : D}
\hypo{[\Theta \mid \Delta \mid \Gamma]\ \Phi \vdash \forall^\Delta x.\varphi(\n x,x)}
\infer2[\Ruleforalltermdelta]{[\Theta \mid \Delta \mid \Gamma]\ \Phi \vdash \varphi(\delta,\delta)}
\end{prooftree}
\\[1.0em]
\begin{prooftree}
  \infer0[\Rulelerefl]{[\Theta \mid \Delta,z:A \mid \Gamma]\ \Phi & \vdash \nz \le z}
\end{prooftree}
\quad
\begin{prooftree}
  \hypo{[\Theta \mid \Delta, z:A \mid \Gamma]\ \takespace{a \le b,\Phi(\n b,\n a)}{\Phi(\nzz)} & \vdash \varphi(\nzz)}
  \infer[double]1[\Rulelefull]{[\Theta \mid \Delta,a:A,b:A \mid \Gamma]\ {a \le b,\Phi(\n b,\n a)} & \vdash \varphi(a,b)}
\end{prooftree}
\\[1.0em]
\begin{prooftree}
  \infer[no rule]0{\hspace{-9.72em}[\Theta\mid\Delta \mid \Gamma,a:A]\ \Phi(a), \varphi(a) \propx}
  \infer[no rule]1{[\Theta \mid \Delta \mid \Gamma, z:A]\ \takespace{\n a \le b,\Phi(a)}{\Phi(z)} & \vdash \varphi(z)}
  \infer[double]1[\Ruleleplus]{[\Theta \mid \Delta,a:A \mid \Gamma,b:A]\ {\n a \le b,\Phi(a)} & \vdash \varphi(b)}
\end{prooftree}
\quad
\begin{prooftree}
  \infer[no rule]0{\hspace{-6.48em}\Delta \vdash \eta : A, \quad \Delta \vdash \rho : A}
  \infer[no rule]1{\takespace{\hspace{13.3em}}{[\Theta,a:A\mid \Delta \mid \Gamma,b:A] {\ \Phi(a,b),\varphi(a,b) \propx}}}
  \infer[no rule]1{[\Theta \mid \Delta, z:A \mid \Gamma]\ {\Phi(\n z,z)} & \vdash \varphi(\nzz)}
  \infer1[\Rulelefull]{[\Theta \mid \Delta \mid \Gamma]\ {\eta \le \rho,\Phi(\rho,\eta)} & \vdash \varphi(\eta,\rho)}
\end{prooftree}
\\[1.0em]
{\begin{prooftree}
\infer[no rule]0{[\Theta,n:A \mid \Delta \mid \Gamma,p:B]\ \Phi(n,p) \vdash \varphi(n,p)}
\infer1[\Ruledinatlift]{[\Theta \mid \Delta,n:A,p:B \mid \Gamma]\ \Phi(n,p)  \vdash \varphi(n,p)}
\end{prooftree}}
\\[1.0em]
{\begin{prooftree}
\infer[no rule]0{[\Theta,n:A \mid \Delta \mid \Gamma,p:A]\ \Phi(n,p) \vdash \varphi(n,p)}
\infer1[\Ruledinatcollapse]{[\Theta \mid \Delta,a:A \mid \Gamma]\ \Phi(\nzz) \vdash \varphi(\nzz)}
\end{prooftree}}
\\[1.0em]
{\begin{prooftree}
\infer[no rule]0{[\Theta',x\!:\!A \mid \Delta \mid \Gamma']\ \takespace[r]{\Phi(x)}{\psi(x)} \propx}
\infer[no rule]1{\hspace{1.3em}[\Theta \mid \Delta \mid \Gamma,x\!:\!A]\ \Phi(x) \propctx}
\infer[no rule]1{[\Theta,\Theta' \mid \Delta,x\!:\!A \mid \Gamma,\Gamma']\ \psi(\n x), \Phi(x) \vdash \varphi(x)}
\infer[double]1[\Ruleimplplusr]{[\Theta,\Gamma' \mid \Delta \mid \Gamma,\Theta',x\!:\!A]\ \Phi(x) \vdash \psi(x) \Rightarrow \varphi(x)}
\end{prooftree}}
\\[1.0em]
{\begin{prooftree}
\infer[no rule]0{[x\!:\!A \mid \Delta \mid y\!:\!B]\ \varphi(x,y) \propx}
\infer[no rule]1{[\Theta,x\!:\!A \mid \Delta \mid \Gamma,y\!:\!B]\ \psi(x,y), \Phi \vdash \varphi}
\infer[double]1[\Ruleimplinvert]{[\Theta,y\!:\!B \mid \Delta \mid \Gamma,x\!:\!A]\ \Phi \vdash \psi(x,y) \Rightarrow \varphi}
\end{prooftree}}
\end{center}
\LinkRuledinatcollapse
\LinkRuledinatlift
  \centering
\caption{Derivable rules in directed first-order logic.}
\label{fig:derivable_rules}
\end{figure*}

\begin{exa}[Directed equality]\label{ex:derivations}
We illustrate properties specific of directed equality, assuming $f \in \Sigma_T, P \in \Sigma_P$ s.t. $A \vdash f : B$ and $\mathsf{pos}(P) := A, \mathsf{neg}(P) := \top$.

\begin{itemize}[leftmargin=1em]
\item Transitivity of directed equality:
\end{itemize}
\[
\begin{prooftree}
\infer0[\Rulehyp]{[ \,z\!:\!A \mid \emptyctx \mid c\!:\!A]\ \takespace{a \le b,\,\n b \le c}{z \le c} & \vdash z \le c}
\infer1[\Ruleleminus]{[a\!:\!A \mid b\!:\!A \mid c\!:\!A]\ a \le b,\,\n b \le c & \vdash a \le c}
\end{prooftree}
\]
\begin{itemize}[leftmargin=1em]
\item Existence of singletons~\cite[1.12.1]{UnivalentFoundationsProgram2013homotopy}:
\end{itemize}
\[
\begin{prooftree}
\infer0[\Rulelerefl]{[\,\emptyctx \mid y\!:\!A \mid {\emptyctx}\,] \  & \vdash \n y \le y}
\infer1[\Ruleexiststermminus]{[\,\emptyctx \mid  y\!:\!A \mid \emptyctx\,] \  & \vdash \exists^- x. x \le y}
\infer1[\Ruleupgrade]{[\,\emptyctx \mid {\emptyctx} \mid y\!:\!A] \  & \vdash \exists^- x. x \le y}
\infer1[\Ruleforall]{[\,\emptyctx \mid {\emptyctx} \mid {\emptyctx}\,] \  & \vdash \forall^+ y.\exists^- x. x \le y}
\end{prooftree}
\]
Note that the above application of $\Ruleexiststermminus$ is justified by the fact that $y:A$ in $x \le y$ is implicitly lifted as dinatural instead of positive: take $\Theta := \emptyctx, \Delta := [y:A], \Gamma := \emptyctx$, and consider the following hypotheses for $\Ruleexiststermminus$:
\[
\begin{array}{rl}
  \phi(n) & := [n:A \mid y:A \mid \emptyctx\,]\ n \le y \propx \\
  \eta & := (\Theta = \emptyctx) , (\Delta = [y:A]) \vdash y:A \\
  \phi(\eta) & := [\,\emptyctx \mid y:A \mid \emptyctx\,]\ \n y \le y \propx
\end{array}
\]
from which the rule $\Ruleexiststermminus$ derives $\exists^- n.\ \phi(n) = \exists^- n.\ n \le y$. We then use $\Ruleupgrade$ to upgrade the dinatural variable $y$ back to a positive one.

\begin{itemize}[leftmargin=1em]
\item Transport\,(monotonicity\,of\,predicates):
\end{itemize}
\[
\begin{prooftree}
\infer0[\Rulehyp]{[\,\emptyctx\mid \emptyctx \mid z\!:\!A \,]\ \takespace{a \le b,\,P(a)}{P(z)} & \vdash P(z)}
\infer1[\Ruleleplus]{ [a\!:\!A\mid \,\emptyctx\mid b\!:\!A]\ a \le b,\,P(a) & \vdash P(b)}
\end{prooftree}
\]
\begin{itemize}[leftmargin=1em]
\item Congruence (monotonicity of terms):
\end{itemize}
\[
\begin{prooftree}
\infer0[\Rulelereflterm]{[\,\emptyctx \mid z\!:\!A \mid \emptyctx\,]\  \vdash f(\n z) \le_B f(z)}
\infer1[\Rulele]{[a\!:\!A \mid \emptyctx \mid b\!:\!A]\ a \le_A b \vdash f (a) \le_B f (b)}
\end{prooftree}
\]
\begin{itemize}[leftmargin=1em]
\item Pair of rewrites:
\end{itemize}
\[
\begin{adjustbox}{max width=1.0\linewidth}
\begin{prooftree}
\infer0[\Rulelereflterm]{[\,\emptyctx\mid x\!:\!A,y\!:\!B \mid \emptyctx\,]\ \vdash (\n x,\n y) \le_{A \times B} (x,y)}
\infer1[\Rulele]{[b\!:\!B \mid x\!:\!A \mid b'\!:\!B]\ b \le_B b' \vdash (\n x,b) \le_{A \times B} (x,b')}
\infer1[\Rulele]{[a\!:\!A,b\!:\!B \mid \,\emptyctx\, \mid a'\!:\!A,b'\!:\!B]\ a \le_A a',\ b \le_B b' \vdash (a,b) \le_{A \times B} (a',b')}
\end{prooftree}
\end{adjustbox}
\]
A converse ``directed injectivity of pairs'' follows by {congruence} with projections.

\begin{itemize}[leftmargin=1em]
\item Higher-order rewriting:
\end{itemize}

\[
\begin{prooftree}
\infer0[\Rulelereflterm]{[\,\emptyctx \mid h:A\Rightarrow B,x:A \mid \emptyctx\,]\ \vdash \n h \cdot \n x \le_B h \cdot x}
\infer1[\Ruleforalldelta]{[\,\emptyctx \mid h:A\Rightarrow B \mid \emptyctx\,]\ \vdash \forall^\Delta x.\ \n h \cdot \n x \le_B h \cdot x}
\infer1[\Rulele]{[f:A\Rightarrow B \mid \emptyctx \mid g:A \Rightarrow B]\ {f \le_{A \Rightarrow B} g} \vdash \forall^\Delta x.\ f \cdot \n x \le_B g \cdot x}
\end{prooftree}
\]
The other direction captures a notion of ``2-cell extensionality''~\cite{Hofmann1997extensional}; such a rule is not derivable in general since, in the $\Preord$ model later introduced in \Cref{ex:dir_doc_preord}, one can interpret $A,B$ can be interpreted as discrete preorders/sets, for which non-extensional models are known~\cite{Pedrot2018failure}.
\end{exa}

\begin{exa}[Failure of symmetry]\label{rem:failure_symmetry}
The rule for directed equality contraction $\Rulele$ cannot be used to derive symmetry, since in the entailment $[ \,\emptyctx \mid a:A,b:A \mid \emptyctx\,]\ \n a \le b \vdash \n b \le a$ both $a$ and $b$ appear \emph{dinaturally}. As we will see in \Cref{ex:dir_doc_preord}, the $\Preord$ model is a countermodel.

\end{exa}
  \begin{exa}[Symmetric equality]\label{thm:bidirectional_symmetric_equality}\LinkRuleiff
  We define an equality formula $a = b := \n a \le b\,\land\,\n b \le a$ in context $[\,\emptyctx \mid a,b:A \mid \emptyctx\,]$. Equality is symmetric since \[[\Theta \mid \Delta,a,b:A \mid \Gamma]\ { \n a \le b \land \n b \le a} \vdash  \n b \le a \land  \n a \le b\] follows by swapping hypotheses. We prove in \Cref{appendix:symmetric_equality_j} that $=$ is precisely the left adjoint to contraction in dinatural contexts.

\end{exa}

\begin{rem}[Natural quantifiers in terms of dinatural ones]\label{rem:forall_plus_minus}
Note that in the presence of dinatural quantifiers $\forall^\Delta,\exists^\Delta$ one can instead define the natural quantifiers $\forall^+,\forall^-,\exists^+,\exists^-$ as an equivalent shorthand. Given a formula $[\Theta \mid \Delta \mid \Gamma,x\!:\!A]\ \varphi(x)$ we prove that $\forall^\Delta.\varphi(x) \iff \forall^+ x.\varphi(x)$ (where on the left-hand side we implicitly reindex $\varphi(x)$ to be in a dinatural context), with the other cases following the same pattern using $\Ruleupgrade$:
\[
\begin{prooftree}
\infer0{[\Theta \mid \Delta \mid \Gamma]\ \Phi & \vdash \forall^\Delta x.\varphi(x) }
\infer1[\Ruleforall]{[\Theta \mid \Delta,x:A \mid \Gamma]\ \Phi & \vdash \varphi(x)}
\infer1[\Ruleupgrade]{[\Theta \mid \Delta \mid \Gamma,x:A]\ \Phi & \vdash \varphi(x)}
\infer1[\Ruleforall]{[\Theta \mid \Delta \mid \Gamma]\ \Phi & \vdash \forall^+ x.\varphi(x)}
\end{prooftree}
\quad
\begin{prooftree}
\infer0{[\Theta \mid \Delta \mid \Gamma]\ \Phi & \vdash \forall^+ x.\varphi(x)}
\infer1[\Ruleforall]{[\Theta \mid \Delta \mid \Gamma,x:A]\ \Phi & \vdash \varphi(x)}
\infer1[\RuledinatliftSPECIALA]{[\Theta \mid \Delta,x:A \mid \Gamma]\ \Phi & \vdash \varphi(x)}
\infer1[\Ruleforall]{[\Theta \mid \Delta,x:A \mid \Gamma]\ \Phi & \vdash \forall^\Delta x.\varphi(x)}
\end{prooftree}
\]
\end{rem}

\begin{exa}[Interaction between $\forall^p,\exists^p$ and $\Rightarrow$]
Note that the formulas $(\exists^- x.x \le y) \Rightarrow P$ and $\forall^+ x.(x \le y \Rightarrow P)$ are both well-formed and are logically equivalent (in the same context). More in general, the following equivalence holds for every $p \in \set{-,\Delta,+}$:
 \[(\exists^p x.P(x)) \Rightarrow Q \iff \forall^{p^\op} x.(P(x) \Rightarrow Q)\]
\end{exa}
\begin{exa}[Classical formulas]\label{classical_formulas}
The following entailment for double-negation elimination
is well-formed for any $[\Theta \mid \Delta \mid \Gamma]\ \varphi \propx$
since we swap $\Theta,\Gamma$ in polarized contexts twice:
\[
\begin{array}{ll}
[\Theta \mid \Delta \mid \Gamma]\ (\varphi \Rightarrow \bot) \Rightarrow \bot \vdash \varphi
\end{array}
\]
The formula for the excluded middle is trickier: given $[x\!:\!N \mid y\!:\!D \mid z\!:\!P]\ \varphi \propx$, we have that $\varphi$ also appears on the left side of $\Rightarrow$ and thus all variables become dinatural:
\[[\,\emptyctx \mid x\!:\!N,y\!:\!D,z\!:\!P \mid \emptyctx\,]\ \vdash \varphi(x,\n y,y,z) \lor (\varphi(x,\n y,y,z) \Rightarrow \bot).\]
Hence, we can define $\neg\varphi := \varphi \Rightarrow \bot$, noting that these two formulas live a priori in different contexts, i.e., if $[\Theta \mid \Delta \mid \Gamma]\ \varphi$ then $[\Gamma \mid \Delta \mid \Theta]\ \neg\varphi$. In the above example we are implicitly reindexing variables as dinatural to ensure that formulas are all in the same context.

We will use the double-negation elimination formula in \Cref{poset_completeness} in the case of classical dFOL.
\end{exa}

\begin{exa}[Classical quantifiers]\label{classical_quantifiers}
\emph{Assuming double negation elimination}, we can prove that for every $p \in \set{-,\Delta,+}$ the following equivalences hold for all $[\,\emptyctx \mid \emptyctx \mid \emptyctx\,], [x :^p A]\ \varphi(x)$ (note the use of $p^\op$ to ensure well-formedness):
\[
\begin{array}{c@{\,}c@{\,}c}
\exists^p x.\varphi(x) & \iff & \neg \forall^{p^\op} x.\neg \varphi(x), \\
\forall^p x.\varphi(x) & \iff & \neg \exists^{p^\op} x.\neg \varphi(x).
\end{array}
\]
\end{exa}

\begin{thm}[Closed formulas]
\label{thm:closed_formulas}
Given a generic entailment with a single variable in each polarized context as follows,
\[
\begin{prooftree}
  \infer0{[x:N \mid y:D \mid z:P]\ \Phi(x,\n y,y,z) \vdash \varphi(x,\n y,y,z),}
\end{prooftree}
\]
we can always construct an equivalent closed formula (i.e., with no free variables) by quantifying over all variables in polarized contexts,
\[
\begin{adjustbox}{max width=\linewidth}
\begin{prooftree}
\infer0{[\,\emptyctx\!\mid\!\emptyctx\!\mid\!\emptyctx\,]\ \vdash
\forall^\Delta x,y,z.\ (\phi_1(\n x,y,\n y,\n z) \land \cdots \land \phi_n(\n x,y,\n y,\n z) \Rightarrow \varphi(x,\n y,y,z))}
\end{prooftree}
\end{adjustbox}
\]
by taking the conjunction of all formulas $\phi_i \in \Phi$. The bottom entailment holds iff the original one holds using $\Ruleimpl$ and $\Ruleforall$. Note that all three quantifiers must be dinatural since, e.g., $z$ appears both negatively (in $\phi_i$) and positively (in $\varphi$). A similar construction can be done in the case for general lists of variables $\Theta, \Delta, \Gamma$ in the intuitive way.
\end{thm}

\begin{rem}[Upgrading is not admissible]
Note that the rule $\Ruleupgrade$ is not an admissible rule because of the axiom case: consider two predicates $[\Theta \mid \Delta \mid \Gamma,x:A]\ P(x),Q(x) \propx$ which depend positively on $x$. One can then add an axiom $\alpha\in\Sigma_A$ in context $\textsf{actx}(\alpha) := [\Theta \mid \Delta,x:A \mid \Gamma]$ where $x$ is reindexed to be dinatural, i.e., $[\Theta \mid \Delta,x:A \mid \Gamma]\ P(x) \vdash Q(x)$, despite the fact that $P(x),Q(x)$ do not depend on $x$ dinaturally.
\end{rem}

\subsection{Examples of signatures}\label{sec:examples_signatures}

\noindent We give examples of signatures to show how dFOL can be used to model directed structure. A suitable extension of dFOL, e.g., a two-level system~\cite{New2023formal}, could be used to show non-trivial theorems.

\begin{exa}[$\lambda$-terms]\label{ex:lambda_terms}
We capture a signature of $\lambda$-terms in HOAS-style~\cite{Fiore1999abstract}:

\begin{itemize}[leftmargin=1em]
  \item $\Sigma_A := \set{T}$ with a type of (untyped) $\lambda$-terms,
  \item $\Sigma_F\!:=\!\set{\llambda,\app}$ for $\lambda$-abstraction and application, with $\dom(\llambda) := T \Rightarrow T$, $\cod(\llambda) :=T$ and $\dom(\app) := T \times T,  \cod(\app) :=T$.
  \item $\Sigma_E := \set{\eta}$, $\Sigma_P := \set{}$, and $\Sigma_A := \set{\beta}$ such that:
  \[
  \begin{array}{c}
  \begin{prooftree}
    \infer0[$(\eta)$]{[f:T \Rightarrow T]\ \left(\lambda x.\app(\llambda(f),x)\right) = f : T \Rightarrow T}
  \end{prooftree}\\[1.0em]
  \begin{prooftree}
   \infer0[$(\beta)$]{[\,\emptyctx\mid s:T\Rightarrow T, t:T \mid\emptyctx\,]\ \app(\llambda(\n s),\n t) \le s \cdot t}
  \end{prooftree}
  \end{array}
  \]
\end{itemize}
We can automatically derive that $\app$, $\lambda$ are congruences w.r.t. $\beta$-reduction:
\[
\begin{prooftree}
  \infer0[\Rulelereflterm]{[\,\emptyctx\mid z:T,t:T \mid\emptyctx\,]\ \vdash \app(\n z,\n t) \le_{T} \app(z,t)}
  \infer1[\Rulele]{[s:T \mid t:T \mid s':T]\ s \le_T s' \vdash \app(s,\n t) \le_{T} \app(s',t)}
\end{prooftree}
\]
The type $T$ can be interpreted by the preorder of $\lambda$-terms ordered by $\beta$-reduction, as an example of the semantics of \Cref{sec:doctrinal_semantics}.
\end{exa}
\begin{exa}[Domains]\label{ex:domain_theory}\!\!\!
We give a signature of $\omega\text{-CPOs}_\bot$ from domain theory~\cite{Fiore1997category}:
\begin{itemize}[leftmargin=1em]
  \item $\Sigma_A := \set{D,\omega}$, with a domain $D$ and the chain $\omega$, i.e., chains in $D$ are just terms $\omega\Rightarrow D$.
  \item $\Sigma_F := \set{0,\mathsf{succ}}$ defined in the intuitive way, $\Sigma_E, \Sigma_P := \set{}$, $\Sigma_A := \set{{\omega}_{\mathsf{axiom}},\bot_{\mathsf{axiom}},\bigsqcup_{\mathsf{axiom}}}$ such that:
\end{itemize}
\[\begin{array}{c}
  \begin{minipage}{0.23\textwidth}\begin{prooftree}
   \infer0[$({\omega}_{\mathsf{axiom}})$]{\vdash \forall^\Delta(i:\omega).\ \n i \le \mathsf{succ}(i)}
  \end{prooftree}\end{minipage}\hspace{6.6em}\begin{minipage}{0.23\textwidth}\begin{prooftree}
   \infer0[$(\bot_{\mathsf{axiom}})$]{\ \vdash \exists^- b.\ \forall^+ x.\ b \le_D x}
  \end{prooftree}\end{minipage}\\[1.7em]
  \begin{prooftree}
   \infer0[$(\bigsqcup_{\mathsf{axiom}})$]{
    \begin{array}{l}
      \vdash \forall^\Delta(c:\omega \Rightarrow D).\ \exists^\Delta (b:D).\ (\forall^-(i:\omega).\ \n c \cdot i \le b) \\
      \hspace{22pt}\land\ (\forall^+(b':D). (\forall^-(i:\omega).\ c \cdot i \le \n b) \Rightarrow \n b \le b')
    \end{array}}
  \end{prooftree}
\end{array}\]
\end{exa}
\begin{exa}[Graph rewriting and temporal logic]\label{ex:graph_rewriting}\!\!\!
We give an example of graph rewriting inspired by the approach used in quantified temporal logics~\cite{Gadducci2021presheaf,Gadducci2023specification}:
\begin{itemize}[leftmargin=1em]
  \item $\Sigma_A := \set{E,N,\omega}$, which are to be thought of as a preorder of edges $N$, a preorder of nodes $E$, and a finite fragment of the chain $\omega$ defined as above.
  \item $\Sigma_F:=\set{\mathsf{src},\mathsf{tgt},\mathsf{idx}_E,\mathsf{idx}_N}$ defined as follows:
  \[\begin{adjustbox}{max width=\linewidth}
    \begin{prooftree}
      \infer0{e\!:\!E \vdash \mathsf{src}(a),\mathsf{src}(a)\!:\!N}
    \end{prooftree}\qquad
    \begin{prooftree}
      \infer0{e\!:\!E \vdash \mathsf{idx}_E(a)\!:\!\omega}
    \end{prooftree}\qquad
    \begin{prooftree}
      \infer0{ n\!:\!N \vdash \mathsf{idx}_N(a)\!:\!\omega}
    \end{prooftree}
\end{adjustbox}\]
  \item Term equations $\Sigma_E = \set{\textsf{coh}_\textsf{src},\textsf{coh}_\textsf{tgt}}$ stating that, respectively,
  \[\begin{adjustbox}{max width=\linewidth}
  \begin{prooftree}
  \infer0[$(\textsf{coh}_{\set{\textsf{src},\textsf{tgt}}})$]{[e\!:\!E]\ \mathsf{idx}_E(\mathsf{src}(e)) = \mathsf{idx}_E(e) = \mathsf{idx}_E(\mathsf{tgt}(e))}
  \end{prooftree}
\end{adjustbox}\]

  \end{itemize}
  The above definitions capture the general settings to talk about graph rewriting. For example, consider the following instance of graph transformation in \Cref{fig:graph_evolution} from~\cite{Gadducci2023specification,Gadducci2021presheaf}, where we show the graph evolving in three successive time steps with edges being deleted/preserved:
\begin{figure}[H]
  \begin{adjustbox}{max width=\linewidth}
  \begin{tikzpicture}[shorten > = 2pt, shorten < = 2pt, auto, node distance = 2cm, semithick, scale=1.0]
    \tikzstyle{vertex} = [circle, draw = black, thick, fill = white, minimum size = 2mm]

    \node[vertex] (n0) [] {$n_0$};
    \node[vertex] (n1) [above right = 0.5cm and 1.2cm of n0] {$n_1$};
    \node[vertex] (n2) [below right = 0.5cm and 1.2cm of n0] {$n_2$};
    \node[vertex] (n4) [right = 3cm of n2] {$n_4$};
    \node[vertex] (n3) [right = 3cm of n1] {$n_3$};

    \path[->] (n4) edge [bend right] node [swap] (e4) {$e_4$} (n3);
    \path[->] (n3) edge [bend right] node [swap] (e3) {$e_3$} (n4);
    \path[->] (n0) edge [bend left]  node        (e0) {$e_0$} (n1);
    \path[->] (n1) edge [bend left]  node        (e1) {$e_1$} (n2);
    \path[->] (n2) edge [bend left]  node        (e2) {$e_2$} (n0);

    \node[vertex] (n5) [right = 3cm of e4] {$n_5$};
    \path[->] (n5) edge [loop left, looseness=10, out=150, in=-150] node (e5) {$e_5$} (n5);

    \path[->, dashed]                                     (n1) edge [bend left] (n3);
    \path[->, dashed]                                     (n2) edge [bend right] (n4);
    \path[->, dashed]                                     (n0) edge [bend right=60, out=270, in=240] (n4);
    \path[->, dotted, shorten < =  6pt, shorten > =  5pt] (e0) edge [bend right, out=-38] (e4);
    \path[->, dotted, shorten < = -2pt, shorten > = -2pt] (e1) edge [bend left]  (e3);
    \path[->, dotted, shorten < =  4pt, shorten > = -2pt, bend left=22pt] (e3) edge [bend left]  (e5);
    \path[->, dashed] (n3) edge [bend left] (n5);
    \path[->, dashed] (n4) edge [bend right] (n5);
\end{tikzpicture}
\end{adjustbox}
\caption{A graph evolving in three separate time steps. We use dashed lines for node rewrites and dotted lines for edge rewrites. Note that $e_2,e_4$ are deleted.}
\label{fig:graph_evolution}
\end{figure}
To capture the temporal evolution of the graph above we enrich the signature of dFOL with constants $\emptyctx \vdash e_i : E, \emptyctx \vdash n_i : N$ for $i \in \{0,\dots,5\}$, and $\emptyctx \vdash \omega_k : \omega$ for $k \in \{0,1,2\}$, and suitable axioms $e_i \le e_j$, $n_i \le n_j$ which capture the situation below, indicating with a solid arrow in each sort $N,E$ whenever an axiom $x \le y$ holds:
  \[
  \begin{tikzpicture}[scale=0.3]
    \begin{scope}[yshift=-1.3cm]
    \node [draw=black, fill={rgb,255: red,0; green,0; blue,0}, shape=circle, minimum size=0.125cm, inner sep=0cm, label={[yshift=-2pt]$\omega_1$}] (w1) at (0.5, -11.75) {};
    \node [draw=black, fill={rgb,255: red,0; green,0; blue,0}, shape=circle, minimum size=0.125cm, inner sep=0cm, label={[yshift=-2pt]$\omega_0$}] (w0) at ([xshift=-4cm]w1) {};
    \node [draw=black, fill={rgb,255: red,0; green,0; blue,0}, shape=circle, minimum size=0.125cm, inner sep=0cm, label={[yshift=-2pt]$\omega_2$}] (w2) at ([xshift=4cm]w1) {};

		\draw [->, shorten >=2pt, shorten <=2pt, black!60] (w0) to (w1);
		\draw [->, shorten >=2pt, shorten <=2pt, black!60] (w1) to (w2);
    \end{scope}

    \begin{scope}[yshift=1.8cm]
\node [draw=black, fill={rgb,255: red,0; green,0; blue,0}, shape=circle,
minimum size=0.125cm, inner sep=0cm, label=left:{$e_0$}] (98) at (-3.5, -0.25) {};

\node [draw=black, fill={rgb,255: red,0; green,0; blue,0}, shape=circle,
minimum size=0.125cm, inner sep=0cm, label=left:{$e_1$}] (100) at (-3.5, -1) {};

\node [draw=black, fill={rgb,255: red,0; green,0; blue,0}, shape=circle,
minimum size=0.125cm, inner sep=0cm, label=left:{$e_2$}] (189) at (-3.5, -1.75) {};

\node [draw=black, fill={rgb,255: red,0; green,0; blue,0}, shape=circle,
minimum size=0.125cm, inner sep=0cm, label=below right:{$e_3$}] (201) at (0.25, -1.5) {};

\node [draw=black, fill={rgb,255: red,0; green,0; blue,0}, shape=circle,
minimum size=0.125cm, inner sep=0cm, label=above right:{$e_4$}] (202) at (0.25, -0.5) {};

\node [draw=black, fill={rgb,255: red,0; green,0; blue,0}, shape=circle,
minimum size=0.125cm, inner sep=0cm, label=below:{$e_5$}] (216) at (4.75, -1) {};
    \node [draw=black, shape=ellipse, minimum width=4.2cm, minimum height=1.7cm, inner sep=0cm] (blob2) at (0.5, -1.0) {};
    \end{scope}

\node [draw=black, fill={rgb,255: red,0; green,0; blue,0}, shape=circle,
minimum size=0.125cm, inner sep=0cm, label=left:{$n_0$}] (191) at (-3.5, -6) {};

\node [draw=black, fill={rgb,255: red,0; green,0; blue,0}, shape=circle,
minimum size=0.125cm, inner sep=0cm, label=left:{$n_1$}] (192) at (-3.5, -6.75) {};

\node [draw=black, fill={rgb,255: red,0; green,0; blue,0}, shape=circle,
minimum size=0.125cm, inner sep=0cm, label=left:{$n_2$}] (193) at (-3.5, -7.5) {};

\node [draw=black, fill={rgb,255: red,0; green,0; blue,0}, shape=circle,
minimum size=0.125cm, inner sep=0cm, label=above right:{$n_3$}] (208) at (0.25, -6.25) {};

\node [draw=black, fill={rgb,255: red,0; green,0; blue,0}, shape=circle,
minimum size=0.125cm, inner sep=0cm, label=below right:{$n_4$}] (209) at (0.25, -7.25) {};

\node [draw=black, fill={rgb,255: red,0; green,0; blue,0}, shape=circle,
minimum size=0.125cm, inner sep=0cm, label=below:{$n_5$}] (223) at (4.75, -6.75) {};

		\draw [<-, black!60, shorten <=2pt, shorten >=2pt] (202) to (98);
		\draw [<-, black!60, shorten <=2pt, shorten >=2pt] (209) to (193);
		\draw [<-, black!60, shorten <=2pt, shorten >=2pt] (223) to (208);
		\draw [<-, black!60, shorten <=2pt, shorten >=2pt] (223) to (209);
		\draw [<-, black!60, shorten <=2pt, shorten >=2pt] (216) to (202);
		\draw [<-, black!60, shorten <=2pt, shorten >=2pt] (201) to (100);
		\draw [<-, black!60, shorten <=2pt, shorten >=2pt] (208) to (192);
		\draw [<-, black!60, shorten <=2pt, shorten >=2pt] (209) to (191);

    \node [draw=black, shape=ellipse, minimum width=3.8cm, minimum height=1.1cm, inner sep=0cm] (blob1) at ([yshift=0.25cm]w1) {};

    \node [draw=black, shape=ellipse, minimum width=4.2cm, minimum height=1.7cm, inner sep=0cm] (blob3) at (0.5, -6.75) {};

		\node at ([xshift=-1cm]blob2.west) {$E$};
		\node at ([xshift=-1cm]blob3.west) {$N$};
		\node at ([xshift=-1cm]blob1.west) {$\omega$};

		\draw [->, shorten >=2pt, shorten <=2pt, transform canvas={xshift=1.0ex}] (blob2) to node [right, midway] {$\textsf{src}$} (blob3);
		\draw [->, shorten >=2pt, shorten <=2pt, transform canvas={xshift=-1.0ex}, below] (blob2) to node [left, midway] {$\textsf{tgt}$} (blob3);

		\draw [->, shorten >=3pt, shorten <=3pt, bend left=-40] (blob3.south west) to node[left, midway] {$\textsf{idx}_E$} (blob1.north west);
		\draw [->, shorten >=3pt, shorten <=3pt, bend left=70] (blob2.south east) to node[right, midway] {$\textsf{idx}_N$} (blob1.north east);
\end{tikzpicture}\]
The intuition is to use $\omega$ to distinguish edges/nodes of the graph at different times, e.g., $\forall i\!\in\!\set{0,1,2}, \mathsf{idx}_E(e_i)=\mathsf{idx}_N(n_i)=0, $ for the graph at the time step $\omega_0$. Similarly, $\forall i\!\in\!\set{3,4}, \mathsf{idx}_N(n_i)=1$ for the elements of the graph at step $\omega_1$, and $\mathsf{idx}_N(n_5)=2$ for the graph at the step $\omega_2$.

We can automatically derive using directed equality induction that rewrites always preserve graph structure, e.g.:
\[\begin{prooftree}

  \infer0[\Rulerefl]{[\,\emptyctx \mid z\!:\!E \mid \emptyctx\,] \quad & \vdash \textsf{src}(\n z) \le_N \textsf{src}(z)}
   \infer1[\RuleJ]{[e\!:\!E \mid \emptyctx \mid e'\!:\!E]\ e \le_E e' & \vdash \textsf{src}(e) \le_N \textsf{src}(e')}
  \end{prooftree}
\]
Note that the same technique cannot be applied in set-theoretic models of first-order logic without trivializing the model.

\end{exa}
\begin{exa}[Petri nets]\label{ex:petri_nets}
Petri nets can be thought of as a way of presenting free symmetric strict monoidal categories~\cite{Meseguer1990petri}; in the preorder case this can be captured using the theory of commutative monoidal preorders, which we can model in our setting as follows:
\begin{itemize}[leftmargin=1em]
  \item $\Sigma_A := \set{T}$, which is to be thought of as a preorder of tokens,
  \item $\Sigma_F:=\set{1,\otimes}$ defined as follows:
  \[\begin{adjustbox}{max width=\linewidth}
    \begin{prooftree}
      \infer0{a\!:\!T,b\!:\!T \vdash a \otimes b\!:\!T}
    \end{prooftree}\qquad
    \begin{prooftree}
      \infer0{\emptyctx \vdash 1\!:\!T}
    \end{prooftree}
\end{adjustbox}\]
  \item Term equations $\Sigma_E = \set{\textsf{sym},\textsf{assoc},\textsf{id}_L,\textsf{id}_R}$ stating that, respectively,
  \[
  \begin{adjustbox}{max width=\linewidth}
  $\begin{array}{cc}
  \begin{prooftree}
  \infer0[$(\textsf{sym})$]{[a,b\!:\!T]\ a \otimes b = b \otimes a : T}
  \end{prooftree}&
  \begin{prooftree}
  \infer0[$(\textsf{assoc})$]{[a,b,c\!:\!T]\ (a \otimes b) \otimes c = a \otimes (b \otimes c) : T}
  \end{prooftree}
  \\[1.0em]
  \begin{prooftree}
  \infer0[$(\textsf{id}_{L})$]{[a,b,c\!:\!T]\ a \otimes 1 = a}
  \end{prooftree}
  &
  \begin{prooftree}
  \infer0[$(\textsf{id}_{R})$]{[a,b,c\!:\!T]\ 1 \otimes a = a}
  \end{prooftree}
  \end{array}$
  \end{adjustbox}\]

  \end{itemize}
  The above definitions capture the general settings for Petri nets, i.e., the theory of (strictly) commutative monoidal preorders. As a concrete example, we consider the following instance of Petri net in \Cref{fig:petri_net}, taken from the context of rewriting logic in \cite{Stehr2001rewriting}:
\begin{figure}[H]
  \begin{adjustbox}{max width=\linewidth}
\begin{tikzpicture}[
    >={Latex[length=7pt, width=3pt]},
    place/.style={circle, draw, minimum size=0.5cm, inner sep=0pt, font=\ttfamily},
    transition/.style={rectangle, draw, minimum width=0.25cm, minimum height=0.25cm},
    label font/.style={font=\ttfamily},
    scale=0.95
]

\node[place] (bank) at (0, 4) {

    \tikz{
      \fill (0,0) circle (0.055);
      \fill (0.14,0) circle (0.055);
      \fill (0.07,0.12) circle (0.055);
    }};
\node[above=0.1cm of bank, font=\ttfamily] {BANK};

\node[transition] (grant1) at (-3.5, 1.5) {};
\node[above left=0.1cm and -0.6cm of grant1, font=\ttfamily] {GRANT-1};

\node[transition] (return1) at (-1, 1.5) {};
\node[above left=0.1cm and -0.4cm of return1, font=\ttfamily] {RETURN-1};

\node[place] (credit1) at (-3.5, -0.3) {};
\node[below=0.1cm of credit1, font=\ttfamily] {CREDIT-1};

\node[place] (claim1) at (-1, -0.3) {
    \tikz{
      \fill (0,0) circle (0.055);
      \fill (0.14,0) circle (0.055);
      \fill (0.07,0.12) circle (0.055);
    }};
\node[below=0.1cm of claim1, font=\ttfamily] {CLAIM-1};

\node[transition] (grant2) at (1, 1.5) {};
\node[above right=0.1cm and -0.4cm of grant2, font=\ttfamily] {GRANT-2};

\node[transition] (return2) at (3.5, 1.5) {};
\node[above right=0.1cm and -0.6cm of return2, font=\ttfamily] {RETURN-2};

\node[place] (credit2) at (1, -0.3) {};
\node[below=0.1cm of credit2, font=\ttfamily] {CREDIT-2};

\node[place] (claim2) at (3.5, -0.3) {
    \tikz{
      \fill (0,0) circle (0.05);
      \fill (0.13,0) circle (0.05);
    }};
\node[below=0.1cm of claim2, font=\ttfamily] {CLAIM-2};

\draw[->] (bank) -- (grant1);
\draw[->] (return1) -- node[near end, below right, font=\ttfamily] {3} (bank);
\draw[->] (grant1) -- (credit1);
\draw[->] (claim1) -- (grant1);
\draw[->] (return1) -- node[right, font=\ttfamily] {3} (claim1);

\draw[->] (bank) -- (grant2);
\draw[->] (return2) -- node[near end, above right, font=\ttfamily] {2} (bank);
\draw[->] (grant2) -- (credit2);
\draw[->] (claim2) -- (grant2);
\draw[->] (return2) -- node[right, font=\ttfamily] {2} (claim2);

\draw[->] (credit1) -- node[near end, above left, inner sep=1pt, font=\ttfamily] {3} (return1);
\draw[->] (credit2) -- node[near end, above left, inner sep=1pt, font=\ttfamily] {2} (return2);
\end{tikzpicture}
\end{adjustbox}
\caption{A Petri net for the banker's problem with two clients~\cite{Stehr2001rewriting}.}
\label{fig:petri_net}
\end{figure}
\noindent We add to the signature the following constant terms, corresponding to places:
  \[\begin{adjustbox}{max width=\linewidth}
    \begin{prooftree}
      \infer0{\emptyctx \vdash \textsf{bank}$, $\textsf{credit-1}$, $\textsf{credit-2}$, $\textsf{claim-1}$, $\textsf{claim-2} : T}
    \end{prooftree}
\end{adjustbox}\]
and add a rewriting axiom for each of the transitions:
\begin{adju}[1.0]$
  \begin{array}{lr}
  \begin{prooftree}
    \infer0[$(\textsf{grant-1})$]{[\,\emptyctx\mid\emptyctx\mid\emptyctx\,]\ \vdash \textsf{bank} \otimes \textsf{claim-1} \le \textsf{credit-1}}
  \end{prooftree}&
  \begin{prooftree}
    \infer0[$(\textsf{return-1})$]{[\,\emptyctx\mid\emptyctx\mid\emptyctx\,]\ \vdash \textsf{credit-1}^3 \le \textsf{bank}^3 \otimes \textsf{claim-1}^3}
  \end{prooftree}\\[1.0em]
  \begin{prooftree}
    \infer0[$(\textsf{grant-2})$]{[\,\emptyctx\mid\emptyctx\mid\emptyctx\,]\ \vdash \textsf{bank} \otimes \textsf{claim-2} \le \textsf{credit-2}}
  \end{prooftree}&
  \begin{prooftree}
    \infer0[$(\textsf{return-2})$]{[\,\emptyctx\mid\emptyctx\mid\emptyctx\,]\ \vdash \textsf{credit-2}^2 \le \textsf{bank}^2 \otimes \textsf{claim-2}^2}
  \end{prooftree}
\end{array}$\end{adju}\vspace{1.0em}
where we define $t^2 := t \otimes t$ and $t^3 := t \otimes t \otimes t$ in the intuitive way.

For instance, one can prove that starting from the initial configuration $\textsf{claim-1}^3 \otimes \textsf{bank}^3$ the system can evolve to the state $\textsf{credit-1} \otimes \textsf{credit-2}$, i.e., that the following entailment is derivable,
\[
\begin{prooftree}
\infer0{\vdash \textsf{claim-1}^3 \otimes \textsf{bank}^3 \le \textsf{credit-1} \otimes \textsf{credit-2}}
\end{prooftree}\vspace{1.0em}
\]
by applying the above axioms and using congruence and transitivity of directed equality.

\end{exa}

\section{Doctrinal Semantics}\label{sec:doctrinal_semantics}
In this section we introduce the doctrinal semantics of dFOL, i.e., directed doctrines. The idea is that reindexing in directed doctrines captures precisely the reindexing of \Cref{def:substitution_in_formulas}. We do this simply by ``augmenting'' the base category of doctrines via a so-called \emph{polarization category}, and then ask for (co)adjunctions to hold with respect to these special reindexings.

\begin{defi}
  We define the \emph{polarization category $\ndp{\C}$} for a cartesian category $\C$, denoting composition diagrammatically as $f \< g$ and pairing as $\ang{f,g}$:
  \begin{itemize}[leftmargin=1em]
  \item Objects: triples of objects $(\Theta \mid \Delta \mid \Gamma) \in \C_0\times\C_0\times\C_0$,
  \item Morphisms $(\Theta \mid \Delta \mid \Gamma) \> (\Theta' \mid \Delta' \mid \Gamma')$ are triples
    \[(n : \Theta \times \Delta \> \Theta' \mid d : \Delta \> \Delta' \mid p : \Gamma \times \Delta \> \Gamma'),\]
  \item Identities are given by $(\pi_1 : \Theta \times \Delta \to \Theta \mid \id_\Delta \mid \pi_1 : \Gamma \times \Delta \to \Gamma)$,
  \item Composition: \[(n \mid d \mid p) \< (n'\mid  d'\mid  p') :=
    (\ang{n,\pi_2 \< d} \< n'   \mid   d \< d'   \mid  \ang{p,\pi_2 \< d} \< p').\]
 \end{itemize}
\end{defi}
\begin{defi}[$\ndp{\C}$ as functor]\label{def:ndp_functor}
The above definition is functorial, i.e., any product-preserving functor $F : \C \> \D$ induces a functor $\ndp{F} : \ndp{\C} \> \ndp{\D}$ defined in the intuitive way $(\Theta\mid \Delta\mid \Gamma) \mapsto (F(\Theta)\mid F(\Delta)\mid F(\Gamma))$, similarly on morphisms.
\end{defi}

In the following, we denote with $\cP$ a functor $\cP : \ndp{\C}^\op \> \Pos$ which plays the role of a doctrine; we refer the reader to \cite{Maietti2023characterization,Jacobs1999categorical} for a standard accounts on doctrines.

There are doctrinal analogues of \Cref{thm:reindex_lift,thm:reindex_collapse}, which we later use in \Cref{def:sem:directed_doctrine} to characterize directed equality:

\begin{defi}[Dinatural lift]\label{thm:sem:reindex_lift}
There is a reindexing functor on $\cP : \ndp{\C}^\op \> \Pos$ as follows:
  \[
  \begin{array}{r@{\,}l}
    \cP({\lift{N,P}}) : & \cP(\Theta \times N \mid \Delta \mid \Gamma \times P) \to \cP(\Theta\mid\Delta \times N\times P\mid\Gamma),\ \text{\textit{where}}\\
    {\lift{N,P}} := ( & \ang{\pi_1,\pi_3} : \Theta\times\Delta\times N \times P \> \Theta\times N \\
    \mid & \takespace{\ang{\pi_1,\pi_3} }{\pi_1} : {\takespace{\Theta\times\Delta\times N \times P}{\Delta\times N \times P} \> \Delta} \\ \mid & \takespace{\ang{\pi_1,\pi_3} }{\ang{\pi_1,\pi_4}} : \takespace{\Theta\times\Delta\times N \times P}{\Gamma\times\Delta\times N \times P} \> \Gamma\times P).
  \end{array}
  \]
  This is the operation used in the top entailment of $\Ruleupgrade$.
\end{defi}
\begin{defi}[Dinatural contract]\label{thm:sem:reindex_collapse}
There is a functor as follows:
  \[
  \begin{array}{r@{\,}l}
    \cP({\contract{A}}) : & \cP(\Theta\times A \mid \Delta \mid \Gamma\times A) \to \cP(\Theta \mid \Delta\times A \mid \Gamma),\ \text{\textit{where}}\\
    {\contract{A}} := ( & \ang{\pi_1,\pi_3} : \Theta\times \Delta\times A \to \Theta\times A \\
    \mid & \takespace{\ang{\pi_1,\pi_3} }{\pi_1} : \takespace{\Theta\times\Delta\times A}{\Delta\times A} \to \Delta \\ \mid & \ang{\pi_1,\pi_3} : \takespace{\Theta\times\Delta\times A}{\Gamma\times\Delta\times A} \to \Gamma\times A).
  \end{array}
  \]$\cP({\contract{A}})$ identifies a positive and a negative variable together into a single dinatural one, and is used to characterize $\le_A$ in \Cref{def:sem:directed_doctrine}.
\end{defi}

A non-standard condition on objects of doctrines is necessary because of the case for base predicates ${P(s \mid t)}$, and we use it for initiality in \Cref{thm:main_theorem}.

\begin{defi}[No-dinatural-variance condition]\label{def:no_dinat_variance}
A functor $\cP : \ndp{\C}^\op \> \Pos$ is said to satisfy the \emph{no-dinatural-variance condition} \emph{(ndv)} if the functor \[\cP(\contract{\Delta}) : \cP(\Theta \times \Delta \mid \top \mid \Gamma \times \Delta) \> \cP(\Theta \mid \Delta \mid \Gamma)\] is a \emph{bijection on objects}, where $\contract{\Delta} := \ang{{\pi}_1, {!}_\Delta, \pi_1}$ is given by
\[
  \begin{array}{r@{\,}l}
    {\contract{\Delta}} := ( & \pi_1 : (\Theta\times \Delta) \times \top  \to \Theta\times \Delta \\
    \mid & \takespace{{\pi}_1}{!_\Delta} : \takespace{(\Theta\times \Delta) \times \top }{\Delta} \to \top \\ \mid & \takespace{{\pi}_1}{{\pi}_1} : \takespace{(\Theta\times \Delta) \times \top }{(\Gamma\times \Delta) \times \top } \to \Gamma\times \Delta),
  \end{array}
\]
and we denote with \[{\eps : \cP(\Theta\mid \Delta\mid\Gamma)_0 \to \cP(\Theta \times \Delta \mid \top \mid \Gamma \times \Delta)_0}\]
the inverse function of sets such that $\cP(\contract{\Delta})(\eps(p)) = p$.
\end{defi}

The intuition for \emph{ndv} is that dinatural variables $x\!:\!A$ in predicates always arise as a dinatural collapse, since in the semantics only $A$ and $A^\op$ are preorders (with no third option). The reason why this condition arises is due to the base case $P(s \mid t)$ (and $s \le_A t$), which only depends on $\textsf{pos}(P)$ and $\textsf{neg}(P)$, without asking for a type ``$\textsf{dinat}(P)$'': if such a type was present in the syntax, \emph{ndv} would not be required in the doctrinal semantics.

The functor $\cP(\contract{\Delta})$ is usually \emph{not} an isomorphism of posets, since $\forall^\Delta d.\psi(a,d,\n d,b)\Rightarrow$ $\varphi(a,\n d,d,b)$ does not imply that ${\forall^+ d. \forall^- d'.\psi(a,d',d,b)\Rightarrow\varphi(a,d',d,b)}$.

\begin{defi}[Polarized doctrine]\label{thm:polarized_doctrine}
A (split) \emph{polarized doctrine} is a cartesian closed category $\C$ with a functor ${\cP : \ndp{\C}^\op \> \Pos}$ that satisfies the \emph{no-dinatural-variance} condition and, moreover, interprets the rule $\Ruleupgrade$: if $\cP(\lift{N,P})(\psi) \le \cP(\lift{N,P})(\varphi)$,  then $\psi \le \varphi$, i.e., the monotone function $\cP(\lift{N,P})$ reflects the order for every $N,P$.
\end{defi}
Closedness of $\C$ is not strictly necessary, and is only used to interpret function types in \Cref{ex:lambda_terms,ex:domain_theory}.
\begin{defi}[Weakenings]\label{thm:sem:weakening}
Given a polarity $p \in \set{-,\Delta,+}$ we shall denote with $\cP(\Theta \mid \Delta \mid \Gamma \mid\mid^p A)$ the fiber obtained by applying the functor $- \times A$ to either $\Theta,\Delta,\Gamma$ depending on $p$ in the intuitive way. We denote weakening functors by $\wk^p_A : \cP(\Theta \mid \Delta \mid \Gamma) \> \cP(\Theta \mid \Delta \mid \Gamma \mid\mid^p A)$.
\end{defi}

\begin{defi}[Logical connectives]
  \label{def:sem:directed_doctrine}
  We define conditions on a polarized doctrine $\cP$. For each definition we require reindexing functors to preserve all structure up-to-equivalence.
  We recall relative (co)adjunctions in \Cref{appendix:relative_adjunction} \cite{Arkor2024formal}.
  \begin{itemize}[leftmargin=1.2em]
  \item \emph{$\cP$ has conjunctions} iff each fiber has finite products (i.e., glbs in preorders).
  \item \emph{$\cP$ has polarized implications \Ruleimpl} iff there is a right $\cP(\lift{\Theta,\Gamma})$-relative coadjoint, denoted as $\psi\,{\Rightarrow}\,-$, to the functor sending $\Phi \mapsto \cP(\lift{\Gamma,\Theta})(\psi) \land \cP(\lift{\Theta,\Gamma})(\Phi)$ for every $\psi \in \cP(\Gamma \mid \Delta \mid \Theta)$. The relative coadjunction can be stated as \[\psi \Rightarrow - \vdash_{\cP(\lift{\Theta,\Gamma})} \cP(\lift{\Gamma,\Theta})(\psi) \land \cP(\lift{\Theta,\Gamma})(-)\]
  and corresponds to the following situation:
  \[\xymatrix@C=1em{
    & {\cP(\Theta \mid \Delta \mid \Gamma)}\ar[dr]^(0.55){\hspace{1.9em}\cP(\lift{\Gamma,\Theta})(\psi) \land \cP(\lift{\Theta,\Gamma})(-)}\ar@{}[d]|(.6){\vdash} & \\
    \cP(\Theta \mid \Delta \mid \Gamma) \ar[ur]^{{\psi\,\Rightarrow\,-}} \ar[rr]_(0.45){\cP(\lift{\Theta,\Gamma})} & & \cP(\top \mid \Delta \times \Theta \times \Gamma \mid \top)
    }\]
Since we take \emph{posetal} doctrines, the above reduces to the following bi-implication ofr $\Phi, \varphi \in \cP(\Theta \mid \Delta \mid \Gamma)$: \[\begin{adjustbox}{max width=\linewidth}\begin{prooftree}
    \infer[no rule]0{[\top \mid \Delta\times \Theta \times \Gamma \mid \top]\ \cP(\lift{\Gamma,\Theta})(\psi) \times \cP(\lift{\Theta,\Gamma})(\Phi) \le \cP(\lift{\Theta,\Gamma})(\varphi)}
    \infer[double]1[\Ruleimpl]{[\Theta \mid \Delta \mid \Gamma]\ \Phi \le \psi \Rightarrow \varphi}
  \end{prooftree}\end{adjustbox}\]
 By a standard argument the above is (contravariantly) functorial in $\psi$, thus inducing a functor \[\Implication : \cP(\Gamma\!\mid\!\Delta\!\mid\!\Theta)^\op \times \cP(\Theta\!\mid\!\Delta\!\mid\!\Gamma) \to \cP(\Theta\!\mid\!\Delta\!\mid\!\Gamma)\]
 which sends two formulas to their implication.
  \item \emph{$\cP$ has polarized quantifiers} iff for every $p\in\set{-,\Delta,+}$ and $A \in \C$ the functor $\wk^p_{A} : \cP(\Theta \mid \Delta \mid \Gamma)\!\>\!\cP(\Theta\mid \Delta\mid \Gamma \mid\mid^p A)$ has a left $\exists^p_{A}$ and a right adjoint $\forall^p_{A}$.

  Moreover, we ask for Beck-Chevalley conditions: for any $f : (\Theta\mid\Delta\mid\Gamma) \> (\Theta'\mid\Delta'\mid\Gamma')$, and $\varphi \in \cP(\Theta' \mid \Delta' \mid \Gamma' \mid\mid^p A)$ the following predicates in $\cP(\Theta \mid \Delta \mid \Gamma)$ must be equal:
  \[\cP(f)(\exists^p_{A[\Theta'\Delta'\Gamma']}(\varphi))
    =
 \exists^p_{A[\Theta\Delta\Gamma]}(\cP(f \mid\mid^p \id_A)(\varphi)),\] where the morphism $(f \mid\mid^p \id_A) : (\Theta\mid\Delta\mid\Gamma\mid\mid^p A) \> (\Theta'\mid\Delta'\mid\Gamma'\mid\mid^p A)$ reindexes with $f$ but leaves $A$ untouched.
  We omit a Frobenius condition as in the standard case, which follows automatically in our case using polarized implications~\cite[1.9.12(i)]{Jacobs1999categorical}.
  \item \emph{$\cP$ has directed equality} iff, given the following functors
  \[\begin{array}{l@{\ }l}
    \cP(\contract{A}) & : \cP(\Theta \times A \mid \Delta \mid \Gamma \times A) \> \cP(\Theta \mid \Delta \times A \mid \Gamma), \\
    \cP(\wk^\Delta_A) & : \cP(\Theta \mid \Delta \mid \Gamma) \> \cP(\Theta \mid \Delta \times A \mid \Gamma),
  \end{array}\]
  there is a $\cP(\wk^\Delta_A)$-relative left adjoint ${{\le_A}\land-}$ to the contraction functor $\cP(\contract{A})$, i.e., $\le_A \land - \dashv_{\cP(\wk^\Delta_A)} \cP(\contract{A})$ holds. This corresponds to the following situation:
    \[\xymatrix@C=-0.3em{
    & \takespace[c]{\cP(\Theta \mid \Delta \mid \Gamma)}{\cP(\Theta \times A \mid \Delta \mid \Gamma \times A)}\ar[dr]^(0.55){\cP(\contract{A})}\ar@{}[d]|(.6){\dashv} & \\
    \cP(\Theta \mid \Delta \mid \Gamma) \ar[ur]^(0.45){{{\le_A}\land-}} \ar[rr]_(0.46){\cP(\wk^\Delta_A)} & & \cP(\Theta \mid \Delta \times A \mid \Gamma)
    }\]
    In the posetal case this is a bi-implication: $\forall \Phi \in \cP(\Theta \mid \Delta \mid \Gamma),\varphi \in \cP(\Theta \times A \mid \Delta \mid \Gamma \times A)$,\vspace{0.8em}\[\begin{prooftree}
      \hypo{\ \takespace{a \le b,\Phi}{[\Theta \mid \Delta \times A \mid \Gamma]\ \cP(\wk^\Delta_A)(\Phi)} & \le \cP(\contract{A})(\varphi)}
      \infer[double]1[\Rulele]{[\Theta \times A \mid \Delta \mid \Gamma \times A]\ {{\le_A}\land \Phi} & \le \varphi}
      \end{prooftree}\vspace{0.8em}\]
  Moreover, we ask for the following Beck-Chevalley condition \cite[3.4.1]{Jacobs1999categorical}; for any map $f := (n \mid d \mid p) : (\Theta\mid\Delta\mid\Gamma) \> (\Theta'\mid\Delta'\mid\Gamma')$ and $\varphi \in \cP(\Theta' \mid \Delta' \mid \Gamma')$, the following predicates in $\cP(\Theta \times A \mid \Delta \mid \Gamma \times A)$ must be equal: \[\le_{A[\Theta\Delta\Gamma]}\!(\cP(f)(\varphi)) = \cP(n \times \id_A \mid d \mid p \times \id_A)(\le_{A[\Theta'\Delta'\Gamma]}\!(\varphi)),\] where $\cP(n \times \id_A \mid d \mid p \times \id_A)$ is the functor that applies $f$ and leaves $A$ unaltered.

  A ``polarized Frobenius'' condition, exemplified in $\Rulelefull$ (i.e., variables are only used negatively in context $\Phi$), follows automatically using polarized implications \cite[3.2.4]{Jacobs1999categorical}.

  \end{itemize}
  \end{defi}

  \noindent Note the similarity between the above relative adjunction and standard accounts of equality in categorical logic \cite{Jacobs1999categorical,Maietti2013quotient,Maietti2023characterization}. In our case we cannot relate directed equality to existentials and their characterization as left adjoints, since the former is given by a \emph{relative} adjunction and the latter by a standard adjunction.

  It is often the case in categorical logic that one wants to study individual connectives in isolation, e.g., to study their completions or to identify specific fragments~\cite{Maietti2023characterization}; in the case above, the notion of directed equality is dependent on the presence of polarized implication, since it allows us to state its rule more clearly with relative adjunctions. We leave it for future work to give a more modular account of directed equality, e.g., by giving an alternative characterization, similar to $\Rulelefull$, which incorporates stability under products in the style of \cite[3.2.4]{Jacobs1999categorical}.

  A binary attribute of mixed variance (closely following the idea of Lawvere in \cite{Lawvere1970equality}) can be recovered in each fiber by taking ${\top \in \cP(\top\mid\top\mid\top)}$ and ${{\le_A} \land {\top} : \cP(A \mid \top \mid A)}$, where indeed the two variables are now separated by their polarity.

  Thanks to the general theory of relative (co)adjunctions~\cite{Arkor2024formal}, both polarized implications and directed equality are \emph{properties} of a polarized doctrine, i.e., they are unique up-to isomorphism~\cite[1.10.4]{Jacobs1999categorical}.

\begin{defi}[Directed doctrine]
A \emph{(split) directed doctrine} is defined to be a $(\top,\land,\Rightarrow,\exists^p,\forall^p,\le)$-polarized doctrine, i.e., it satisfies all above conditions.
\end{defi}

\subsection{Examples of directed doctrines}
\label{posetal_semantics}
We present the main model of dFOL where types are preorders:
  \begin{exa}[Directed doctrine of preorders]\label{ex:dir_doc_preord}
    The directed doctrine $\CPreord : \ndp{\Preord}^\op \> \Pos$ of preorders has as base category (i.e. types/terms) the category $\mathsf{Preord}$ of preorders and monotone functions, and is defined by
    \[\begin{array}{r@{\,}l}
      \CPreord(\Theta \mid \Delta \mid \Gamma) := [\sem{\Theta}^\op \times (\sem{\Delta}^\op \times \sem{\Delta}) \times \sem{\Gamma}, \I]_{{\Delta}}, \\[0.7em]
      \alpha \le \beta := \forall n \in \sem{\Theta},d \in \sem{\Delta},p \in \sem{\Gamma}.\ \alpha(n,d,d,p) \le \beta(n,d,d,p)
    \end{array}
    \]
    where objects are monotone maps $\alpha, \beta \in \sem{\Theta}^\op \times (\sem{\Delta}^\op \times \sem{\Delta}) \times \sem{\Gamma} \> \I$ and the poset relation between these is defined as above. We omit the inductive definition of how types are interpreted, since they simply use the product and internal hom of $\mathsf{Preord}$ in the intuitive way; we will often omit semantics brackets to distinguish between a type $A$ and their interpretation as a preorder $\sem{A}$ . The idea for entailments is reminiscent of dinatural transformations \cite{Dubuc1970dinatural}, since we use the fact that $\sem{\Delta}_0 = \sem{\Delta}^\op_0$ have the same objects.
    \begin{itemize}[leftmargin=1em]
\item  The \emph{ndv} condition holds, since \[\sem{\Theta}^\op \times (\sem{\Delta}^\op \times \sem{\Delta}) \times \sem{\Gamma} \iso (\sem{\Theta} \times \sem{\Delta})^\op \times \top \times (\sem{\Gamma} \times \sem{\Delta}).\]

  \item  Rule $\Ruleupgrade$ holds: if ${\psi},\varphi$ do not depend on a dinatural variable $\sem{D}$ or $\sem{D}^\op$ one can just rephrase any entailment $\sem{\psi}\le\sem{\varphi}$ to have, respectively, $(\sem{\Theta}\!\times\!\sem{D})^\op$ or $\sem{\Gamma}\!\times\!\sem{D}$ in the domain.
  \item Truth $\top$ is interpreted by the constant monotone function into $1 \in \I$; similarly, falsity $\bot$ is the constant function into $0 \in \I$.
  \item Conjunction $\psi \land \varphi$ is interpreted by the pointwise product (i.e., glb) of functions in $\I$, and disjunction $\psi \lor \varphi$ is interpreted by the pointwise coproduct (i.e., lub) of functions in $\I$. Implication $\psi \Rightarrow \varphi$ is given by postcomposing $\ang{\psi^\op,\varphi}$ with $\le_\I : \I^\op \times \I \> \I$.
 \item  Reindexing on formulas is given by precomposition of monotone functions: given terms $\eta,\rho,\delta$ and a formula $\varphi$ such that
 \[\begin{array}{r@{\,}l}
  \sem{\eta}   & : \sem{\Theta} \times \sem{\Delta} \> \sem{N}, \\
  \sem{\delta} & : \takespace{\sem{\Theta} \times \sem{\Delta}}{\sem{\Delta}} \> \sem{D}, \\
  \sem{\rho} & : \takespace{\sem{\Theta} \times \sem{\Delta}}{\sem{\Gamma} \times \sem{\Delta}} \> \sem{P} \text{ and} \\
  \sem{\varphi} & : [\sem{N}^\op \times (\sem{D}^\op \times \sem{D}) \times \sem{P}, \I]
 \end{array}
 \]
 reindexing corresponds in the semantics to the following monotone function:
 \[\cP(\eta,\delta,\rho)(\varphi)(n,d'\!,\!d,p)\!:=\!\varphi(\eta^\op(n,d'),\delta^\op(d),\delta(d),\rho(p,d)).\]
 \item   Polarized quantifiers are given using indexed products and coproducts in $\I$ (i.e., glb and lub respectively). In the case of $\forall^\Delta$ (and other dinatural quantifiers) we rely on $\sem{A}^{\op}_0 = \sem{A}_0$, in a way similar to a decategorification of ends \cite{Loregian2021coend}.

  \item Directed equality is given by the following monotone function:
  \[
  \begin{array}{l@{\,}l}
    \le_A & : [(\Theta \times A)^\op \times (\Delta^\op \times \Delta) \times (\Gamma \times A), \I] \\
    {\le_A} & := ((n,a),(d',d),(p,a')) \mapsto a \le_A a' \in \I
  \end{array}\]
  \item Rule $\Rulelerefl$ corresponds to $\forall a.\top \le_\I (a \le_A a)$.
  \item The relative adjunction $\RuleJ$ for directed equality is interpreted by the following construction: assume $f : \CPreord(\Theta,\Delta,\Gamma), g : \CPreord(\Theta \times A,\Delta,\Gamma \times A)$, such that \[\forall a:A,n\!:\!N,d\!:\!D,p\!:\!P. f(n,d,d,p) \le g((n,a),d,d,(p,a)).\] Note that dinatural collapse is applied on $g$. We show \[\forall (n,a)\!:\!N\times A,d\!:\!D,(p,a')\!:\!P\times A.\ (a \le_A a') \land f(n,d,d,p) \le_\I g((n,a),d,d,(p,a')).\] Assuming $a \le_A a'$ and $f(n,d,d,p)$, we use monotonicity of $g$ to obtain \[f(n,d,d,p) \le g((n,a),d,d,(p,a)) \le g((n,a),d,d,(p,a'))\] as desired.
  This proof motivates the syntactic restriction for $a,b$ to be natural in the conclusion, since we use monotonicity of either $a$ or $a'$, \emph{but not both}.
    \end{itemize}
    We spell out the semantics of formulas for the reader's convenience since it plays a crucial role in \Cref{posetal_semantics}. We show for simplicity only the cases for $\exists^+$ and $\forall^\Delta$. The semantic interpretation of formulas is given by induction in the intuitive way:
    \[\begin{array}{r@{\,}l}
      \sem{-} : & \set{[\Theta \mid \Delta \mid \Gamma] \propx} \\ \to & [\sem{\Theta}^\op \times (\sem{\Delta}^\op \times \sem{\Delta}) \times \sem{\Gamma}, \I]_{{\Delta}} \\
      \sem{P(s \mid t)} & := \lambda n,d',d,p \mapsto \sem{P}(\sem{s}_f^\op(n,d'), \sem{t}_f(d,p)) \\
      \sem{\bot} & := \lambda n,d',d,p \mapsto 0_\I \\
      \sem{\top} & := \lambda n,d',d,p \mapsto 1_\I \\
      \sem{\varphi \land \psi} & := \lambda n,d,d',p \mapsto \sem{\varphi}(p,d',d,n) \sqcap \sem{\psi}(n,d',d,p) \\
      \sem{\varphi \lor \psi} & := \lambda n,d,d',p \mapsto \sem{\varphi}(p,d',d,n) \sqcup \sem{\psi}(n,d',d,p) \\
      \sem{\varphi \Rightarrow \psi} & := \lambda n,d,d',p \mapsto \sem{\varphi}(p,d',d,n) \le_\I \sem{\psi}(n,d',d,p) \\
      \sem{\exists^+(\varphi)} & := \lambda n,d,d',p \mapsto \bigsqcup_{a \in A} \sem{\varphi}(n,d',d,p,a) \in \I \\
      \sem{\forall^\Delta(\psi)} & := \lambda n,d,d',p \mapsto \bigsqcap_{a \in A} \sem{\psi}(n,(d',a),(d,a),p) \in \I \\
      \sem{a \le_A b} & := \lambda n,d',d,p \mapsto \sem{s}_f^\op(n,d') \le_A \sem{t}_f(d,p) \in \I
    \end{array}
    \]
  \end{exa}
  \begin{thm}[$\CPreord$ is boolean]
  The doctrine of preorders $\CPreord$ is boolean, in the sense for every closed formula $[\,\emptyctx \mid \emptyctx \mid \emptyctx\,]\ \varphi$ the entailment $[\,\emptyctx \mid \emptyctx \mid \emptyctx\,]\ \vdash \varphi \lor (\varphi \Rightarrow \bot)$ is validated. Note that $\varphi$ can appear on both sides of the implication because it is a closed formula. In the semantics, this corresponds with the fact that $\forall p \in \I, p \sqcup (p \le 0) = 1$ in $\I$.

  \end{thm}
  \begin{defi}[Syntactic directed doctrine]\label{def:syntactic_directed_doctrine}
    Given a theory $\Sigma$, we define the \emph{syntactic doctrine} $\Syn(\Sigma) : \ndp{\Ctx}^\op \> \Pos$ as the syntactic directed doctrine inductively generated from $\Sigma$, as follows: $\Ctx$ is the category where objects are contexts $\Gamma,\Delta$, morphisms are term substitutions $\Gamma \vdash \Delta$, and ${\cP(\Theta\!\mid\!\Delta\!\mid\!\Gamma) := \set{[\Theta\!\mid\!\Delta\!\mid\!\Gamma]\ \Phi \propctx}}$, where the poset relation is given by multi-entailment judgements $\Phi \vdash \Psi$, defined just like term substitutions.
  \end{defi}

  \begin{thm}[$\Syn$ has no-dinatural-variance]
    $\Syn(\Sigma)$ satisfies the \emph{ndv} condition, i.e., there is a bijection $\set{[\Theta\mid\Delta\mid\Gamma]\ \varphi\propx} \iso \set{[\Theta,\Delta \mid\,\emptyctx\,\mid\Gamma,\Delta]\ \varphi\propx}.$
  \end{thm}
  \begin{proof}
  By induction on $\varphi$, using the fact that ${\set{{(\Theta,\Delta),{\emptyctx}} \vdash A}\!\iso\!\set{\Theta,\Delta \vdash A}}$ in the two base cases $\le_A$ and $P{(s \mid t)}$.
\end{proof}

Despite the fact that dFOL does not have a judgement for equality of entailments one can also give proof-relevant semantics by using categories as models; in this case entailments are interpreted as \emph{non-dinatural} transformations, i.e., indexed families of functions on which we impose no dinaturality condition. The rest of the structure is defined in the intuitive way, using limits and colimits in $\Cat$. This allows one to recover the semantics in symmetric monoidal categories described in \cite{Meseguer1990petri,Stehr2001rewriting} for Petri nets.

\begin{exa}[$\Cat$ doctrines]
  There is a directed doctrine $\Psh_{\textsf{no-dinat}} : \ndp{\Cat}^\op \> \CAT$ defined on the category of (small) categories $\Cat$ as follows:
    \[\begin{array}{r@{\,}l}
      \Psh_{\textsf{no-dinat}}(\Theta \mid \Delta \mid \Gamma) := [\sem{\Theta}^\op \times (\sem{\Delta}^\op \times \sem{\Delta}) \times \sem{\Gamma}, \Set]_{\textsf{no-dinat}}, \\[0.7em]
      \hom(\alpha, \beta) := \displaystyle \prod_{n \in \sem{\Theta}}\prod_{d \in \sem{\Delta}}\prod_{p \in \sem{\Gamma}}\ \alpha(n,d,d,p) \longrightarrow_\Set \beta(n,d,d,p)
    \end{array}
    \]
  by sending $(\Theta,\Delta,\Gamma)$ to the category where objects are (co)presheaves of shape $\sem{\Theta}^\op \times (\sem{\Delta}^\op \times \sem{\Delta}) \times \sem{\Gamma} \> \Set$, and morphisms between them are \emph{non-dinatural transformations}, i.e., indexed families of functions defined as above which are not required to satisfy any dinaturality condition. The lack of dinaturality is necessary because of the rule $\Rulecut$, which requires morphisms to be closed under composition and form a category. The rest of the structure (and the conditions for a directed doctrine) follows easily from the same ideas of the preorder case, using products and coproducts in $\Set$ to define quantifiers~\cite{Loregian2021coend}.
\end{exa}

\begin{exa}[$\Set$  doctrines]
  The $\Set$ doctrine of sets and subsets~\cite{Maietti2013quotient} can be lifted to a directed one by suitably precomposing the discrete poset functor $\Set \> \Pos$ and using functoriality of $\textsf{ndp}$ (\Cref{def:ndp_functor}).
\end{exa}

\begin{thm}
A directed doctrine $\cP : \ndp{\C}^\op\!\>\!\Pos$ induces a doctrine $\cP' : \C^\op\!\>\!\Pos$ by precomposing with the functor ${\Downarrow}\!:\!\C\!\to\!\ndp{\C}$ given by $C \mapsto (\top\!\mid\!C\!\mid\!\top)$; intuitively, this captures the non-directed ``sub-logic'' inside dinatural contexts. Such doctrine will have $\land$, $\Rightarrow$, $\forall$, $\exists$, $=$ (defined as in \Cref{thm:bidirectional_symmetric_equality}) if the directed doctrine is equipped with such structure (only dinatural quantifiers are needed).
A doctrine can be lifted to a directed one by precomposing with ${\Uparrow} : \ndp{\C} \to \C$ given by $(\Theta\!\mid\!\Delta\!\mid\!\Gamma) \mapsto \Theta \times \Delta \times \Delta \times \Gamma$, satisfying the \emph{ndv} condition.
\end{thm}

\section{Interpretation}\label{sec:interpretation}

We now describe the functorial semantics of dFOL, mostly following the   standard approach of functorial semantics à-la-Lawvere~\cite{Lawvere1963functorial,Jacobs1999categorical}.

\begin{defi}
  We denote the (2-)category of directed doctrines and their morphisms as $\mathsf{DDoctrine}$, as in \cite{Maietti2015unifying}: a \emph{morphism of directed doctrines} $\cP\!\to\!\cP'$ for $\cP\!:\!\ndp{\C}^\op\!\rightarrow\!\Pos$, $\cP'\!:\!\ndp{\D}^\op \rightarrow \Pos$ is defined as a pair $(F,\alpha)$, where $F\!:\!\C\!\to\!\D$ is a functor that preserves CCC structure, and $\alpha\!:\!\cP \Rightarrow \ndp{F}^\op \< \cP'$ is a natural transformation such that each functor \[\alpha_{\Theta,\Delta,\Gamma} : \cP(\Theta\mid\Delta\mid\Gamma) \> \cP'(F(\Theta),F(\Delta),F(\Gamma))\]
  preserves all the structure (i.e., terminals, products, directed equality, polarized implications and quantifiers) present in each fiber~\cite{Jacobs1999categorical}. For instance, preservation of directed equality means that \[\alpha_{\Theta,\Delta,\Gamma}({\le_A} {\land} \varphi) = \cP'({\iso_{\Theta A}} \mid \id \mid {\iso_{\Gamma A}})({\le'_{F(A)}}{\land} F(\varphi))\] where ${\iso_{AB} := \ang{F(\pi_1),F(\pi_2)}: F(A \times B) \to F(A) \times F(B)}$.

  A \emph{2-cell} ${(F,\alpha) \Rightarrow (G,\beta)}$ is defined by a natural transformation ${\theta : F \Rightarrow G}$ such that \[{\alpha_{\Theta,\Delta,\Gamma}(p) \le \cP'(\theta_{\Theta,\Delta}\mid\theta_{\Delta}\mid\theta_{\Gamma,\Delta})(\beta_{\Theta,\Delta,\Gamma}(p))},\] where $\theta_{\Theta,\Delta},\theta_{\Gamma,\Delta}$ are given in the intuitive way.
  \end{defi}
  \begin{defi}\label{def:theory_morphism}
    We define the \emph{1-category of theories} $\Theory$ where objects are theories $\Sigma$ and \emph{morphisms of theories} $M : {\Sigma \to \Sigma'}$ are defined as tuples of set functions $M := ({b : \Sigma_B \to \Sigma'_B},{p:\Sigma_P\to\Sigma'_P},$\-${f:\Sigma_F\to\Sigma'_F})$ such that
    \begin{itemize}[leftmargin=1em]
    \item $b \< \dom' = \dom \< \sem{-}_b$ and $b \< \cod' = \cod \< \sem{-}_b$,
    \item $\forall e \in \Sigma_E,\sem{\mathsf{actx}(e)}_{b\mathsf{ctx}} \vdash \sem{\mathsf{eqL}(e)}_f = \sem{\mathsf{eqR}(e)}_f : \sem{\mathsf{eqt(e)}}_b$ is derivable in $\Sigma'$,
    \item $p \< \mathsf{pos}' = \mathsf{pos} \< \sem{-}_p$ and $p \< \mathsf{neg}' = \mathsf{neg} \< \sem{-}_p$,
    \item $\forall a \in \Sigma_A,[\sem{\mathsf{actx}(a)}_{b\mathsf{ctx}}]\ \sem{\mathsf{hyp}(a)}_{p\mathsf{ctx}} \vdash \sem{\mathsf{conc}(a)}_{p\mathsf{ctx}}$ is derivable in $\Sigma'$,
    \end{itemize}where $\sem{-}_b$, $\sem{-}_{b\mathsf{ctx}}$, $\sem{-}_f$,  $\sem{-}_p$, $\sem{-}_{p\mathsf{ctx}}$ denote the translations on types induced by the functions above on types, contexts, terms, formulas, and propositional contexts respectively.
  \end{defi}
  \begin{thm}[Initial theory]\label{thm:initiality_empty_theory}
  There is a theory $\emptyset$ defined by always choosing the empty set in \Cref{def:theory}. Clearly $\emptyset$ is an initial object in $\Theory$.
  \end{thm}
  \begin{defi}[Underlying theory]\label{def:underlying_theory}
    Given a directed doctrine $\cP : \ndp{\C}^\op \> \Pos$, we define the \emph{underlying theory} $\Lang(\cP) \in \Theory$ as follows:
    \begin{itemize}[leftmargin=1em]
    \item base types $\Sigma_B := \C_0$ is the set of objects of $\C$,
    \item base terms $\Sigma_F := \C_1$ is the set of arrows of $\C$, with their $\dom,\cod$ functions;
    \item term equalities $\Sigma_E := \set{(A,\Gamma,s,t)\mid s,t \in \set{\Gamma\!\vdash\!t\!:\!A}, \sem{t}_f = \sem{s}_f}$ are terms (formed with $\Sigma_B/\Sigma_F$) for which their interpretation $\sem{-}_f$ is equal in $\C$.
    \item base formulas $\Sigma_P\!:=\!\coprod_{k\in C_0\times C_0 \times C_0} \cP(k)_0$ is the set of all objects in all fibers ${\cP(\Theta\mid\Gamma\mid\Delta)}$, with ${\mathsf{pos}(\Theta,\Delta,\Gamma,p) := \Theta\times\Delta}$, and $\mathsf{neg}(\Theta,\Delta,\Gamma,p) := \Gamma\times\Delta$. Such choice of $\textsf{pos},\textsf{neg}$ relies intuitively on \emph{ndv}, since specifically choosing the products $\Theta \times \Delta$ and $\Gamma \times \Delta$ covers \emph{every} predicate bijectively.
    \item the set $\Sigma_A$ has a symbol whenever $\psi \le \varphi$ holds for some $\psi,\varphi$ in any poset.
    \end{itemize}
  \end{defi}

  \begin{defi}[Model]
    A \emph{model} of a theory $\Sigma$ in $\cP$ is a doctrine morphism $\Syn(\Sigma) \to \cP$. When $\cP := \CPreord$, a model corresponds precisely with the classical notion of model one would expect, i.e., a preorder for each base type in $\Sigma_B$, a monotone function for each base term in $\Sigma_T$, etc.

  \end{defi}

We now establish that dFOL is the internal language of directed doctrines, which allows one to use the logic to reason about the properties of directed doctrines, with doctrine morphisms serving as a notion of model. We use the term ``internal language'' in the sense of \cite{Jacobs1999categorical,Pitts1995categorical,Barr1990category}.

\begin{thm}[Internal language correspondence]\label{thm:main_theorem}
    The above constructions are functorial, forming 2-functors $\Syn : \Theory \rightleftarrows \mathsf{DDoctrine} : \Lang$. Moreover, they form a bi-adjunction between 2-categories~\cite[2.2.5]{Jacobs1999categorical}, i.e., there is an equivalence of categories between the category $\Syn(\Sigma) \longrightarrow \cP \text{ in $\mathsf{DDoctrine}$}$ and the set (viewed as a discrete category) $\Sigma \longrightarrow \Lang(\cP) \text{ in $\Theory$}$ for any $\Sigma$ and $\cP$, as follows:
\[
\begin{prooftree}
  \hypo{\Syn(\Sigma) \longrightarrow \cP \text{ in $\mathsf{DDoctrine}$}}
  \infer[double]1{\Sigma \longrightarrow \Lang(\cP) \text{ in $\Theory$}}
\end{prooftree}
\]
    Additionally, we have that every directed doctrine $\cP$ is equivalent to the syntactic doctrine built from the internal language of $\cP$, i.e., $\Lang \< \Syn \cong \id_{\mathsf{DDoctrine}}$.
  \end{thm}
  \begin{proof}
    Detailed in \Cref{appendix:internal_language_correspondence}, using \emph{ndv} condition.
  \end{proof}

The above is not an adjoint equivalence, as in, e.g., \cite{Maietti2005modular,Maietti2005relating}, because we do not require the notion of theory and theory morphism in \Cref{def:theory} to be closed under the operators of the logic, and indeed the syntactic doctrine has strictly more predicates than the base predicates of the underlying theory.

Soundness and completeness for the syntax (i.e., initiality \cite{Pitts1995categorical}) follow as immediate corollaries from the 2-adjunction.
\begin{cor}[Soundness]
If an entailment is derivable in directed first-order logic in the empty theory then it holds in every directed doctrine $\cP$.
\end{cor}
\begin{proof}
By \Cref{thm:initiality_empty_theory}, $\emptyset$ is initial in $\Theory$, hence $\Syn(\emptyset)$ is also initial since $\Syn$ is a left adjoint. From \Cref{thm:main_theorem} we have a doctrine morphism $\Syn(\emptyset) \longrightarrow \cP$ which corresponds precisely with a model of the empty theory in $\cP$. Hence, we can apply this doctrine morphism to the derivation tree of the entailment given to obtain the corresponding internal statement in $\cP$.
\end{proof}
\begin{cor}[Completeness]
If an entailment holds in every directed doctrine $\cP$ then it is derivable in directed first-order logic in the empty theory.
\end{cor}
\begin{proof}
Since $\Syn(\emptyset)$ is a directed doctrine, the statement is also true in $\Syn(\emptyset)$. By definition of $\Syn(\emptyset)$, a statement is true internally to such doctrine precisely when it is derivable in the empty theory $\emptyset$.
\end{proof}

\section{$\CPreord$ completeness for classical dFOL}\label{poset_completeness}

We now prove that a classical version of dFOL is complete with respect to models in $\CPreord$, i.e., that every entailment that holds in all models in preorders is derivable in the syntax. The proof follows precisely the standard approach of building a canonical Henkin-style model from the syntax itself, as in~\cite{Dalen2013logic}; the presence of polarized contexts does not play a significant role since closed terms are always the same regardless of whether they are substituted for a positive, negative or dinatural variable. In the classical Henkin proof, derivability of equality formulas induces a quotient on the syntactically constructed base types: in our setting, the quotient construction is simply replaced by the preorder relation in the model, which is precisely defined to be derivability of directed equalities $\le_A$.

In this section we consider the following restricted version of dFOL, which we refer to as \emph{classical dFOL}:
\begin{itemize}[leftmargin=1em]
\item We add the classical axiom schema for double-negation elimination (\Cref{classical_formulas}) for every formula $[\Theta \mid \Delta \mid \Gamma]\ \varphi$: \[[\Theta \mid \Delta \mid \Gamma]\ (\varphi \Rightarrow \bot) \Rightarrow \bot \vdash \varphi.\]
\item We consider only $\times$ and $\top$ as type constructors and restrict ourselves only to a single sort $A$ in $\Sigma_B$, called \emph{domain}, which corresponds to the usual domain of first-order logic~\cite{Dalen2013logic}. Using the properties of products and singletons one can prove that, up-to-isomorphism, all contexts are isomorphic to the number $n\in \mathbb{N}$ of free variables that they declare. We assume w.l.o.g. that every function symbol has $A$ as codomain, since a function symbol which return a products is the same as a pair of functions, and function symbols returning $\top$ are trivial. Hence, terms are simply indicated as $i \vdash f:A$ with $i$ free variables and polarized contexts correspond to triples $[i \mid j \mid k]$ indicating the number of free variables for each of the three kinds of variables.

\item By a slight abuse of notation, we identify the set of axioms $\Sigma_A$ given in \Cref{def:sign_judg} with a set of \emph{closed} formulas, called \emph{sentences}, which we ask to be closed under syntactic entailment. This can be done w.l.o.g. because of \Cref{thm:closed_formulas}.
\end{itemize}
We report here the main definitions to highlight the differences with respect to the standard Henkin proof, although the general structure is precisely the same as the classical case (e.g., the use of Zorn's lemma in \Cref{thm:lindebaum}~\cite{Dalen2013logic}). We refer the reader to \Cref{appendix:classical_completeness} for details and proofs, reporting here only the general idea.

\begin{defi}[Henkin theory]
A theory $(\Sigma_L,\Sigma_A)$ is called a \emph{Henkin theory} if for every $p\in \set{-,\Delta,+}$ and formula $\varphi$ in context $[\,\emptyctx \mid \emptyctx \mid \emptyctx\,], [x\!:^p\!A]\ \varphi(x)$ there is a constant term $\top \vdash c\!:\!A$ in $\Sigma_F$ such that $\exists^{p} x.\varphi(x) \Rightarrow \varphi(c) \in \Sigma_A$.

\end{defi}

\begin{defi}[Henkin extension]
The \emph{Henkin extension} of a theory $(\Sigma_L,\Sigma_A)$ is a theory $\Sigma^* := (\Sigma_L^*,\Sigma^*_A)$ defined by
\begin{itemize}[leftmargin=1em]
\item adding a constant symbol $\top \vdash c_\varphi:A$ to $\Sigma_F^*$ for each sentence of the form $\exists^p x.\varphi(x)$ in $\Sigma_F$ for some $p\in\set{-,\Delta,+}$, and
\item adding a corresponding Henkin axiom $\exists^{p} x.\varphi(x) \Rightarrow \varphi(c_\varphi)$ to $\Sigma_A^*$.
\end{itemize}
\end{defi}

\begin{thm}[Henkin extension is conservative]
$(\Sigma_L^*,\Sigma^*_A)$ is conservative over $\Sigma$, in the sense that every $\varphi \in \Sigma^*_A$ which does not use the new constant symbols is already contained in $\Sigma_A$.
\end{thm}

Note that $(\Sigma_L^*,\Sigma^*_A)$ is not necessarily a Henkin theory, since new existential formulas that do not yet have witnesses. The standard technique is to iterate this process countably many times and take the union of the chain.

\begin{defi}[Henkin completion]
The \emph{Henkin completion} of a theory $(\Sigma_L,\Sigma_A)$ is the (Henkin) theory $\Sigma^\omega := (\bigcup_i \Sigma^i_L, \bigcup_i \Sigma^i_A)$ where $\Sigma^0 := (\Sigma_L,\Sigma_A)$ and $\Sigma^{n+1} := (\Sigma^n)^*$ is the Henkin extension of $\Sigma^n$.
\end{defi}

In order to build the model we must be able to decide whether an entailment is derivable or not; for this we extend the Henkin completion to a maximally consistent theory.

\begin{defi}[Maximally consistent extensions]
A theory $\Sigma$ is said to be \emph{maximally consistent} if for every closed formula $[\,\emptyctx \mid \emptyctx \mid \emptyctx\,]\ \varphi \propx$, either $\varphi \in \Sigma_A$ or $\varphi \Rightarrow \bot \in \Sigma_A$.
\end{defi}

\begin{thm}[Lindenbaum]\label{thm:lindebaum}
Every consistent theory $\Sigma := (\Sigma_L,\Sigma_A)$ has a maximally consistent extension $\Sigma^\textsf{max} := (\Sigma_L,\Sigma^\textsf{max}_A)$.
\end{thm}
\begin{proof}
Straightforward by Zorn's lemma~\cite[3.1.9]{Dalen2013logic}.
\end{proof}

\begin{lem}[Maximally consistent exts. preserve Henkin]
If $\Sigma$ is Henkin, then $\Sigma^{\textsf{max}}$ is again Henkin.
\end{lem}

\begin{thm}[Model existence lemma]\label{model_existence_lemma}
If a theory $\Sigma$ is consistent, then there is a model $\Syn(\Sigma) \longrightarrow  \CPreord$ of $\Sigma$.
\end{thm}
\begin{proof}
We report here only the model construction. We consider the Henkin completion $\Sigma^\omega$ of $\Sigma$, and then take its maximally consistent extension $\mathcal{S} := (\Sigma^{\omega})^{\textsf{max}}$, which is still consistent and Henkin. Hereafter $\vdash$ refers to derivability in $\mathcal{S}$ as theory. We build a syntactic model as follows:
\begin{itemize}[leftmargin=1em]
\item The domain $\sem{A} := \set{\top \vdash t:A}$ is defined as the preorder with the set of closed terms $\top \vdash t : A$ as objects and the relation $s \le_A t$ holds iff ${[\,\emptyctx \mid \emptyctx \mid \emptyctx\,]\ \vdash s \le_A t}$ is derivable in $\mathcal{S}$; since this relation coincides with directed equality, it is a preorder relation since it is provably reflexive and transitive, as shown in \Cref{ex:derivations}.
\item Each function symbol $f \in \mathcal{S}_F$ such that $\textsf{dom}(f) = n$ is interpreted as the monotone function $\sem{f} : \sem{A}^n \to \sem{A}$ simply defined by ${\sem{f}(t_1,\ldots,t_n) := f(t_1,\ldots,t_n)}$. Monotonicity is the fact that directed equality is monotone on terms, i.e., congruence in \Cref{ex:derivations}.
\item Each predicate symbol $P(p \mid n) \in \mathcal{S}_P$ is interpreted as the monotone function $\sem{P} : \sem{A}^\op \times \sem{A} \to \I$ defined by sending $P(s \mid t)$ to $1 \in \I$ iff ${[\,\emptyctx \mid \emptyctx \mid \emptyctx\,]\ \vdash P(s \mid t)}$ is derivable in $\mathcal{S}$, and to $0_\I$ otherwise. Monotonicity is the fact that directed equality is monotone on predicates, i.e., transport in \Cref{ex:derivations}.
\end{itemize}
Finally, we prove in \Cref{appendix:classical_completeness} that $1 \le \sem{\varphi} \iff [\,\emptyctx \mid \emptyctx \mid \emptyctx\,]\ \vdash \varphi$ by induction on $\varphi$.
\end{proof}

From the model existence lemma we derive completeness for $\CPreord$ in classical dFOL.

\begin{thm}[$\CPreord$ completeness]
For any theory $\Sigma$, if an entailment $[\Theta \mid \Delta \mid \Gamma]\ \Phi \vdash \varphi$ holds in every model $\Syn(\Sigma) \longrightarrow \CPreord$ of $\Sigma$, then it is derivable in classical dFOL.
\end{thm}

\section{Conclusion and Future Work}\label{sec:conclusions}
In this paper we introduced a sound and complete syntax for a directed first-order proof-irrelevant type theory with polarities and a ``directed'' notion of equality $\le_A$, which we characterized as a left relative adjoint. Our analysis of the universal property of directed equality further enlightens the original question by Lawvere on the precise role of $\hom$ and variance in logic~\cite{Lawvere1970equality}, which we analyzed here for the \emph{logic of preorders} and its categorical semantics.

The idea of polarized contexts is a further step towards a satisfactory syntax for directed type theory, which even in the first-order case with preorder semantics requires a careful treatment of polarity. Type dependency is particularly non-trivial, since it is not obvious what role variance should play when variables appear in the types themselves.

We saw how dFOL is only a slight departure from first-order logic, carrying over much of the standard results and allowing for it to be expanded in the same way that the latter is the basis of many logical systems. The notion of directed doctrine could be studied more precisely under the 2-categorical perspective, e.g., by giving algebraic characterizations of directed equality~\cite{Emmenegger2020elementary} or the study of their completions~\cite{Maietti2013quotient,Maietti2023characterization}.

Following the geometric intuition for HoTT, the theory of directed spaces and directed algebraic topology might prove to be a suitable model for our logic, enabling reasoning about concurrent systems~\cite{Fajstrup2016directed,North2019towards} with directed equality acting as a notion of reachability, or hemimetric spaces~\cite{Duque2023fixed}, possibly with a directed quantitative extension in the style of~\cite{Dagnino2025quantitative,Mardare2016quantitative}.

Another approach to capturing variances could be based on a multicategorical ``graded'' approach \cite{Levy2019locally}, tagging variables with polarities and using a binary operation that combines the polarity of variables together such that $- \otimes + = \Delta$.

A relatively straightfoward extension to this work is the addition of $A^\op$ types, as in \cite{Licata20112}: our work lays the foundation to capture this idea precisely, since we can state how positive occurrences of $A^\textsf{op}$ are equivalent to negative ones of $A$ in a polarized context; hence $\op$-types become a ``representable'' way (in the sense of \cite{Shulman2018contravariance}) to capture such swap between positive/negative variables.

\bibliographystyle{alphaurl}
\bibliography{iwilare}

\appendix

\section{Standard rules for symmetric equality}
\label{appendix:symmetric_equality}

We illustrate the rules of symmetric equality in first-order logic for the reader to compare them to our directed rules. We use the typical type theoretical formulation using $\Rulerefl$ and the $J$-rule, which characterize equality using the following introduction and elimination rules:
\[
\begin{adjustbox}{max width=\linewidth}
\begin{prooftree}
\infer0[\Rulerefl]{[\Gamma, x : A]\ \Phi & \vdash x = x}
\end{prooftree}
\begin{prooftree}
\hypo{[\Gamma, z : A]\ \takespace{a = b, \, \Phi(b, a) }{\Phi(z,z)} \vdash & P(z,z)}
\infer1[\RuleJ]{[\Gamma, a : A, b : A]\ a = b, \, \Phi(a,b) \vdash & P(a,b)}
\end{prooftree}
\end{adjustbox}
\]
\LinkRuleJ
\LinkRulerefl
The intuition behind the $J$-rule is the following: if one wants to prove a property $P(a,b)$ for any $a,b:A$ assuming that ${a = b}$, then it is enough to prove $P$ for the case $P(z,z)$ where $P$ is instantiated with the same variable $z:A$, and similarly for the context $\Phi(z,z)$.

Using these two rules one can automatically derive the usual properties of equality, e.g., that equality is symmetric, for $\Phi := [\,], P(a,b) := (b = a)$. Transitivity of symmetric equality follows similarly:
\[
\begin{adjustbox}{max width=\linewidth}
\begin{prooftree}
\infer0[\Rulerefl]{[z:A]\ \takespace{a = b}{ } & \vdash z = z}
\infer1[\RuleJ]{[a:A,b:A]\ a = b & \vdash b = a}
\end{prooftree}
\begin{prooftree}
\infer0[\Rulehyp]{[z:A]\ \takespace{a = b,\,b = c}{ z = c} & \vdash z = c}
\infer1[\RuleJ]{[a:A,b:A,c:A]\ a = b,\,b = c & \vdash a = c}
\end{prooftree}
\end{adjustbox}
\]

\section{Derived rules}
\label{appendix:derived_rules}
We show derivations for the rules given in \Cref{fig:derivable_rules} as well as additional ones.

\subsection{Derived rules for quantifiers}

\begin{itemize}[leftmargin=1em]
\item The rule $\Ruleexiststermminus$ follows by $\Rulecut$ the hypothesis with a generic propositional context on the left, assuming to have a term $\Theta,\Delta \vdash \eta : N$:
  \[
\begin{adjustbox}{max width=\linewidth}
\begin{prooftree}
\hypo{[\Theta \mid \Delta \mid \Gamma]\ \Phi \vdash \varphi(\eta)}
\infer0[\Rulehyp]{[\Theta \mid \Delta \mid \Gamma]\ \exists^- x.\varphi(x),\Phi \vdash \exists^- x.\varphi(x)}
\infer1[\Ruleexists]{[\Theta,x\!:\!N \mid \Delta \mid \Gamma]\ \varphi(x),\Phi \vdash \exists^- x.\varphi(x)}
\infer1[\Rulereindex]{[\Theta \mid \Delta \mid \Gamma]\ \varphi(\eta),\Phi \vdash \exists^- x.\varphi(x)}
\infer2[\Rulecut]{[\Theta \mid \Delta \mid \Gamma]\ \Phi \vdash \exists^- x.\varphi(x)}
\end{prooftree}
\end{adjustbox}
\]
\item The rule $\Ruleforalltermdelta$ follows by reindexing with an arbitrary term $\Delta \vdash \delta : A$:
  \[
  \begin{prooftree}
  \hypo{\Delta \vdash \delta : A}
  \hypo{[\Theta \mid \Delta \mid \Gamma]\ \Phi \vdash \forall^\Delta x.\varphi(\n x,x)}
  \infer1[\Ruleforall]{[\Theta \mid \Delta,x\!:\!A \mid \Gamma]\ \Phi \vdash \varphi(\n x,x)}
  \infer2[\Rulereindex]{[\Theta \mid \Delta \mid \Gamma]\ \Phi \vdash \varphi(\delta,\delta)}
  \end{prooftree}
  \]
\item The following rules for quantifiers follow precisely the same idea:
\[
\begin{array}{c}
\begin{prooftree}
\hypo{\Delta,\Gamma \vdash \rho : P}
\hypo{[\Theta \mid \Delta \mid \Gamma]\ \Phi \vdash \varphi(\rho)}
\infer2[\Ruleexiststermplus]{[\Theta \mid \Delta \mid \Gamma]\ \Phi \vdash \exists^+ x.\varphi(x)}
\end{prooftree}\\[1.25em]
\begin{prooftree}
\hypo{\Delta,\Gamma \vdash \rho : P}
\hypo{[\Theta \mid \Delta \mid \Gamma]\ \Phi \vdash \forall^+ x.\varphi(x)}
\infer2[\Ruleforalltermplus]{[\Theta \mid \Delta \mid \Gamma]\ \Phi \vdash \varphi(\rho)}
\end{prooftree}\\[1.25em]
\begin{prooftree}
\hypo{\Theta,\Delta \vdash \eta : N}
\hypo{[\Theta \mid \Delta \mid \Gamma]\ \Phi \vdash \varphi(\eta)}
\infer2[\Ruleexiststermminus]{[\Theta \mid \Delta \mid \Gamma]\ \Phi \vdash \exists^- x.\varphi(x)}
\end{prooftree}\\[1.25em]
\begin{prooftree}
\hypo{\Theta,\Delta \vdash \eta : N}
\hypo{[\Theta \mid \Delta \mid \Gamma]\ \Phi \vdash \forall^- x.\varphi(x)}
\infer2[\Ruleforalltermminus]{[\Theta \mid \Delta \mid \Gamma]\ \Phi \vdash \varphi(\eta)}
\end{prooftree}\\[1.25em]
\begin{prooftree}
\hypo{\Delta \vdash \delta : D}
\hypo{[\Theta \mid \Delta \mid \Gamma]\ \Phi \vdash \varphi(\delta,\delta)}
\infer2[\Ruleexiststermdelta]{[\Theta \mid \Delta \mid \Gamma]\ \Phi \vdash \exists^\Delta x.\varphi(\n x,x)}
\end{prooftree}\\[1.25em]
\begin{prooftree}
\hypo{\Delta \vdash \delta : D}
\hypo{[\Theta \mid \Delta \mid \Gamma]\ \Phi \vdash \forall^\delta x.\varphi(x)}
\infer2[\Ruleforalltermdelta]{[\Theta \mid \Delta \mid \Gamma]\ \Phi \vdash \varphi(\delta,\delta)}
\end{prooftree}
\end{array}
\]
\LinkRuleexiststermplus
\LinkRuleforalltermplus
\LinkRuleforalltermminus
\LinkRuleexiststermdelta
\LinkRuleexiststermminus
\LinkRuleforalltermdelta
\LinkRuleforalldelta
\end{itemize}

\subsection{Derived rules for polarized implication}

In this section we present a series of rules regarding polarized implication. We show several cases explicitly to convince the reader that the formal rules of polarized implication, when combined with $\Rulereindex$ and $\Ruleupgrade$, follow precisely the intuition that variables invert variance. In the rest of the examples here we occasionally do not specify which specific derived rule for $\Ruleimpl$ is being used.

\begin{itemize}[leftmargin=1em]
\item We explicitly show the derivation for $\Ruleimplplusr$, where $x$ appears dinaturally on the top and $[\Theta \mid \Delta \mid \Gamma,x\!:\!A]\ \Phi(x) \propctx$, $[\Theta',x\!:\!A \mid \Delta \mid \Gamma']\ \takespace[r]{\Phi(x)}{\psi(x)} \propx$:
\[
\begin{prooftree}
\infer[no rule]0{[\Theta,\Theta' \mid \Delta,x\!:\!A \mid \Gamma,\Gamma']\ \psi(\n x), \Phi(x) \vdash \varphi(x)}
\infer[double]1[\Ruleimplplusr]{[\Theta,\Gamma' \mid \Delta \mid \Gamma,\Theta',x\!:\!A]\ \Phi(x) \vdash \psi(x) \Rightarrow \varphi(x)}
\end{prooftree}
\]
\LinkRuleimplplusr
We show both directions separately.

Top-to-bottom:
\[
\begin{prooftree}
\infer[no rule]0{[\Theta,\Theta' \mid \Delta,x\!:\!A \mid \Gamma,\Gamma']\ \psi(\n x), \Phi(x) \vdash \varphi(x)}
\infer1[\Rulereindex]{[\,\emptyctx \mid \Delta,\Theta,\Theta',\Gamma,\Gamma',x\!:\!A \mid \emptyctx\,]\ \psi(\n x), \Phi(x) \vdash \varphi(x)}
\infer1[\Ruleimpl]{[\Theta,\Gamma' \mid \Delta \mid \Gamma,\Theta',x\!:\!A]\ \Phi(x) \vdash \psi(x) \Rightarrow \varphi(x)}
\end{prooftree}
\]
Bottom-to-top:
\[
\begin{prooftree}
\infer[no rule]0{[\Theta,\Gamma' \mid \Delta \mid \Gamma,\Theta',x\!:\!A]\ \Phi(x) \vdash \psi(x) \Rightarrow \varphi(x)}
\infer1[\Ruleimpl]{[\,\emptyctx \mid \Delta,\Theta,\Theta',\Gamma,\Gamma',x\!:\!A \mid \emptyctx\,]\ \psi(\n x), \Phi(x) \vdash \varphi(x)}
\infer1[\Ruleupgrade]{[\Theta,\Theta' \mid \Delta,x\!:\!A \mid \Gamma,\Gamma']\ \psi(\n x), \Phi(x) \vdash \varphi(x)}
\end{prooftree}
\]

In the top-to-bottom direction, the rule $\Ruleimpl$ can be applied because both $\psi$ and $\varphi$ are reindexed with the same reindexing operation that is used implicitly in the top side of $\Ruleimpl$.

Rule $\Ruleupgrade$ can be used because the variables of $\Theta,\Gamma$ in $\Phi,\varphi$ are implicitly lifted to be dinatural, which is precisely the operation used in the top side of $\Ruleupgrade$.

\item Other variants $\Ruleimplminusl,\Ruleimplminusr$ with a negative variable in the context are defined identically.
\LinkRuleimplminusl
\LinkRuleimplminusr
\item We show the rule $\Ruleimplinvert$ that inverts singular appearances of variables in $\psi$, where $x,y$ do not appear in $\Phi$ and $\varphi$, i.e., we assume
${[\Theta \mid \Delta \mid \Gamma]\ \Phi \propctx}$ and
${[x\!:\!A \mid \Delta \mid y\!:\!B]\ \psi(x,y) \propx}$. Note that $\psi$ does not depend on $\Theta,\Gamma$, since we want occurrences to switch polarity from negative to positive directly, hence if any variable appeared in $\Phi,\varphi$ they would appear as dinatural in either top or bottom side of the rule.
\[
\begin{prooftree}
\infer[no rule]0{[\Theta,x\!:\!A \mid \Delta \mid \Gamma,y\!:\!B]\ \psi(x,y), \Phi \vdash \varphi}
\infer[double]1[\Ruleimplinvert]{[\Theta,y\!:\!B \mid \Delta \mid \Gamma,x\!:\!A]\ \Phi \vdash \psi(x,y) \Rightarrow \varphi}
\end{prooftree}
\]
\LinkRuleimplinvert
we show both directions separately.

Top-to-bottom:
\[
\begin{prooftree}
\infer[no rule]0{[\Theta,x\!:\!A \mid \Delta \mid \Gamma,y\!:\!B]\ \psi(x,y), \Phi & \vdash \varphi}
\infer1[\Rulereindex]{[\,\emptyctx \mid \Delta,\Theta,\Gamma,x\!:\!A,y\!:\!B \mid \emptyctx\,]\ \psi(x,y), \Phi & \vdash \varphi}
\infer1[\Ruleimpl]{[\Theta,y\!:\!B \mid \Delta \mid \Gamma,x\!:\!A]\ \takespace[r]{\psi(x,y),\Phi}{\Phi} &\vdash \psi(x,y) \Rightarrow \varphi}
\end{prooftree}
\]
Bottom-to-top:
\[
\begin{prooftree}
\infer[no rule]0{[\Theta,y\!:\!B \mid \Delta \mid \Gamma,x\!:\!A]\ \takespace[r]{\psi(x,y),\Phi}{\Phi} & \vdash \psi(x,y) \Rightarrow \varphi}
\infer1[\Ruleimpl]{[\,\emptyctx \mid \Delta,\Theta,\Gamma,x\!:\!A,y\!:\!B \mid \emptyctx\,]\ \psi(x,y), \Phi & \vdash \varphi}
\infer1[\Ruleupgrade]{[\Theta,x\!:\!A \mid \Delta \mid \Gamma,y\!:\!B]\ \psi(x,y), \Phi & \vdash \varphi}
\end{prooftree}
\]

\item We prove more general rules $\RuleimplL,\RuleimplR$ for $\Rightarrow$ in which $\Phi,\varphi$ do not use $N$ or $P$, respectively (of which $\Ruleimplplusr,\Ruleimplplusl$ and $\Ruleimplminusr,\Ruleimplminusl$ are special cases):
\[
\begin{array}{c}
\begin{prooftree}
\infer[no rule]0{\takespace{[\Theta,N \mid \Delta \mid \Gamma,P]}{[N \mid \Delta \mid P]}\ \takespace[l]{\Phi \propctx,\,\varphi \propx}{\psi \propx}}
\infer[no rule]1{[\Theta,N \mid \Delta \mid \Gamma,P]\ \Phi \propctx,\,\varphi \propx}
\infer[no rule]1{[\Theta,N \mid \Delta \mid \Gamma,P]\ \psi, \Phi & \vdash \varphi}
\infer[double]1[\RuleimplL]{[\Theta  \mid \Delta,N,P \mid \Gamma]\ \takespace{\psi, \Phi}{\Phi} & \vdash \psi \Rightarrow \varphi}
\end{prooftree}
\\[2.0em]
\begin{prooftree}
\infer[no rule]0{\takespace{[\Theta,N \mid \Delta \mid \Gamma,P]}{[P \mid \Delta \mid N]}\ \takespace[l]{\Phi \propctx,\,\varphi \propx}{\psi \propx}}
\infer[no rule]1{[\Theta,N \mid \Delta \mid \Gamma,P]\ \Phi \propctx,\,\varphi \propx}
\infer[no rule]1{[\Theta \mid \Delta,N,P \mid \Gamma]\ \psi, \Phi & \vdash \varphi}
\infer[double]1[\RuleimplR]{[\Theta,N \mid \Delta \mid \Gamma,P]\ \takespace{\psi, \Phi}{\Phi} & \vdash \psi \Rightarrow \varphi}
\end{prooftree}
\end{array}
\]
\LinkRuleimplplusr
\LinkRuleimplplusl
We show $\RuleimplL$, showing both directions for ${[N \mid \Delta \mid P]\ \psi}$, and ${[\Theta,N \mid \Delta \mid \Gamma,P]}\ {\Phi,\,\varphi}$:
\[
\begin{array}{c}
\begin{prooftree}
\infer[no rule]0{[\Theta,N \mid \Delta \mid \Gamma,P]\ \psi, \Phi & \vdash \varphi}
\infer1[\Rulereindex]{[\,\emptyctx \mid \Delta,N,P,\Theta,\Gamma \mid \emptyctx\,]\ \psi, \Phi & \vdash \varphi}
\infer1[\Ruleimpl]{[\Theta \mid \Delta,N,P \mid \Gamma]\ \takespace{\psi, \Phi}{\Phi} & \vdash \psi \Rightarrow \varphi}
\end{prooftree}
\\[2.0em]
\begin{prooftree}
\infer[no rule]0{[\Theta \mid \Delta,N,P \mid \Gamma]\ \takespace{\psi, \Phi}{\Phi} & \vdash \psi \Rightarrow \varphi}
\infer1[\Ruleimpl]{[\,\emptyctx \mid \Delta,N,P,\Theta,\Gamma \mid \emptyctx\,]\ {\psi, \Phi} & \vdash \varphi}
\infer1[\Ruleupgrade]{[\Theta,N \mid \Delta \mid \Gamma,P]\ \psi, \Phi & \vdash \varphi}
\end{prooftree}
\end{array}
\]
\LinkRuleimplL
We show $\RuleimplR$, showing both directions for ${[P \mid \Delta \mid N]\ \psi}$, and ${[\Theta,N \mid \Delta \mid \Gamma,P]}\ {\Phi,\,\varphi}$:
\[
\begin{array}{c}
\begin{prooftree}
\infer[no rule]0{[\Theta \mid \Delta,N,P \mid \Gamma]\ \psi, \Phi & \vdash \varphi}
\infer1[\Rulereindex]{[\,\emptyctx \mid \Delta,N,P,\Theta,\Gamma \mid \emptyctx\,]\ \psi, \Phi & \vdash \varphi}
\infer1[\Ruleimpl]{[\Theta,N \mid \Delta \mid \Gamma,P]\ \takespace{\psi, \Phi}{\Phi} & \vdash \psi \Rightarrow \varphi}
\end{prooftree}
\\[2.0em]
\begin{prooftree}
\infer[no rule]0{[\Theta,N \mid \Delta \mid \Gamma,P]\ \takespace{\psi, \Phi}{\Phi} & \vdash \psi \Rightarrow \varphi}
\infer1[\Ruleimpl]{[\,\emptyctx \mid \Delta,N,P,\Theta,\Gamma \mid \emptyctx\,]\ {\psi, \Phi} & \vdash \varphi}
\infer1[\Ruleupgrade]{[\Theta \mid \Delta,N,P \mid \Gamma]\ \psi, \Phi & \vdash \varphi}
\end{prooftree}
\end{array}
\]
\LinkRuleimplR
In both derivations above, the rule $\Ruleupgrade$ can be used because $\psi$ does not use $\Theta,\Gamma$, hence both $\Theta$ and $\Gamma$ are implicitly weakened/contracted to dinatural. On the other hand, $N,P$ already appear naturally in $\Phi,\varphi$.

Similarly, the rule $\Ruleimpl$ can only be applied (in both cases) because both $\Phi$ and $\varphi$ are reindexed with the same weakening/contraction operation that is used implicitly in the top side of $\Ruleimpl$.
\end{itemize}

\subsection{Derived rules for directed equality}

\LinkRulelefull
\LinkRulelerefl
\LinkRuleleplus
\LinkRulelereflterm
\LinkRuleleterm
\LinkRuleleminus

\begin{itemize}[leftmargin=1em]
\item The rule $\Rulelereflterm$ is obtained by $\Rulereindex$-ing the only dinatural variable in $\Rulelerefl$ with an arbitrary term $\Delta \vdash \delta : A$:
\[
\begin{prooftree}
\infer0[\Rulelerefl]{[\Theta \mid \Delta, z:A \mid \Gamma]\ {\Phi} & \vdash \n z \le_A z}
\infer1[\Rulereindex]{[\Theta \mid \Delta \mid \Gamma]\ {\Phi} & \vdash \delta \le_A \delta}
\end{prooftree}
\]
\item The rule $\Rulelefull$  is a version $\Rulele$ that allows for variables to appear \emph{negatively} in the context hypothesis $\Phi$, using polarized exponentials for $[\Theta,a:A \mid \Delta \mid \Gamma,b:A]\  \Phi(a,b) \propctx$, $\varphi(a,b) \propx$ in the same context:
\[
\begin{prooftree}
  \hypo{[\Theta \mid \Delta, z:A \mid \Gamma]\ {\Phi(\n z,z)} & \vdash \varphi(\nzz)}
  \infer[double]1[\RuleimplL]{[\,\emptyctx \mid \Delta,\Theta,\Gamma,z:A \mid \emptyctx\,]\  & \vdash \Phi(z,\n z) \Rightarrow \varphi(\n z,z)}
  \infer[double]1[\Rulele]{[a:A \mid \Delta,\Theta,\Gamma \mid b:A]\ {a \le b} & \vdash \Phi(b,a) \Rightarrow \varphi(a,b)}
  \infer[double]1[\RuleimplR]{[\Theta \mid \Delta,a:A,b:A \mid \Gamma]\ {a \le b,\Phi(\n b,\n a)} & \vdash \varphi(a,b)}
\end{prooftree}
\]
\item The rule $\Ruleleplus$ follows precisely the same idea as $\Rulele$ and $\Rulelefull$, but allows for positive variables to be directly kept positively (rather than dinatural, as it would be the case in $\Rulele$) whenever they appear as such.
\[
\begin{prooftree}
  \hypo{[\Theta\mid\Delta \mid \Gamma,a:A]\ \Phi \propctx}
  \infer[no rule]1{[\Theta\mid\Delta \mid \Gamma, b:A]\ \varphi \propctx}
  \infer[no rule]1{[\Theta \mid \Delta \mid \Gamma, z:A]\ \takespace{a \le b}{\Phi(z)} & \vdash \varphi(z)}
  \infer[double]1[\RuleimplL]{[\,\emptyctx \mid \Delta,\Theta,\Gamma,z:A \mid \emptyctx\,]\ \takespace{a \le b}{ } & \vdash \Phi(\n z) \Rightarrow \varphi(z)}
  \infer[double]1[\Rulele]{[a:A \mid \Delta,\Theta,\Gamma \mid b:A]\ {a \le b} & \vdash \Phi(a) \Rightarrow \varphi(b)}
  \infer[double]1[\RuleimplR]{[\Theta \mid \Delta,a:A \mid \Gamma,b:A]\ {\n a \le b,\Phi(a)} & \vdash \varphi(b)}
\end{prooftree}
\]
\item The following rule $\Ruleleminus$ is identical but using directed equality on ``contravariant'' predicates which have negative positions, instead.
\[
\begin{prooftree}
  \hypo{[\Theta,b:A\mid\Delta \mid \Gamma]\ \Phi \propctx}
  \infer[no rule]1{[\Theta,a:A\mid\Delta \mid \Gamma]\ \varphi \propctx}
  \infer[no rule]1{[\Theta, z:A \mid \Delta \mid \Gamma]\ \takespace{a \le b}{\Phi(z)} & \vdash \varphi(z)}
  \infer[double]1[\RuleimplL]{[\,\emptyctx \mid \Delta,\Theta,\Gamma,z:A \mid \emptyctx\,]\ \takespace{a \le b}{ } & \vdash \Phi(z) \Rightarrow \varphi(\n z)}
  \infer[double]1[\Rulele]{[a:A \mid \Delta,\Theta,\Gamma \mid b:A]\ {a \le b} & \vdash \Phi(b) \Rightarrow \varphi(a)}
  \infer[double]1[\RuleimplR]{[\Theta,a:A \mid \Delta,b:A \mid \Gamma]\ {a \le b,\Phi(\n b)} & \vdash \varphi(a)}
\end{prooftree}
\]
\item Rule $\Ruleleterm$ follows from $\Rulelefull$ by $\Rulereindex$ing with terms $\Delta \vdash \eta : A$ and $\Delta \vdash \rho : A$.
\[
\begin{prooftree}
  \infer[no rule]0{\hspace{-9.48em}\Delta \vdash \eta : A, \quad \Delta \vdash \rho : A}
  \infer[no rule]1{\hspace{-9.48em}[\Theta,a:A\mid \Delta \mid \Gamma,b:A] \takespace[l]{\ \Phi(\n z,z)}{\ \Phi(a,b) \propctx,\varphi(a,b) \propx} & }
  \infer[no rule]1{[\Theta \mid \Delta, z:A \mid \Gamma]\ {\Phi(\n z,z)} & \vdash \varphi(\nzz)}
  \infer1[\Rulelefull]{[\Theta \mid \Delta,a:A,b:A \mid \Gamma]\ {a \le b,\Phi(\n b,\n a)} & \vdash \varphi(a,b)}
  \infer1[\Rulereindex]{[\Theta \mid \Delta \mid \Gamma]\ {\eta \le \rho,\Phi(\rho,\eta)} & \vdash \varphi(\eta,\rho)}
\end{prooftree}
  \]

\end{itemize}

\subsection{Other examples}\label{appendix:other_examples}

\begin{itemize}[leftmargin=1em]
\item The following formulas involving quantifiers and polarized implication are equivalent, for some predicate $P$ not depending on either $x,y$:
\[
\begin{prooftree}
\hypo{[\Theta,y:A \mid \Delta \mid \Gamma]\ \Phi & \vdash (\exists^- x.x \le y) \Rightarrow P}
\infer[double]1[\Ruleimplinvert]{[\Theta \mid \Delta \mid \Gamma,y:A]\ (\exists^- x.x \le y), \Phi & \vdash P}
\infer[double]1[\Ruleexiststermminus]{[\Theta,x:A \mid \Delta \mid \Gamma,y:A]\ x \le y, \Phi & \vdash P}
\infer[double]1[\Ruleimplinvert]{[\Theta,y:A \mid \Delta \mid \Gamma,x:A]\ \Phi & \vdash x \le y \Rightarrow P}
\infer[double]1[\Ruleforalltermplus]{[\Theta,y:A \mid \Delta \mid \Gamma]\ \Phi & \vdash \forall^+ x. (x \le y \Rightarrow P)}
\end{prooftree}
\]
\item We show that the rule for double negation elimination (DNE) is equivalent to the law of excluded middle (LEM) even in the presence of polarized variables. We show explicitly the DNE to LEM direction to illustrate how the variance is still tracked in these classical proofs, and that it does not pose any technical problems. The proofs are essentially the same as the standard ones~\cite{Dalen2013logic,Mimram2020program}. For convenience, let $h := (\varphi(x,\n y,y,z) \lor (\varphi(\n x,y,\n y,\n z) \Rightarrow \bot)) \Rightarrow \bot$ and $\n \varphi := \varphi(\n x,y,\n y,\n z)$, and similarly with $\n h$.
\[
\begin{adjustbox}{max width=\linewidth}
\begin{prooftree}
  \infer0[\Rulehyp]{[...]\ h,\varphi \vdash h}
  \infer0[\Rulehyp]{[...]\ h,\varphi \vdash \varphi}
  \infer1[\Ruleor]{[...]\ h,\varphi \vdash \varphi \lor (\n \varphi \Rightarrow \bot)}
  \infer2[\Rulecut]{[...]\ h,\varphi \vdash \bot}
  \infer1{[...]\ h \vdash \n \varphi \Rightarrow \bot}
  \infer1[\Ruleor]{[...]\ h \vdash \varphi \lor (\n \varphi \Rightarrow \bot)}

  \infer0[\Rulehyp]{[...]\ h \vdash h}

  \infer2[\Rulecut]{[...]\ h \vdash \bot}
  \infer1[\Ruleimplinvert]{[...]\ \vdash (\n h \Rightarrow \bot)}
  \infer1[\seqrule{dne}]{[\,\emptyctx \mid x\!:\!N,y\!:\!D,z\!:\!P \mid \emptyctx\,]\ \vdash \varphi \lor (\n \varphi \Rightarrow \bot)}
\end{prooftree}
\end{adjustbox}
\]
\item We prove that, assuming the rule for double negation elimination, $\exists^p x.\varphi(x)$ is equivalent to $\neg \forall^{p^\op} x.\neg \varphi(x)$, for $p \in \{+,-\}$ and $p^\op$ its opposite polarity. Assume for simplicity that $[\,\emptyctx \mid \emptyctx \mid \emptyctx\,], [x :^p A]\ \varphi(x) \propx$ in order to avoid having to deal with fully-dinatural contexts, although the equivalence can be proven similarly.

We first prove that $\neg \exists^p x.\varphi(x)$ is equivalent to $\forall^{p^\op} x.\neg \varphi(x)$ (note that these are bijections):
\[
\begin{prooftree}
\hypo{[\,\emptyctx \mid \emptyctx \mid \emptyctx\,]\ \Phi \vdash \neg \exists^p x, \varphi(x)}
\infer[double]1[\Ruleimpl]{[\,\emptyctx \mid \emptyctx \mid \emptyctx\,]\ \Phi, (\exists^p x, \varphi(x)) \vdash \bot}
\infer[double]1[\Ruleexists]{[\,\emptyctx \mid \emptyctx \mid \emptyctx\,], [x:^p A]\ \Phi, \varphi(x) \vdash \bot}
\infer[double]1[\Ruleimpl]{[\,\emptyctx \mid \emptyctx \mid \emptyctx\,], [x:^{p^\op}A]\ \Phi \vdash \neg \varphi(x)}
\infer[double]1[\Ruleforall]{[\,\emptyctx \mid \emptyctx \mid \emptyctx\,]\ \Phi \vdash \forall^{p^\op} x.\neg \varphi(x)}
\end{prooftree}
\]
Finally, we prove using LEM (which, as shown above follows from DNE just like in the classical case) that for any closed formula $\neg \varphi \Rightarrow \neg \psi$ implies $\psi \Rightarrow \varphi$, from which the initial claim easily follows using (anti)monotonicity of $\neg$. We report the proof for completeness although this is precisely like the classical case since it is enough to prove it in the empty context:
\[
\begin{prooftree}
\infer0[\Rulehyp]{[\,\emptyctx \mid \emptyctx \mid \emptyctx\,]\ \Phi, \psi, \varphi \vdash  \varphi}
\infer1[\Ruleimpl]{[\,\emptyctx \mid \emptyctx \mid \emptyctx\,]\ \Phi, \varphi \vdash \psi \Rightarrow \varphi}

\infer0[\Ruleimpl]{[\,\emptyctx \mid \emptyctx \mid \emptyctx\,]\ \Phi, \neg \psi, \psi \vdash \bot}
\infer1[\Rulebot+\Rulecut]{[\,\emptyctx \mid \emptyctx \mid \emptyctx\,]\ \Phi, \neg \psi, \psi \vdash \varphi}
\infer1[\seqrule{hyp}]{[\,\emptyctx \mid \emptyctx \mid \emptyctx\,]\ \Phi, \neg \varphi, \psi \vdash \varphi}
\infer1[\Ruleimpl]{[\,\emptyctx \mid \emptyctx \mid \emptyctx\,]\ \Phi, \neg \varphi \vdash \psi \Rightarrow \varphi}
\infer2[\Ruleor]{[\,\emptyctx \mid \emptyctx \mid \emptyctx\,]\ \Phi, \psi \lor \neg \varphi \vdash \psi \Rightarrow \varphi}
\end{prooftree}
\]
\end{itemize}

\section{Relative (co)adjunctions}
\label{appendix:relative_adjunction}

We recall here for the reader's convenience an elementary description of (co)relative adjunctions and (co)relative adjoints~\cite{Arkor2024formal,Ulmer1968properties}.

\begin{defi}[Relative adjunction {\cite[5.1]{Arkor2024formal}}]
  \label{def:relative_adjunction}
      Consider the following arrangement of categories and functors:
      \[\xymatrix{
      & \D \ar[dr]^-R \ar@{}[d]|(.6){\dashv} & \\
      \C \ar[ur]^-L\ar[rr]_-J && \X
      }\]
      In this situation, we say that \emph{$L$ is the left $J$-relative left adjoint to $R$}, written $L \dashv_J R$ and indicated in the above diagram with a central `\,$\dashv$\,', if there is a bijection
      \[\D(L(x),y)\cong \X(J(x),R(y))\]
      natural in both arguments $x : \C , y : \D$.
\end{defi}
\begin{defi}[Relative coadjunction {\cite[7.3]{Arkor2024formal}}]
  \label{def:relative_coadjunction}
      Consider the following arrangement of categories and functors:
      \[\xymatrix{
      & \D \ar[dr]^-L \ar@{}[d]|(.6){\vdash} & \\
      \C \ar[ur]^-R\ar[rr]_-J && \X
      }\]
      In this situation, we say that \emph{$L$ is the left $J$-relative coadjoint to $R$}, written $R \vdash_J L$ and indicated in the above diagram with a central `\,$\vdash$\,', if there is a bijection
      \[\X(L(x),J(y))\cong \D(x,R(y))\]
      natural in both arguments $x : \D , y : \C$.
\end{defi}
One obtains the standard definition of adjunction when $\X=\C$, $J = \id_\C$.
In the case of posets, the above adjointness condition simply becomes a bi-implication of conditions.

\section{Directed equality as isomorphism}
\label{appendix:rule_j_equivalent}

\begin{thm}
We show that the rules for directed equality $\Rulelerefl$ and $\Rulele$ can be equivalently captured by the top-to-bottom direction of $\Rulele$ and its bottom-to-top $\Ruleleinv$, i.e., $\Rulele$ being an isomorphism fully characterizes directed equality.
\end{thm}
\begin{proof}
In one direction, we assume the following rule:
\[
\begin{prooftree}
\hypo{[\Theta,a:A \mid \Delta \mid \Gamma,b:A]\ {a \le b,\Phi} & \vdash \varphi(a,b)}
\infer1[\Ruleleinv]{[\Theta \mid \Delta, z:A \mid \Gamma]\ \takespace{a \le b,\Phi}{\Phi} & \vdash \varphi(\nzz)}
\end{prooftree}
\]
\LinkRuleleinv
we derive $\Rulelerefl$ simply by picking $\varphi(a,b) := a \le b$ in context $[\Theta,a:A \mid \Delta \mid \Gamma,b:A]$:
\[
\begin{prooftree}
\infer0[\Rulehyp]{[\Theta,a:A \mid \Delta \mid \Gamma,b:A]\ {a \le b,\Phi} & \vdash a \le b}
\infer1[\Ruleleinv]{[\Theta \mid \Delta, z:A \mid \Gamma]\ \takespace{a \le b,\Phi}{\Phi} & \vdash \n z \le z}
\end{prooftree}
\]
In the other direction, we assume the rule $\Rulelerefl$ and derive $\Ruleleinv$:
\[
\begin{adjustbox}{max width=\linewidth}
\begin{prooftree}
\infer0[\Rulelerefl]{[\Theta \mid \Delta,z:A \mid \Gamma]\ {\Phi} & \vdash \n z \le z}
\hypo{[\Theta,a:A \mid \Delta \mid \Gamma,b:A]\ {a \le b,\Phi} & \vdash \varphi(a,b)}
\infer1[\Rulereindex]{[\Theta \mid \Delta,z:A \mid \Gamma]\ {\n z \le z,\Phi} & \vdash \varphi(\n z,z)}
\infer2[\Rulecut]{[\Theta \mid \Delta, z:A \mid \Gamma]\ {\Phi} & \vdash \varphi(\nzz)}
\end{prooftree}
\end{adjustbox}
\]

\end{proof}

\section{Symmetric equality via directed equality}
\label{appendix:symmetric_equality_j}
\begin{thm}[Symmetric equality via directed equality]
We show how the notion of symmetric equality using directed equality $a = b := \n a \le b \land \n b \le a$ in context $[\,\emptyctx \mid a,b:A \mid \emptyctx\,]$, defined in \Cref{thm:bidirectional_symmetric_equality}, satisfies the characterization of symmetric equality as left adjoint to reindexing, i.e., we prove the following statement,
\[
\begin{prooftree}
\infer0{[z:A]\ \Phi & \vdash P(z,z)}
\infer1{[a:A,b:A]\ a = b, \Phi & \vdash P(a,b)}
\end{prooftree}
\]
which in our directed case looks like this, with no restrictions on the appearance of variables,
\[
\begin{prooftree}
\infer0{[\Theta \mid \Delta,z:A \mid \Gamma]\ \Phi & \vdash P(\n z,z,z,\n z)}
\infer1{[\Theta \mid \Delta,a:A,b:A \mid \Gamma]\ \n a \le b \land \n b \le a, \Phi & \vdash P(\n a,a,\n b,b)}
\end{prooftree}
\]
assuming $[\Theta,b:A,c:A \mid \Delta \mid \Gamma,a:A,d:A]\ P(a,b,c,d) \propx$. Note that this corresponds precisely with the characterization of symmetric equality as left adjoint to contraction \emph{but in the dinatural context}~\cite{Jacobs1999categorical}.

\end{thm}
\begin{proof}
We prove this using a version of transport (as in \Cref{ex:derivations}) with two variables (one positive and one negative), and then reindex these variables by collapsing indices together: $x,w$ are contracted into $a$, and $y,z$ are contracted into $b$.
\[
\begin{adjustbox}{max width=\linewidth}
\begin{prooftree}
\infer0[\Rulehyp]{[\Theta \mid \Delta,p:A,r:A \mid \Gamma]\ \Phi & \vdash P(a,\n a,p,\n r) \Rightarrow P(\n a,a,\n p,r)}
\infer1[\Rulele]{[\Theta,z:A \mid \Delta,r:A \mid \Gamma,w:A]\ z \le w, \Phi & \vdash P(a,\n a,w,\n r) \Rightarrow P(\n a,a,z,r)}
\infer1[\Rulele]{[\Theta,x,z:A \mid \Delta \mid \Gamma,y,w:A]\ x \le y \land z \le w, \Phi & \vdash P(a,\n a,w,x) \Rightarrow P(\n a,a,z,y)}
\infer1[\Ruleimpl]{[\Theta,z:A \mid \Delta,x,w:A \mid \Gamma,y:A]\ \n x \le y \land z \le \n w, P(\n a,a,w,x), \Phi & \vdash P(\n a,a,z,y)}
\infer1[\Rulereindex]{[\Theta \mid \Delta,a:A,b:A \mid \Gamma]\ \n a \le b \land \n b \le a, P(\n a,a,a,\n a), \Phi & \vdash P(\n a,a,\n b,b)}
\infer1[\Rulecut]{[\Theta \mid \Delta,a:A,b:A \mid \Gamma]\ \n a \le b \land \n b \le a, \Phi & \vdash P(\n a,a,\n b,b)}
\end{prooftree}
\end{adjustbox}
\]
where in the last step we cut with the hypothesis $[\Theta \mid \Delta,z:A \mid \Gamma]\ \Phi \vdash P(\n z,z,z,\n z)$.

We use $\Ruleimpl$ and $\Rulele$ just for clarity purposes, to highlight the natural occurrences of variables (instead of using $\Rulelefull$). Note that, following the polarity of $P$, $x \le y$ is used to transport covariantly (as in $\Ruleleplus$) and $z \le w$ is used contravariantly (as in $\Ruleleminus$).
\end{proof}

\section{Details for \Cref{thm:main_theorem} for doctrinal semantics}

\label{appendix:internal_language_correspondence}
\begin{thm}[Internal language correspondence, proof]
The constructions $\Syn$ and $\Lang$ in \Cref{thm:main_theorem} form a (bi-)adjunction between the categories of theories and directed doctrines, i.e., there are equivalences of categories between the above category and the set below for any theory $\Sigma$ and directed doctrine $\cP : \ndp{\C}^\op \> \Pos$:
\[
\begin{prooftree}
  \hypo{\Syn(\Sigma) \longrightarrow \cP \text{ in $\mathsf{DDoctrine}$}}
  \infer[double]1{\Sigma \longrightarrow \Lang(\cP) \text{ in $\Theory$}}
\end{prooftree}
\]
Moreover, we prove that $\Lang \< \Syn \cong \id_{\mathsf{DDoctrine}}$.
\end{thm}
\begin{proof}
  We start by describing the two constructions, focusing in particular on the case of base predicates, since it is where we need \Cref{def:no_dinat_variance}.
  \begin{itemize}[leftmargin=1em]
    \item{(\emph{Construction $\Downarrow$}).} One evaluates the doctrine morphism \[{(F,\alpha) : \Syn(\Sigma) \longrightarrow \cP}\] at the judgements involving $\Sigma$ to obtain the actions of the following theory morphism:
    \begin{itemize}[leftmargin=1em]
    \item $b(\sigma) := F_0(\sigma \type \in \Obj(\Syn(\Sigma)))  \in (\Obj(\C) \equiv \Lang(\cP)_B).$
    \item $f(\sigma) := F_1(\sem{\dom(\sigma)}_b \vdash \sigma(\id) : \sem{\cod(\sigma)}_b) \in \Lang(\cP)_F$. For the induced interpretation one has that ${\sem{A}_b = F(A)}$, $\sem{t}_f = F(t)$ hold by induction on $A,t$, since $F_0,F_1$ pre\-ser\-ve products, implications, etc., in each inductive step.
    \item Term equality condition: given $e \in \Sigma_E$, its associated base equality holds in $\Syn(\Sigma)$ therefore the terms/arrows are equal. Functors $F$ preserve equalities of arrows into $\cP$, hence their $\sem{-}_f$ is equal in $\Lang(\cP)_E$ by definition.
    \item For predicates, assume we are given a predicate $\sigma \in \Sigma_F$; we pick a corresponding predicate $p(\sigma) \in \Lang(\cP)_F$ as follows: \[{p(\sigma) := (F(\mathsf{pos}(\sigma)),F(\top),F(\mathsf{neg}(\sigma)),\tilde p)}\in \Lang(\cP)_F,\]
    where the point
    \[\tilde p :=\alpha(\sigma(\pi_1,\pi_1) \propx) \in {\cP(F(\mathsf{pos}(\sigma)) \mid F(\top) \mid F(\mathsf{neg}(\sigma)))_0}\] is obtained by applying $\alpha$ to the base case $P(s \mid t) \propx$, and picking $\Delta := \top$ and the syntactic terms $s,t$ as
    \[
    \begin{array}{ll}
      {\sem{\textsf{neg}(\sigma)}_b, \top \vdash s := \pi_1 : \sem{\textsf{neg}(\sigma)}_b}, \\
      {\sem{\textsf{pos}(\sigma)}_b, \top \vdash t := \pi_1 : \sem{\textsf{pos}(\sigma)}_b}.
    \end{array}\]
    Note that the choice above corresponds to $\contract{\Delta}$ in \Cref{def:no_dinat_variance}.
    Moreover, \[
    \begin{array}{ll}
      {\textsf{pos}(\tilde p) := F(\mathsf{pos}(\sigma)) \times F(\top) \iso F(\textsf{pos}(\sigma)) = \sem{\textsf{pos}(\sigma)}_b}, \\
      {\textsf{neg}(\tilde p) := F(\mathsf{neg}(\sigma)) \times F(\top) \iso F(\textsf{neg}(\sigma)) = \sem{\textsf{neg}(\sigma)}_b}.
    \end{array}\]
    \item For axioms, we apply monotonicity of $\alpha$ on the relation $\mathsf{hyp}(a) \vdash \mathsf{conc}(a)$ in $\Syn$, which holds by the axiom case for $\sigma \in \Sigma_F$. From this we obtain that the desired $\le$ relation in $\cP$, which is exactly how $\Lang(\Sigma)_A$ was defined.
    \end{itemize}
    \item{(\emph{Construction $\Uparrow$}).} Each component of the doctrine morphism $(\sem{-},\sem{-}_\varphi)$ is given by induction on derivations, using the theory morphism $M := (b,p,f) : \Sigma \to \Lang(\cP)$ for the base cases and the structure of $\cP$ for the inductive steps. The functor $\sem{-} : \set{A \type}_\Sigma \to \C_0$ is defined as:
    \begin{itemize}[leftmargin=1em]
    \item $\sem{-}_0$ acts by induction on types, using the product functor ${{-\times-}} : \Cop\x\C\>\C$ to interpret product types, etc.; for base types ${A \in \Sigma_B}$ one uses the action $M_b(A) \in \C_0$.
    \item $\sem{-}_1$ acts by induction on terms, similarly as above. Functoriality is ensured by a substitution lemma.
    \item The components of the natural transformation \[{\sem{\Theta\mid\Delta\mid\Gamma}_\varphi : \Syn(\Theta\mid\Delta\mid\Gamma) \> \cP(\sem{\Theta}\mid\sem{\Delta}\mid\sem{\Gamma})}\] are the functors which, on objects, act by induction on formula derivations in $\Syn$ and use the structure of $\cP$, e.g., for $\Conjunction$, implication $\Implication$, and directed equality in $\cP$, given in \Cref{def:sem:directed_doctrine}, in order to combine objects in the inductive step.

    For the base judgement $\sigma(s \mid t) \propx$ with two terms $\Theta,\Delta \vdash s : {\mathsf{neg}(\sigma)}$,  $\Gamma,\Delta \vdash s : {\mathsf{pos}(\sigma)}$, first we consider
    \[p(\sigma \in \Sigma_P) \in \cP(\Theta' \mid \Delta' \mid \Gamma'),\] and then recover the ``original'' non-collapsed predicate using the no-dinatural-variance condition,
    \[
    \begin{array}{r@{\,}l}
      \eps(p(\sigma))\in\cP(& \Theta'\times\Delta'=: \mathsf{neg}(p(\sigma)) = \sem{\mathsf{neg}(\sigma)} \\
       \mid & \top \\
       \mid & \Gamma'\times\Delta'=: \mathsf{pos}(p(\sigma)) = \sem{\mathsf{pos}(\sigma)}),
    \end{array}\]
    where the last two equalities
    \[
    \begin{array}{r@{\,}l}
      \forall \sigma, \mathsf{neg}(p(\sigma)) & = \sem{\mathsf{neg}(\sigma)}, \\
      \forall \sigma, \mathsf{pos}(p(\sigma)) & = \sem{\mathsf{pos}(\sigma)},
    \end{array}\]
    come as part of the conditions of morphism of theories from \Cref{def:theory_morphism}.

    Finally, we send $\sigma(s \mid t) \propx$ to the predicate obtained by substituting the (interpretation of) terms as follows,
    \[\begin{adjustbox}{max width=\linewidth}$
      \sem{\sigma(s \mid t) \propx}_{\varphi} := \cP(\sem{s}_f \mid {!}_\Delta \mid \sem{t}_f)(\varepsilon(p(\sigma)))\in\cP(\sem{\Theta}\mid\sem{\Delta}\mid\sem{\Gamma}).$\end{adjustbox}\]
using the following reindexing (omitting semantic brackets on contexts):
 \[
  \begin{array}{r@{\,}l@{\,}ll}
    (\sem{s}_f \mid {!_\Delta} \mid \sem{t}_f) & : & \nowidth{(\Theta \mid \Delta \mid \Gamma) \to (\Theta' \times \Delta' \mid \top \mid \Gamma' \times \Delta')} & \\
    := & ( & \sem{s}_f & : \takespace{\Theta \times \Delta}{\Theta \times \Delta} \to (\sem{\textsf{neg}(p)} := \Theta' \times \Delta'),\\
       & \mid & {!_\Delta} & : \takespace{\Theta \times \Delta}{\Delta} \to \top\\
       & \mid & \sem{t}_f & : \takespace{\Theta \times \Delta}{\Gamma \times \Delta} \to (\sem{\textsf{pos}(p)} := \Gamma' \times \Delta').
  \end{array}
\]

    Naturality is a semantic substitution lemma, which follows by induction.
    \item The action on morphisms (entailments) is given by interpreting each rule with the properties of $\cP$.
    \end{itemize}
  \end{itemize}
  We show the equivalence with details for base predicates.
  \begin{itemize}[leftmargin=1em]
      \item \emph{(${{\Downarrow}\<{\Uparrow}={\id}}$).}
      Suppose we are given a morphism of doctrines $(F,\alpha)$; recall that we construct the component on predicates for the corresponding morphism of theories as
\[
\begin{array}{l}
  b(\sigma) := \alpha(\sigma \type) \\
  p(\sigma) := (F(\mathsf{pos}(\sigma)),\top,F(\mathsf{neg}(\sigma)),\alpha(\sigma(\pi_1,\pi_1) \propx))
\end{array}
\] by applying $F$ on the base predicate formula.

In the other direction, recall that in general the constructed doctrine morphism $(\sem{-},\sem{-}_\varphi)$ sends judgements $\sigma(s \mid t) \propx$ to \[\cP(\sem{s}_f \mid {!}_\Delta \mid \sem{t}_f)(\eps(p(\sigma \in \Sigma_P))),\]
where $p(\sigma) := p(\sigma)_4 \in \cP(p(\sigma)_1 \mid p(\sigma)_2 \mid p(\sigma)_3)$.

To show that we end up with the original doctrine we need to show that $\alpha$ has the same action on predicates. In particular, it suffices to prove it for the base cases (since the rest of the structure is fixed and pertains to the structure of the doctrine).  Hence, we need to show
 \[
  \alpha(\sigma(s \mid t) \propx) = \cP(\sem{s}_f \mid {!}_\Delta \mid \sem{t}_f)(\eps(\alpha(\sigma(\pi_1,\pi_1) \propx))).
 \]

 The key idea is that $\varepsilon$ in this context corresponds to the identity functor, since we are given a predicate $\alpha(\sigma(\pi_1,\pi_1))$ in the fiber $\cP(\Theta \times \Delta \mid \top \mid \Gamma \times \Delta)$ which already picks the dinatural context to be $\top$. The desired equality
 \[
  \alpha(\sigma(s \mid t) \propx) = \cP(\sem{s}_f \mid {!}_\Delta \mid \sem{t}_f)(\alpha(\sigma(\pi_1,\pi_1) \propx))\]
  follows from the fact that $\alpha$ is natural, i.e., it commutes with reindexings. In particular, reindexing with $\Syn$ corresponds precisely with substitution of the terms in the formula, obtaining $\alpha(\sigma(s \mid t) \propx)$.

\item \emph{(${{\Uparrow}\<{\Downarrow}\!=\!{\id}}$).}
Suppose we are given a theory morphism $M := (b,f,p)$; recall that we construct $\sem{-}$ and the syntactic doctrine by induction, defining $\sem{-}_\varphi$ on the base cases $\sigma(s \mid t) \propx$ to \[\sem{\sigma(s \mid t) \propx}_\varphi :=
\cP(\sem{s}_f \mid {!} \mid \sem{t}_f)(\eps(p(\sigma \in \Sigma_P)))
\]
where \[
\begin{array}{r@{\,}l}
  p(\sigma) \in & \cP(p(\sigma)_1 \mid p(\sigma)_2 \mid p(\sigma)_3),\\
  \varepsilon(p(\sigma)) \in & \cP(\sem{\mathsf{pos}(\sigma)} \mid \top  \mid  \sem{\mathsf{neg}(\sigma)}) \\
  \iso & \cP(p(\sigma)_1 \times p(\sigma)_2 \mid  \top  \mid  p(\sigma)_3 \times p(\sigma)_2).\end{array}\]
Recall that the theory morphism reconstructed from the above is given in general by
\[
p'(\sigma) := (F(\mathsf{pos}(\sigma)),\top,F(\mathsf{neg}(\sigma)),\alpha(\sigma(\pi_1,\pi_1) \propx)).
\]
In our case $\alpha$ is $\sem{-}_{\varphi}$, hence
\[
\begin{array}{l}
p'(\sigma) := (\sem{\mathsf{pos}(\sigma)}  \mid  \top  \mid  \sem{\mathsf{neg}(\sigma)}, \\
\takespace{p'(\sigma) := a aaaaaaaaaaaaaaaaaaaaaaaaaa}{\cP(\sem{\pi_1} \mid {!} \mid \sem{\pi_1})(\eps(p(\sigma \in \Sigma_P)))},
\end{array}
\]
where the reindexing $\sem{\pi_1} : \sem{\mathsf{neg}(\sigma)} \times \top \> \sem{\mathsf{neg}(\sigma)}$ applied to to $\varepsilon(p(\sigma))$ is precisely the one described in \Cref{def:no_dinat_variance}, hence we immediately obtain that $p'(\sigma) = p(\sigma)$.

\emph{($\Lang \< \Syn \cong \id_{\mathsf{DDoctrine}}$).} Finally, we show that there is an equivalence between a directed doctrine $\cP$ and the resulting doctrine $\Syn(\Lang(\cP))$.
The structure followed here is precisely the same as that of the 2-adjunction shown above; we describe the proof explicitly here for clarity on how the \emph{ndv} condition is used.

Since the equivalence is relatively standard, we only sketch the main detail of the proof at the level of predicates to justify the need for the \emph{ndv} condition from \Cref{def:no_dinat_variance}. In particular, we will leave implicit the equivalence between the base category $\C$ and the base category $\Ctx_\Sigma$ of the resulting doctrine $\Syn(\Lang(\cP)) : \ndp{\textsf{Ctx}_\Sigma}^\op \> \Pos$, and instead focus on showing that the posetal fibers of $\cP$ are equivalent to those of $\Syn(\Lang(\cP)).$

We first prove that for every $(\Theta \mid \Delta \mid \Gamma) \in \textsf{Ctx}_\Sigma \iso \C$ the poset $\Syn(\Lang(\cP))(\Theta \mid \Delta \mid \Gamma)$ and the poset $\cP(\Theta \mid \Delta \mid \Gamma)$ are isomorphic \emph{as sets}, since it is where we use \emph{ndv}. The equivalence of posets in the proper sense follows from the fact that the definition of syntactic doctrine and that of underlying language capture (and reconstruct) \emph{all} of the $\le$ relations of the original doctrine.

\begin{itemize}
\item First, we show that given a predicate $p \in \cP(\Theta \mid \Delta \mid \Gamma)$, we show that there exists a predicate $\varphi  \in \Syn(\Lang(\cP))(\Theta \mid \Delta \mid \Gamma)$ which is equivalent to the former.

By definition of underlying theory, the theory $\Lang(\cP)$ will have a corresponding predicate symbol $p \in \Sigma_P$ with $\textsf{neg}(p) := \Theta \times \Delta$ and $\textsf{pos}(p) := \Gamma \times \Delta$. Hence we pick the syntactic predicate \[\varphi_p := p(\textsf{id}_{\Theta\times\Delta},\textsf{id}_{\Gamma\times\Delta}) \in \Syn(\Lang(\cP))(\Theta \mid \Delta \mid \Gamma)\] in the syntax of dFOL, where
\[
\begin{array}{l@{\,}l}
\takespace{\Theta}{\Gamma} \times \Delta \vdash \textsf{id}_{\Gamma \times \Delta} & : (\textsf{neg}(p) := \takespace{\Theta}{\Gamma} \times \Delta) \\
{\Theta} \times \Delta  \vdash \textsf{id}_{\Theta \times \Delta} & : (\textsf{pos}(p) := {\Theta} \times \Delta)
\end{array}
\] are the projection terms in the syntax.

\item In the opposite direction, given a predicate $\varphi \in \Syn(\Lang(\cP))(\Theta \mid \Delta \mid \Gamma)$ we obtain a predicate of $\cP$ by interpreting $\varphi$ recursively, i.e., applying the structure of a directed doctrine for the connectives, and in the base cases we interpret the terms and (pre)compose them to the interpretation of base predicates which appear in $\varphi$ (which exists because we use $\Lang(\cP)$ as the underlying theory).

In detail, the base case $\varphi := p(s \mid t)$ for terms $(\Theta, \Delta \vdash s : \textsf{neg}(p))$, $(\Gamma, \Delta \vdash t : \textsf{pos}(p))$ and some base predicate $p \in \Sigma_{\cP}$ is obtained as follows: since $p$ comes from the signature $\textsf{Lang}(\cP)$ it hence corresponds to some element $p \in \cP(\Theta' \mid \Delta' \mid \Gamma')$ and is such that $\textsf{neg}(p) := \Theta' \times \Delta'$ and $\textsf{pos}(p) := \Gamma' \times \Delta'$.
We construct the following predicate in $\cP(\Theta \mid \Delta \mid \Gamma)$: first, we use $\varepsilon$ from \Cref{def:no_dinat_variance} to obtain an element $\varepsilon(p) \in \cP(\Theta' \times \Delta' \mid \top \mid \Gamma' \times \Delta')$ and then apply the doctrine reindexing to obtain
\[\sem{\varphi := p(s \mid t)} := \cP(\sem{s} \mid {!_\Delta} \mid \sem{t})(\varepsilon(p)) \in \cP(\Theta \mid \Delta \mid \Gamma),\]
where the reindexing is defined as follows, using the interpretation of terms in the semantics:
 \[
  \begin{array}{r@{\,}l@{\,}ll}
    (\sem{s} \mid {!_\Delta} \mid \sem{t}) & : & \nowidth{(\Theta \mid \Delta \mid \Gamma) \to (\Theta' \times \Delta' \mid \top \mid \Gamma' \times \Delta')} & \\
    := & ( & \sem{s} & : \takespace{\Theta \times \Delta}{\Theta \times \Delta} \to (\textsf{neg}(p) := \Theta' \times \Delta'),\\
       & \mid & {!_\Delta} & : \takespace{\Theta \times \Delta}{\Delta} \to \top\\
       & \mid & \sem{t} & : \takespace{\Theta \times \Delta}{\Gamma \times \Delta} \to (\textsf{pos}(p) := \Gamma' \times \Delta').
  \end{array}
\]

\end{itemize}
We prove that the above constructions are mutually inverse.
\begin{itemize}
\item Given a point $p \in \cP(\Theta \mid \Delta \mid \Gamma)$, we obtain a formula \[\varphi_p := {p}(\pi_1,\pi_1) \in \Syn(\Lang(\cP))(\Theta \mid \Delta \mid \Gamma)\]
as above. Then, we interpret $\varphi_p$ using the case of base predicates, i.e., $\sem{s} := \pi_1$ and $\sem{t} := \pi_1$ and the resulting predicate in $\cP$ is given by
\[\sem{{p}(\pi_1,\pi_1)} := \cP(\pi_1 \mid {!_\Delta} \mid \pi_1)(\varepsilon(p)).\] But the statement that $\cP(\pi_1 \mid {!_\Delta} \mid \pi_1)(\varepsilon(p)) = p$ is precisely the \emph{ndv} condition given in \Cref{def:no_dinat_variance}, i.e., the fact that $\varepsilon$ is a bijection.

\item Given a formula $\varphi \in \Syn(\Lang(\cP))(\Theta \mid \Delta \mid \Gamma)$, we obtain a predicate of $\cP$ by interpreting its structure recursively, i.e., interpreting the terms which compose it and (pre)composing them to the interpretation of base predicates which appear in $p'$ (which exists because we use $\Lang(\cP)$ as the underlying theory). Since $\sem{\varphi}$ is defined by induction, the proof similarly follows by induction. In the base case $\varphi := P(s \mid t)$, we obtain an interpretation
\[\sem{\varphi := P(s \mid t)} := \cP(\sem{s} \mid {!_\Delta} \mid \sem{t})(\varepsilon(P)) \in \cP(\Theta \mid \Delta \mid \Gamma),\]
for which the original formula is reconstructed in the syntax as
\[\varphi_p := (\cP(\sem{s} \mid {!_\Delta} \mid \sem{t})(\varepsilon(P)))(\textsf{id}_{\Theta\times\Delta},\textsf{id}_{\Gamma\times\Delta}).\]

One then shows that this formula, although not syntactically equal to the original one $P(s \mid t)$, is equivalent to it in the syntax: \[P(s \mid t) \iff (\cP(\sem{s} \mid {!_\Delta} \mid \sem{t})(\varepsilon(P)))(x,\n y,y,z)\] and hence these are equal in the \emph{poset} $\Syn(\Lang(\cP))$ (i.e., where equivalent objects are identified with the same point). In particular, this follows by noticing that both of these have the same semantics, i.e., are equal in the semantics of the doctrine, and hence are equivalent in the syntax because all $\le$ relations in $\cP$ (i.e., including equivalences) are captured as axioms in the syntax.

\end{itemize}
\end{itemize}

\end{proof}

\section{Details for completeness in $\CPreord$ of classical directed first-order logic}
\label{appendix:classical_completeness}

\begin{defi}[Henkin extension]
Take a theory $(\Sigma_L,\Sigma_A)$ and consider the set $\textsf{OneVarF} := \set{[\,\emptyctx\!\mid\!\emptyctx\!\mid\!\emptyctx\,], [x\!:^p\!A]\,\varphi(x)\!\propx \mid p \in \set{-,\Delta,+}}$. Then, we define the \emph{Henkin extension} as the following theory $\Sigma^* := (\Sigma_L^*,\Sigma_A^*)$:
\[\begin{array}{r@{}l}
  \Sigma^* := (&\takespace{\Sigma_A}{\Sigma_L} \cup \set{\takespace[l]{\exists^{p} x.\varphi(x) \Rightarrow \varphi(c_\varphi)}{c_\varphi} \mid \varphi \in \textsf{OneVarF}}, \\ & \Sigma_A \cup \set{\exists^{p} x.\varphi(x) \Rightarrow \varphi(c_\varphi) \mid \varphi \in \textsf{OneVarF}}).
\end{array}\]
\end{defi}

\begin{thm}[Henkin extension is conservative]
$(\Sigma_L^*,\Sigma^*_A)$ is conservative over $\Sigma$, in the sense that every $\varphi \in \Sigma^*_A$ which does not use the new constant symbols is already contained in $\Sigma_A$.
\end{thm}
\begin{proof}
Immediate by induction on syntactic derivations of the form $\Phi, (\exists^p x.\varphi(x) \Rightarrow \varphi(c_\varphi)) \vdash \psi$.
\end{proof}

\begin{lem}[Henkin completion is Henkin and equiconsistent]\label{lem:henkin_completion}
The Henkin completion $\Sigma^\omega$ of some $\Sigma$ is a Henkin theory. Moreover, if $\Sigma$ is consistent (i.e., $\bot \notin \Sigma_A$) then $\Sigma^\omega$ is consistent.
\end{lem}
\begin{proof}
To show that $\Sigma^\omega$ is Henkin, consider some formula $\psi := \exists^p x.\varphi(x) \in \Sigma^\omega$. Then $\psi \in \Sigma^n_A$ for some $n$, and since its Henkin axiom is added at step $n+1$ it must also be present at step $\omega$. For consistency, suppose by contradiction that $\bot \in \Sigma^\omega_A$: then $\bot \in \Sigma^n_A$ for some $n$, which is conservative over the original $\Sigma$; since each Henkin extension preserves conservativity, contradiction.
\end{proof}

\begin{lem}[Maximally consistent exts. preserve Henkin]
If $\Sigma$ is Henkin, then $\Sigma^{\textsf{max}}$ is again Henkin.
\end{lem}
\begin{proof}
If $\exists^p x.\varphi(x) \in \Sigma_A$, then still $\exists^p x.\varphi(x) \in \Sigma^{\textsf{max}}_A$.
\end{proof}

\begin{thm}[Model existence lemma]
If a theory $\Sigma$ is consistent, then there is a model $\Syn(\Sigma) \longrightarrow  \CPreord$ of $\Sigma$.
\end{thm}
\begin{proof}
We refer to \Cref{model_existence_lemma} for the definition of the model.
We prove that $1 \le \sem{\varphi} \Rightarrow [\,\emptyctx \mid \emptyctx \mid \emptyctx\,] \vdash \varphi$ by induction on $\varphi$:
\begin{itemize}[]
  \item If $\varphi$ is either $s \le_A t$ or $P(s \mid t)$, then there is nothing to do since we precisely defined $\le_A$ and $\sem{P}$ to coincide with derivability.
  \item $\varphi := \top$: trivial.
  \item $\varphi := \bot$: impossible since $1 \not \le 0$.
  \item $\varphi := \varphi_1 \land \varphi_2$: \[
  \begin{array}{r@{\,}l@{\ \ \ \ }l}
         & 1 \le \sem{\varphi_1 \land \varphi_2} & \\
    \iff & 1 \le \sem{\varphi_1} \text{ and } 1 \le \sem{\varphi_2} & \text{(i.h.)} \\
    \iff & [...] \vdash \varphi_1 \text{ and } [...] \vdash \varphi_2 & \\
    \iff & [...] \vdash \varphi_1 \land \varphi_2 & \\
  \end{array}
  \]
  \item $\varphi := \varphi_1 \lor \varphi_2$: \[
  \begin{array}{r@{\,}l@{\ \ \ \ }l}
         & 1 \le \sem{\varphi_1 \lor \varphi_2} & \\
    \iff & 1 \le \sem{\varphi_1} \text{ or } 1 \le \sem{\varphi_2} & \text{(i.h.)} \\
    \iff & [...] \vdash \varphi_1 \text{ or } [...] \vdash \varphi_2 & \\
    \iff & [...] \vdash \varphi_1 \lor \varphi_2 & \\
  \end{array}
  \]
  \item $\varphi := \varphi_1 \Rightarrow \varphi_2$: \[
  \begin{array}{r@{\,}l@{\ \ \ \ }l}
         & 1 \le \sem{\varphi_1 \Rightarrow \varphi_2} & \\
    \iff & 1 \le (\sem{\varphi_1} \le_\I \sem{\varphi_2}) & \\
    \iff & 1 \le \sem{\varphi_1} \text{ implies } 1 \le \sem{\varphi_2} & \text{(i.h.)} \\
    \iff & [...] \vdash \varphi_1 \text{ implies } [...] \vdash \varphi_2 & \text{$(*)$} \\
    \iff & [...] \vdash \varphi_1 \Rightarrow \varphi_2 & \\
  \end{array}
  \]
  In the last $(*)$ step, we bottom-to-top direction follows by assuming $[...] \vdash \varphi_1 \Rightarrow \varphi_2$ and $[...] \vdash \varphi_1$ one derives $[...] \vdash \varphi_2$ using $\Rulecut$. In the other top-to-bottom direction, we use the fact that $\mathcal{S}$ is maximally consistent; there are two cases:
  \begin{enumerate}
    \item if $\varphi_2 \in \mathcal{S}$ then clearly $[...] \vdash \varphi_2$, and we are done since $[...] \vdash \varphi_1 \Rightarrow \varphi_2$ is derivable by ignoring the hypothesis.
    \item if $\varphi_2 \not \in \mathcal{S}$ then $\varphi \Rightarrow \bot \in \mathcal{S}$ and hence $[...] \vdash \varphi_1 \Rightarrow \bot$ is derivable; since $[...] \vdash \bot \Rightarrow \varphi_2$ we obtain $[...] \vdash \varphi_1 \Rightarrow \varphi_2$ by $\Rulecut$.
  \end{enumerate}
  \item $\varphi := \exists^p x. \varphi'(x)$:
    \begin{itemize}
      \item[($\Rightarrow$)] Assume that $1 \le \sem{\exists^p x. \varphi'(x)}$; then there exists some $v \in \sem{A}$ such that $1 \le \sem{\varphi'}(v)$, hence $1 \le \sem{\varphi'(v)}$ by a simple induction on formulas. By i.h., we have that $[\,\emptyctx \mid \emptyctx \mid \emptyctx\,] \vdash \varphi'(v)$; by $\Ruleexists$ we obtain $[\,\emptyctx \mid \emptyctx \mid \emptyctx\,]\ \vdash \exists^p x.\varphi'(x)$.
      \item[($\Leftarrow$)] Assume that $[\,\emptyctx \mid \emptyctx \mid \emptyctx\,] \vdash \exists^p x. \varphi'(x)$; then by the fact that we are working in a Henkin theory we have that $[\,\emptyctx \mid \emptyctx \mid \emptyctx\,]\ \vdash \exists^p x. \varphi'(x) \Rightarrow \varphi'(c_{\varphi'})$, hence $[\,\emptyctx \mid \emptyctx \mid \emptyctx\,]\ \varphi'(c_{\varphi'})$ is derivable using the rules of $\Ruleimpl$. Then, by inductive hypothesis $1 \le [\,\emptyctx \mid \emptyctx \mid \emptyctx\,]\ \varphi'(c_{\varphi'})$, and hence $1 \le \sem{\varphi'}(\sem{c_{\varphi'}})$; finally, we obtain that $1 \le \sem{\exists^p x. \varphi'(x)}$ since exists is defined using lub.
    \end{itemize}

  \item $\varphi := \forall^p x. \varphi'(x)$: follows from the previous case for exists and the fact that in the classical case $\forall$ can be defined in terms of $\exists,\bot$, and $\Rightarrow$, following \Cref{classical_quantifiers}.
    \end{itemize}

\end{proof}

Before proving the main completeness theorem we need a simple deduction theorem in the style of~\cite[2.7.3.2]{Mimram2020program}.
\begin{thm}[Deduction theorem for \text{dFOL}]\label{deduction_theorem}
Consider a theory $\Sigma := (\Sigma_L, \Sigma_A)$ and let $\Sigma' := (\Sigma_L, \Sigma_A \cup \set{\psi})$ for some closed formula $\psi$. If an entailment $[\Theta \mid \Delta \mid \Gamma]\ \Phi \vdash_{\Sigma'} \varphi$ is derivable in $\Sigma'$, then $[\Theta \mid \Delta \mid \Gamma]\ \Phi \vdash_\Sigma \psi \Rightarrow \varphi$ is also derivable in $\Sigma$.
\end{thm}
\begin{proof}
Straightforward proof by induction on derivations.
\end{proof}

\begin{thm}[$\CPreord$ completeness]
For any theory $\Sigma$, if an entailment $[\Theta \mid \Delta \mid \Gamma]\ \Phi \vdash \varphi$ holds in every model $\Syn(\Sigma) \longrightarrow \CPreord$ of $\Sigma$, then it is derivable in classical dFOL.
\end{thm}
\begin{proof}
The model existence lemma in \Cref{model_existence_lemma} states that there exists a model $M : \Syn(\mathcal{S}) \longrightarrow \CPreord$ for every consistent theory $\mathcal{S}$, where in particular consistency is defined as $[\,\emptyctx \mid \emptyctx \mid \emptyctx\,]\not\vdash_\mathcal{S} \bot$ using $\mathcal{S}$ as theory.
We will use its contrapositive, stating that if there are no models of $\mathcal{S}$ then $[\,\emptyctx \mid \emptyctx \mid \emptyctx\,] \vdash_\mathcal{S} \bot$.

Suppose that the above entailment is captured by a closed formula $\psi$, using \Cref{thm:closed_formulas}, and assume that $\psi$ is derivable in every model of the theory $\Sigma$.

Then, the theory $\Sigma \cup \set{\neg \psi}$ defined by adding $\neg \psi$ as axiom to $\Sigma_A$ has no models, since all models with $\Sigma$ already derive $\psi$.

By the contrapositive of the model existence lemma, \[[\,\emptyctx \mid \emptyctx \mid \emptyctx\,] \vdash_{\Sigma \cup \set{\neg \psi}} \bot,\] and by the deduction theorem in \Cref{deduction_theorem} we obtain that $[\,\emptyctx \mid \emptyctx \mid \emptyctx\,]\ \vdash_\Sigma \neg \psi \Rightarrow \bot$; since we assume to be working in classical dFOL we have access to double negation elimination (c.f., \Cref{poset_completeness}), hence derive that $[\,\emptyctx \mid \emptyctx \mid \emptyctx\,] \vdash_\Sigma \psi$, as desired.
\end{proof}

\end{document}